\documentclass[11pt,a4paper]{article}

\usepackage{epsfig}
\usepackage[T1]{fontenc}    
\usepackage{graphics}
\usepackage{graphicx}
\usepackage{pstricks,pst-coil,pst-fill,pst-plot}
\usepackage[fleqn]{amsmath}    
\usepackage{amssymb,amsthm}    
\usepackage{amsfonts}   
\usepackage{verbatim}   
\usepackage{mathrsfs}   
\usepackage{dsfont}
\usepackage{enumerate}   
\newcommand{\be}[1]{\begin{equation}\label{#1}}
\newcommand{\ee}{\end{equation}}

\newcommand{\num}{\\\rule{0pt}{20pt}}
\newcommand{\numa}[1]{\\\rule{0pt}{#1pt}}
\newcommand{\dis}{\displaystyle}
\newcommand{\eq}[1]{(\ref{#1})}
\newcommand{\tr}{\mathop{\rm tr}}

\newtheorem{prop}{Proposition}[section]
\newtheorem{lemma}{Lemma}[section]

\newtheorem{Def}{Definition}[section]
{\theoremstyle{remark}
\newtheorem{rem}{Remark}[section]}

\def\Proof{\medskip\noindent {\it Proof --- \ }}

\let\qed=\cqfd
\makeatletter \@addtoreset{equation}{section} \makeatother

\newcommand\beq{\begin{equation}}
\newcommand\enq{\end{equation}}

\def\beqa{\begin{eqnarray}}
\def\eeqa{\end{eqnarray}}
\def\barr{\begin{array}}
\def\ea{\end{array}}

\def\det{\operatorname{det}}
\def\eps{\epsilon}
\def\ga{\gamma}
\def\la{\lambda}
\def\sg{\sigma}

\newcommand{\f}[2]{{\ensuremath{%
    \mathchoice%
    {\dfrac{#1}{#2}}
    {\dfrac{#1}{#2}}
    {\frac{#1}{#2}}
    {\frac{#1}{#2}}
}}}
\newcommand{\pa}[1]{\ensuremath{\left(#1\right)}}
\newcommand{\paa}[1]{\ensuremath{\left\{#1\right\}}}
\newcommand{\pac}[1]{\ensuremath{\left[#1\right]}}
\newcommand{\paf}[2]{\ensuremath{\left(\f{#1}{#2}\right)}}
\newcommand{\pab}[2]{\ensuremath{\pa{\barr{c} #1 \\ #2 \ea }}}

\def\eps{\epsilon}
\def\ga{\gamma}

\def\la{\lambda}
\def\sg{\sigma}
\def\Sg{\Sigma}
\def\om{\omega}

\def\wt{\widetilde}

\newcommand{\mc}[1]{\ensuremath{\mathcal{#1}}}
\newcommand{\mf}[1]{\ensuremath{\mathfrak{#1}}}
\newcommand{\msc}[1]{\ensuremath{\mathscr{#1}}}

\newcommand{\Int}[2]{\ensuremath{\int\limits_{#1}^{#2}}}

\newcommand{\sul}[2]{\ensuremath{\sum\limits_{#1}^{#2}}}
\newcommand{\pl}[2]{\ensuremath{\prod\limits_{#1}^{#2}}}

\newcommand{\Cx}{\ensuremath{\mathbb{C}}}

\let\tend=\rightarrow

\newcommand{\s}[1]{\ensuremath{\sinh\pa{#1}}}

\newcommand{\ex}[1]{e^{#1}}

\newcommand{\id}{\ensuremath{\mf{id}}}

\newcommand{\tbl}[2]{ \ensuremath{ \pa{\barr{c}  \paa{#1} \\ \paa{#2} \ea }  } }

\newcommand{\ddet}[2]{\ensuremath{\underset{#1}{\mathrm{det}}\pac{#2}}}

\newcommand{\abs}[1]{\ensuremath{\mid #1 \mid}}

\newcommand{\moy}[1]{\ensuremath{\langle #1 \rangle}}
\newcommand{\Mmoy}[1]{\ensuremath{\langle\hspace{-1mm}\langle #1 \rangle\hspace{-1mm}\rangle}}

\newcommand{\Dp}[1]{\ensuremath{\partial_{#1}}}

\newcommand{\e}[1]{\ensuremath{\mathrm{#1}}}
\newcommand{\dd}{\e{d} }

\begin{document}
\begin{flushright}
LPENSL-TH-12-08
\end{flushright}
\par \vskip .1in \noindent

\vspace{24pt}

\begin{center}
\begin{LARGE}
\vspace*{1cm}
  {Algebraic Bethe ansatz approach to the asymptotic
behavior of correlation functions }
\end{LARGE}

\vspace{50pt}

\begin{large}

{\bf N.~Kitanine}\footnote[1]{LPTM, CNRS UMR 8089, Universit\'e
de Cergy-Pontoise, France, kitanine@ptm.u-cergy.fr},~~
{\bf K.~K.~Kozlowski}\footnote[2]{ Laboratoire de Physique, 
CNRS UMR 5672, ENS Lyon,  France,
 karol.kozlowski@ens-lyon.fr},~~
{\bf J.~M.~Maillet}\footnote[3]{ Laboratoire de Physique, CNRS UMR 5672, ENS Lyon,  France,
 maillet@ens-lyon.fr},\\
{\bf N.~A.~Slavnov}\footnote[4]{ Steklov Mathematical Institute,
Moscow, Russia, nslavnov@mi.ras.ru},~~
{\bf V.~Terras}\footnote[5]{ Laboratoire de Physique, CNRS UMR 5672, ENS Lyon,  France, veronique.terras@ens-lyon.fr, on leave of
absence from LPTA, UMR 5207 du CNRS, Universit\'e Montpellier II}
\par

\end{large}

\vspace{80pt}

\centerline{\bf Abstract} \vspace{1cm}
\parbox{12cm}{\small  We describe a method to derive, from first principles, the long-distance asymptotic
behavior of correlation functions of integrable models in the
framework of the algebraic Bethe ansatz. We apply this approach to
the longitudinal spin-spin correlation function of the XXZ
Heisenberg spin-1/2 chain (with magnetic field) in the disordered
regime as well as to the density-density correlation function of the
interacting one-dimensional Bose gas. At leading order, the results
confirm the Luttinger liquid and conformal field theory predictions.
}
\end{center}

\newpage

\tableofcontents

\newpage

\section{Introduction\label{INT}}

The aim of this article is to derive the long-distance asymptotic
behavior of correlation functions of  integrable models ``from first
principles''. To reach this goal, we will use the framework of the Bethe ansatz \cite{Bet31,Orb58,Wal59,YanY66,YanY66a} in its algebraic version \cite{FadST79,FadT79}. For simplicity, we shall explain the main
features of our approach using the example of the XXZ
spin-$\frac{1}{2}$  Heisenberg model \cite{Hei28,Bet31}.
However, it will become apparent
that the method we propose here is quite general and applies, for
example, even to continuous integrable models like the interacting
one-dimensional Bose gas \cite{LieL63,Lie63}.

The starting point of our approach is an exact expression, that we called the master
equation representation, for the generating
function of the  two-point correlation functions of these models that we obtained in our previous works \cite{KitMST02a,KitMST05me,KitMST05k,KitKMST07}.
In the case of the XXZ spin-$\frac{1}{2}$
Heisenberg model (at non zero magnetic field and in the disordered
regime), this means that we will begin our investigation from the exact expressions for
the correlation functions on a finite lattice, and then take successively
the thermodynamic limit and  the large distance limit. The
long-distance asymptotic behavior that we obtain confirms, at the
leading order, the predictions given from bosonization \cite{LutP75,Hal80,Hal81a,Hal81b} and
conformal field theories \cite{Car84,Aff85,Car86,BloCN86}.

\subsection{Historical context: a brief survey}

Bethe ansatz and especially its algebraic versions
\cite{Bet31,Orb58,Wal59,YanY66,YanY66a,FadST79,FadT79,Bax82L,Gau83L,LieM66L,BogIK93L}
provide a powerful framework for the construction and the resolution
of a wide class of quantum integrable models in low dimension. The
central object of this approach is the $R$-matrix satisfying the
(cubic) Yang-Baxter equation and providing the structure constants
of the associated (quadratic) Yang-Baxter algebra for the operator
entries of the quantum monodromy matrix $T(\lambda)$. The representation
theory for this algebra leads to the construction of integrable
models operator algebras including in a natural way the Hamiltonian,
its associated commuting conserved charges generated by the
transfer matrix ${\cal T}(\lambda)=\tr T(\lambda)$ together with the
creation-annihilation operators determining their common spectrum.

The archetype of such models is provided by the XXZ Heisenberg
spin-$\frac{1}{2}$ chain in a magnetic field with periodic boundary
conditions \cite{Hei28}:

 \begin{equation}\label{0-HamXXZ}
 H=\sum_{k=1}^{M}\left(
 \sigma^x_{k}\sigma^x_{k+1}+\sigma^y_{k}\sigma^y_{k+1}
 +\Delta(\sigma^z_{k}\sigma^z_{k+1}-1)\right)-hS_z,
 \end{equation}
where
 \begin{equation}\label{0-Sz}
 S_z=\frac{1}{2}\sum_{k=1}^{M}\sigma^z_{k},\qquad
 [H,S_z]=0.
 \end{equation}
Here $\Delta$ is the anisotropy parameter, $h$ an external classical
magnetic field, and the length $M$ of the chain is assumed even. The
quantum space of states is ${\cal H}={\otimes}_{k=1}^M {\cal H}_k$,
where ${\cal H}_k\sim \mathbb{C}^2$ is called the local quantum
space at site $k$. The operators $\sigma^{x,y,z}_{k}$ act as the corresponding Pauli matrices in the space ${\cal H}_k$ and as the identity operator elsewhere.

The description of the full dynamical properties of such a model
amounts to the determination of its correlation functions and
associated structure factors. This was first achieved in the case
equivalent to free fermions, namely at  $\Delta = 0$ for the XXZ
chain. We stress that these free fermion results already required
tremendous efforts to get satisfactory answers, in particular for
the spin-spin correlation functions
\cite{McCW67,Mcc68,BarM71,MccBA71,MccW73,BarMTW76,McCTW77a,SatMJ79,SatMJ80,JimMMS80,McCPW81,ColIKT92,ColIKT93,ColIT97}.
Going beyond this free fermion point turned out to be very involved.
The first attempts to compute the correlation functions at arbitrary
anisotropy $\Delta$ came from the algebraic Bethe ansatz
\cite{IzeK85} (see also \cite{IzeK84} for the one-dimensional Bose
gas). They led however to rather implicit representations. This was
due to the complicated combinatorics involved in the description of
the Bethe states, in the action of the local operators on them and
in the computation of  their scalar products. The notion of dual
fields (which are quantum operators)  was introduced in \cite{Kor87}
to try to overcome such combinatorial difficulties. In this approach
(see \cite{BogIK93L} and references therein), the  correlation
functions are obtained in terms of expectation values (with respect
to the dual fields vacuum)  of Fredholm determinants depending on
several such dual fields. Unfortunately, these auxiliary quantum
fields can hardly be eliminated from the final answers, making the
use of such results quite problematic. More explicit representations
were obtained later on in terms of  multiple integrals for the
elementary blocks of the correlation functions; it uses a completely
different approach, working directly with infinite chains (hence
with several hypothesis), based on the representation theory of
quantum affine algebras and their associated $q$-vertex operators
\cite{JimMMN92,JimM96,JimM95L}. More algebraic representations of
these elementary blocks have been obtained recently along similar
lines together with a link to fermionic operator expressions of the
corresponding correlation functions
\cite{BooJMST06,BooJMST06a,BooJMST06b,BooJMST07}. Extensions of this
scheme to non zero temperature have been considered in
\cite{GohKS04,GohKS05}.

The resolution of the so-called quantum inverse scattering problem
for this model (namely the reconstruction of the local spin
operators in terms of the operator entries of the monodromy matrix)
opened the way to  the actual computation of the form factors (the
matrix elements of the local spin operators in the eigenstates
basis) and correlation functions \cite{KitMT99,KitMT00,MaiT00} in
the framework of the algebraic Bethe ansatz. With the help of previous
results (like determinant representations for the partition function with domain wall boundary conditions, and for the norms and scalar
products of Bethe states \cite{GauMcCW81,Kor82,IzeK84,Ize87,Sla89}),
this approach led to explicit determinant representations of the form
factors of the Heisenberg spin chains in a magnetic field as well as to
multiple integral expressions for the elementary blocks (or reduced density matrix) for their correlation functions in the thermodynamic limit. Remarkably, these
integral representations  coincide (in the zero
magnetic field limit) with those obtained previously directly
for the infinite chains using $q$-vertex operator methods and the
representation theory of the quantum affine algebras
\cite{JimMMN92,JimM96,JimM95L}.

These advances immediately raised the next challenge, namely the question of
obtaining manageable expressions for physical
correlation functions such as the spin-spin correlation functions.
This is a central problem in the field of  integrable models, on the
one hand from the viewpoint of effective applications of these
models to the realm of condensed matter physics (the associated dynamical
structure factors are measurable quantities), and on the other hand
concerning the more fundamental question of the long distance
asymptotic behavior of the correlation functions. While the practical determination of the dynamical structure factors can effectively  be addressed using the above form factors expressions in terms of determinants combined with powerful numerical techniques to sum up their corresponding series (see \cite{CauM05,CauHM05}),  answering the second question from purely analytical techniques looks like a fantastic theoretical problem. It amounts to understanding how the microscopic nearest neighbors
interactions, for example in Heisenberg spin chains,  integrate to
produce an effective long distance behavior of the spin correlations. In
particular, for models having no gap in the spectrum, the spin-spin
correlation functions are  believed to have, besides their possible
trivial constant value, a power law decay with the distance.
For generic integrable models, in particular not  equivalent to free
fermions, extracting such an asymptotic behavior ``from first
principles'' has been a challenge for many years.

In the case of fully interacting gapless models (like the XXZ
Heisenberg spin-$\frac{1}{2}$ chain for anisotropy $ -1 <\Delta <
1$), the first predictions for this power law behavior came from a
conjectured correspondence between such integrable lattice models with
nearest neighbors interactions  and a continuum theory having
long-range interactions, the Luttinger model. Using this hypothesis,
Luther and Peschel \cite{LutP75} succeeded to predict  the exact
value for the XXZ critical exponents at zero magnetic field as
functions of the anisotropy parameter $\Delta$. The above
correspondence was further enlighten by Haldane's development of the
Luttinger liquid concept, leading in turn to predictions for  the
critical exponents of the XXZ model in a magnetic field
\cite{Hal80,Hal81a,Hal81b}. The Bethe ansatz techniques were used
there to compute the parameters describing the low energy excitation
spectrum  of the XXZ model and to show that they indeed satisfy
relations that are characteristic of the  Luttinger liquid
universality class. This conjecture was further studied in a series
of papers \cite{BogIK86,BogIR87,IzeKR89} where, in particular, the
long-distance behavior of the longitudinal spin-spin correlation
function was computed  in perturbation to the second order in
$\Delta$ around the free fermion point $\Delta = 0$, with results in
agreement with the above predictions.

Another approach to this problem stemmed  from the hypothesis that
critical statistical systems with short range interactions should be
described by a conformal field theory. In this framework,  the
mapping of a conformal field theory defined on the plane into one
on a strip of finite width $l$ led to the determination of the
critical exponents in terms of the eigenvalues of the transfer
matrix along the strip \cite{Car84,Aff85,Car86,BloCN86}.
In particular, for large width $l$, the
central charge and the scaling dimensions of scaling operators of
the theory are given by the $l^{-1}$ behavior of respectively the
ground state and the different excited states energy levels. The
possibility to compute the finite size corrections to the spectrum
of integrable models in the framework of Bethe ansatz methods gave
rise to a prediction for the corresponding critical exponents, in
full agreement with the above bosonization approach \cite{DeVW85,Woy87,WoyE87, DesDV88,WoyET89, KluB90,KluBP91, KluWZ93,DesDV95}.

Further works in these directions (based however on several
conjectures) sharpened this picture by providing, in addition to the
critical exponents, predictions (at zero magnetic
field) for the amplitudes of the spin-spin correlation functions asymptotic behavior, which are  not directly accessible from the bosonization or conformal field theory approaches \cite{Luk98,Luk99, LukT03}.

\subsection{Overview of the method}

As already mentioned, the purpose of the present article is to
describe an effective method allowing us to derive  the
long-distance asymptotic behavior of the correlation functions
``from first principles''. We will use the algebraic Bethe ansatz
framework and our previous results \cite{KitMST02a,
KitMST05me,KitKMST07} giving an exact (master equation) formula for
the spin-spin correlation functions of a finite XXZ chain. Starting
from this formula, we will explain how to take the thermodynamic
limit and extract the large distance asymptotic behavior of the
correlation functions step by step from their lattice expressions.
As the master equation representation also applies to other
integrable models \cite{KitKMST07}, we will briefly describe how to
implement the XXZ related analysis to the computation of the
asymptotic behavior of the density-density correlation functions in
the interacting Bose gas in one dimension. Let us now  briefly
explain the main features of the  {\em master equation
representation} (we refer the reader to the original articles
\cite{KitMST05me,KitKMST07} for more details).

\medskip

In the case of physical correlation functions, the dependence on the
distance $m$ appears in the form of the $m^{th}$ power of the
elementary shift operator. This operator in its turn, is some
function of the transfer matrix of the integrable model at hand. The
first  step towards the master equation is  to consider an {\em
integrable deformation (twist) of the transfer matrix} (and hence of
the shift operator) depending on some continuous parameter $\kappa$.
This parameter is chosen in such a way that one recovers the usual
correlation function at say $\kappa = 1$, and such that the spectrum
of the twisted transfer matrix for $\kappa$ around some $\kappa_0$
is simple and disjoint from the one for $\kappa=1$. These properties
allow us to sum up exactly  the form factor type expansion of the
$\kappa$-deformed correlation function with respect to the
eigenstates of this $\kappa$-twisted transfer matrix (for any
$\kappa$ in some neighborhood of $\kappa_0$)  in terms of a single
multiple contour integral that we call the master equation
representation. The original correlation function one started with
(corresponding to the point $\kappa = 1$)  is then reconstructed at
the end of the computation from the knowledge of its values in an
open neighborhood of $\kappa_0$.

Originally, such a representation was first derived in
\cite{KitMST05me} for the {\em generating function of the
longitudinal spin-spin correlation functions} of the finite XXZ
chain.  This generating function is given by the ground state
average value of the operator $e^{\beta\mc{Q}_m}$ \cite{IzeK85},
where $\beta$ is an arbitrary complex number and
 \begin{equation}\label{0-Q1}
 \mc{Q}_m=\frac12\sum_{n=1}^{m}(1-\sigma_n^z)\, ,
 \end{equation}
so that we have
 \begin{equation}\label{0-corr-funct}
 \langle\sigma_1^z\sigma_{m+1}^z\rangle=2D^2_m\left.\frac{\partial^2}{\partial\beta^2}
 \langle e^{\beta
 \mc{Q}_m}\rangle\right|_{\beta=0}+2\langle\sigma^z\rangle-1\, ,
 \end{equation}
where the symbol $D^2_m$ stands for the second lattice derivative :
\begin{equation*}
D^2_m\, f(m)=f(m+1)+f(m-1)-2f(m).
\end{equation*}
Setting $\kappa=e^\beta$,  the operator $e^{\beta\mc{Q}_m}$ can be rewritten as
 \begin{equation}\label{0-defQ}
 e^{\beta \mc{Q}_m}
=
\prod_{n=1}^m\bigg(\frac{1+\kappa}2+\frac{1-\kappa}2\,
 \sigma_n^z\bigg)
\, ,
 \end{equation}
and, thanks to the solution of the quantum inverse problem, it is shown to be equal to the $m^{th}$-power of the $\kappa$-twisted transfer matrix multiplied by the $m^{th}$-power of the inverse of the usual transfer matrix. Hence it is the simplest example for which the master equation representation can be obtained.

Analogous representations were obtained for other correlation functions of the XXZ chain in \cite{KitMST05k}. Moreover, the time dependent case can even be considered along the same lines \cite{KitMST05d}. Later on, we also derived  similar representations for correlation functions of a wide class of integrable systems (including continuous theories) possessing the $R$-matrix of the six-vertex model \cite{KitKMST07}.

\medskip

One of the first problem to solve in this approach is that it is {\it a priori} not easy to take the thermodynamic limit directly in the master equation, for example for
$\langle e^{\beta \mc{Q}_m}\rangle$. This is due to the fact that, in this limit, the number of integration variables involved into the master equation tends to infinity with the size of the lattice. However, starting from this representation, one can obtain various
series expansions for the above master equation representation for the generating function. In these series representations the thermodynamic limit can be taken rather easily
\cite{KitMST05me,KitKMST07}.

In this article we give a {\em new expansion of
the master equation}, appropriate for the study of the long-distance
asymptotic behavior of the generating function $\langle e^{\beta
\mc{Q}_m}\rangle$. The key idea is to consider the master equation written  as a multiple contour integral such that the only poles of the integrand inside the contour are at the Bethe parameters characterizing the ground state. It turns out that the algebraic structure of these poles is given by the square of a Cauchy type determinant, in complete analogy with the free fermion case. In fact this feature is quite general and will appear for other correlation functions and other models as well, although with various minor modifications.

\medskip

Then, the next step of the asymptotic analysis also takes its roots in the free fermion case. There, the obtained series is the expansion of a
Fredholm determinant of an integral operator $I+V_0$ (see \eq{4-kernel}),
 \begin{equation}\label{0-FD-exp0}
 \langle e^{\beta
 \mc{Q}_m}\rangle_{{}_{\Delta=0}}
 =
 \sum_{n=0}^\infty\frac{1}{n!}
 \int\dd^n\lambda\;\det_n V_0(\lambda_j,\lambda_k)\, ,
 \end{equation}
its kernel $V_0$ depending on the distance $m$.
It is known  from the analysis of such kernel that the main contribution to the
$m\to\infty$ asymptotic behavior is generated by the traces of
powers of the kernel $V_0$,  $\tr(V_0^k)$.  Independently of the power $k$ involved,  every
trace behaves as $O(m)$. Therefore the natural idea would be to
reorder the series \eq{0-FD-exp0} in such a way as to obtain an
expansion with respect to $\tr(V_0^k)$. This can be easily done by
presenting the determinants in \eq{0-FD-exp0} as a sums over
permutations provided the latter are ordered with respect to their {\em cycles decomposition}
(see for example \cite{Bon04L}).

In the case of a general anisotropy parameter $\Delta$, the series of multiple integrals for
the generating function cannot be reduced to the form
\eq{0-FD-exp0}. Nevertheless, it can still be reordered in a way similar to
the case of free fermions. Such a reordering leads to the appearance
of multiple integrals of a special type that we call {\em cycle integrals}
(see Section~\ref{sec-cycles}). These cycle integrals play the role of
analogs of $\tr(V_0^k)$. Each of them has a computable long-distance
asymptotic behavior that we recently obtained using Riemann-Hilbert
techniques applied to the Fredholm determinant of an integral
operator with generalized sine kernel \cite{KitKMST08a}. Our strategy is then
rather simple: {\em using the asymptotic behavior for each cycle
integral, we can sum up asymptotically the multiple series}
corresponding to the generating function  $\langle e^{\beta
\mc{Q}_m}\rangle$. It so happens that such asymptotic series exponentiates in
a natural way that mimic a Fredholm determinant expansion.
However, in our case, all integrals are coupled in a non trivial
way, and the correlation function itself is not a Fredholm
determinant (except in the free fermion point).  We believe that this strategy  can be applied to quite general cases although the details should be adjusted accordingly.

\medskip

As a result of this procedure, we are able to find the desired long-distance
asymptotic behavior for $\langle e^{\beta
\mc{Q}_m}\rangle$:
 \begin{equation}\label{ebQm-intro}
 \moy{\ex{\beta
 \mc{Q}_m}}
 = \sum_{\sigma=0,\pm}{\widehat G}^{(0)}(\beta+2\pi i\sigma,m)
 \big[1+\e{o}(1)\big]\, ,
 \end{equation}
where we have defined the function
\begin{equation}\label{G0-intro}
 {\widehat G}^{(0)}(\beta,m)={\cal C}(\beta)\cdot
 e^{\beta mD} m^{\frac{\beta^2 {\cal Z}^2}{2\pi^2}}\, .
 %
 \end{equation}
Here ${\cal C}(\beta)$, $D$ and ${\cal Z}$ are constants depending on the  thermodynamic quantities that can be computed in terms of the low
energy spectrum properties of the XXZ chain. This spectrum  can be characterized by the solutions of the system
of Bethe equations \eq{1-BE_Y} \cite{Bet31,Orb58}. In the
thermodynamic limit, the ground state of the model  appears to be the
Dirac sea in the momentum space: the spectral parameters of the
particles occupy the interval $[-q,q]$ with a density $\rho
(\lambda)$. The Fermi boundary $q$ of the interval depends on the anisotropy parameter
$\Delta$ and on the magnetic field $h$.
Note that $q$ remains finite for a non zero magnetic
field, while it tends to infinity in the vanishing
magnetic field limit. In the present article, {\em $q$ will
be kept finite in all computations}. However, the final expressions that we obtain are well defined in the zero magnetic field limit as well. In the Bethe ansatz framework,
$q$ and $\rho(\la)$ can be obtained in the following way.
Let $\eps(\la)$ be the dressed energy, that is to say the solution of the integral equation
\begin{equation}\label{epsilon}
\eps(\la)
+\f{1}{2\pi}\Int{-q}{q}K(\la-\mu) \, \eps(\mu)\, \dd\mu
=h-2p'_0(\la)\sin \zeta ,
\end{equation}
where
 \begin{equation}\label{0-K-XXZ}
 K(\lambda)=\frac{\sin2\zeta}{\sinh(\lambda+i\zeta)\sinh(\lambda-i\zeta)},
 \qquad \cos\zeta=\Delta,\qquad 0<\zeta<\pi,
 \end{equation}
and $p_0(\lambda)$ is the bare momentum of the particle (see
\eq{1-p0}). Then $q$ is such that $\eps(q)=0$. Similarly, the ground
state density $\rho(\lambda)$ \cite{LieSM61,LieL63} satisfies a
linear integral equation
 \begin{equation}\label{0-rho}
 \rho(\lambda)
+\frac1{2\pi}\int\limits_{-q}^q
K(\lambda-\mu)\,\rho(\mu)\,\dd\mu
=\frac1{2\pi}p'_0(\lambda),
 \end{equation}
 The dressed momentum is closely related to the density as
 \begin{equation}\label{0-Dmom}
 p(\lambda)=2\pi\int_0^\lambda \rho(\mu)\,\dd\mu.
 \end{equation}
The Fermi momentum, denoted as $p_{{}_F}$, is given by its value at the Fermi boundary, $p_{{}_F}=p(q)$.
The average density $D$ is related to the ground state magnetization
$\langle\sigma^z\rangle$ and is expressed in terms of the Fermi momentum as
 \begin{equation}\label{0-Dens}
 \langle\sigma^z\rangle=1-2D,\qquad
 D=\int\limits_{-q}^q \rho(\mu)\,\dd\mu
  =\frac{p_{{}_F}}\pi \, .
 \end{equation}
Note that $D=\frac12$ in the case of a zero magnetic field hence implying a vanishing
magnetization.

Another constant appearing in the asymptotic behavior \eq{G0-intro} is the quantity ${\cal Z}$. It can be identified  with the value of the dressed
charge $Z(\lambda)$ at the Fermi boundary, ${\cal Z}=Z(\pm q)$.
We recall that $Z(\lambda)$ satisfies an integral equation similar to
\eq{0-rho}
 \begin{equation}\label{0-DC}
 Z(\lambda)
 +\frac1{2\pi}\int\limits_{-q}^qK(\lambda-\mu)\,Z(\mu)\,\dd\mu
 =1 ,
 \end{equation}
and can be interpreted in the XXZ model as the intrinsic magnetic
moment of the elementary excitations \cite{BogIK93L}.

Finally, the constant ${\cal C}(\beta)$ in \eq{G0-intro} can be computed in terms of a ratio of four Fredholm determinants. The associated integral operators have compact support on contours surrounding the interval $[-q, q]$. The explicit formulas are given in Section \ref{AEebQ}.  At this stage, we would like to make two important remarks.

The first one concerns the fact that all the Fredholm
determinants involved in our computations, for example those appearing in the
constant ${\cal C}(\beta)$, are associated to {\em bounded integral operators
acting on some compact contours surrounding the finite interval $[-q, q]$}. Due to the finiteness of $q$, all these determinants take finite values, which is a direct
consequence of working at a non zero value of the magnetic field. When $q \to \infty$ the contours are no more compact and the corresponding Fredholm determinants diverge.
It can nevertheless be checked that, in the  zero magnetic
field limit, the overall constant  ${\cal C}(\beta)$ appearing in the final
answer \eq{G0-intro} remains finite although each of its individual parts (i.e.
the four above mentioned  Fredholm determinants) diverges in this limit.

The second remark is the following: since the operator
$e^{\beta \mc{Q}_m}$
\eq{0-defQ} is a polynomial in $\kappa=e^\beta$, the generating function is a $2\pi i$-periodic
function of $\beta$. Of course, the leading term of the long-distance
asymptotic behavior  is itself not necessarily a periodical
function of $\beta$.
Nonetheless, our analysis shows that {\em this a priori broken periodicity is at least partly restored by the
first oscillating correction to the leading term}, as it can be seen from the equation
\eq{ebQm-intro}.
More precisely, the result of the direct
computation of the first oscillating correction to the asymptotics
\eq{ebQm-intro} can be reproduced by a simple shift
$\beta\to\beta\pm2\pi i$ from the non-oscillating term. This property, that we will prove here, has been recently used in \cite{BorSS07} to get predictions for the asymptotic behavior of $\langle e^{\beta \mc{Q}_m}\rangle$ at zero magnetic field.

The long-distance asymptotic behavior of the longitudinal spin-spin
correlation function then can  be extracted from the one of
$\langle\ex{\beta \mc{Q}_m}\rangle$ using \eqref{0-corr-funct}.
At the first leading orders, it reads,
 \begin{equation}\label{9-corr-funct}
 \langle\sigma_1^z\sigma_{m+1}^z\rangle_{{}_{\e{leading}}}
=\langle\sigma^{z} \rangle ^{2} -
 \frac{2{\cal Z}^2}{\pi^2 m^2}+2|F_\sigma|^2\cdot
 \frac{\cos(2mp_{{}_F}) }{m^{2{\cal Z}^2}}\, ,
 \end{equation}
where $F_\sigma$ is related to the properly normalized form factor
of the operator $\sigma^z$ between the ground state and  an excited
state containing one particle and one hole sited at the two
different boundaries of the Fermi sphere, namely at $q$ and $-q$;
hence, it corresponds to an umklapp process \cite{EmeLP76,Hal81b,Hal81a}. The factor  $2$ in front of $|F_\sigma|^2$ just corresponds to the two possible states of this kind. This result agrees with  bosonization
and  conformal field theory analysis. Note however that, in the
thermodynamic limit, all form factors scale to zero as some power of the
size of the lattice. The precise coefficient and proper
scaling behavior of the form factor which results from our
computations is therefore rather non-trivial. We have nevertheless checked that, at zero
magnetic field, this constant $F_\sigma$ goes to a finite value, and that it agrees to second order in $\Delta$ around
the point $\Delta = 0$ with the predictions given  in \cite{Luk98,Luk99,LukT03}.

\bigskip


\bigskip

This article is organized as follows.

In Section~\ref{sec-master}, we explain how to obtain, in the case
of the XXZ spin chain, a new series expansion for the master
equation representation of $\langle\ex{\beta \mc{Q}_m}\rangle$, that
is suitable first for taking the thermodynamic limit and then for
extracting the asymptotic behavior of this correlation function.

This expansion is written in Section~\ref{sec-cycles} in terms of
cycle integrals. Then, we recall the results of~\cite{KitKMST08a}
and apply them to get an asymptotic expansion of the cycle integrals
in a form adapted for their asymptotic summation.

The asymptotic summation of these cycle integrals is described in
Section~\ref{sec-sum}. It is divided into several steps according to
the different nature of the terms to sum up. In particular one of
the steps uses a generalization of the Lagrange series (the details
are given in an appendix).

This procedure enables us, in Section~\ref{sec-results}, to describe the asymptotic behavior of the correlation function  $\langle\ex{\beta \mc{Q}_m}\rangle$ at large distances $m$. We show the existence of  both oscillating and non-oscillating terms that we compute at the leading orders. The asymptotic behavior of the longitudinal spin-spin correlation function follows and is also given in the same section. It confirms at the leading order the predictions from bosonization and conformal field theory.

In Section~\ref{sec-QNLS}, the whole procedure is applied to another model, the one-dimensional interacting Bose gas, for which we compute the asymptotic behavior of the density-density correlation function.

Finally, various technicalities are gathered in the appendices. We would like in particular to draw the reader's attention to Appendix~\ref{ap-LSD} which contains rather essential ingredients for the summation described in Section~\ref{sec-sum}.



\section{Master equation and its thermodynamic limit}
\label{sec-master}

As already mentionned in the Introduction, the main object of study of this article (at least up to Section~\ref{sec-results}) is the ground state
expectation value of the operator $e^{\beta \mc{Q}_m}$ \eq{0-Q1} for the XXZ model \eqref{0-HamXXZ} in an external magnetic field.
In this section, we recall the master
equation representation obtained in \cite{KitMST05me,KitKMST07} for this operator in the finite chain, and show how such a master equation can be expanded into some series suitable both for the thermodynamic limit and for the asymptotic analysis performed in the following sections.

\subsection{Master equation for $\langle e^{\beta \mc{Q}_m}\rangle$} \label{ND}

The master equation gives a multiple integral representation for the
expectation value of the operator $e^{\beta \mc{Q}_m}$ with respect
to an $N$-particles Bethe eigenstate
$|\psi(\{\lambda\})\rangle$ of the Hamiltonian. It reads,
\begin{align}
 \langle e^{\beta \mc{Q}_m}\rangle
&=\frac{\langle\psi(\{\lambda\})|e^{\beta \mc{Q}_m}
|\psi(\{\lambda\})\rangle}{\langle\psi(\{\lambda\})|\psi(\{\lambda\})\rangle}
        \nonumber\\
&=\frac{(-1)^N}{N!}
 \hspace{-1mm}
 \oint\limits_{\Gamma(\{\lambda\})}\prod_{j=1}^N
 \left(\frac{\dd z_j}{2\pi i}\;
       \frac{l(z_j)\,
             d(z_j)}{l(\lambda_j)\,
             d(\lambda_j)}\right)
 \frac{\Bigl[\det_N\Omega_\kappa(\{z\},\{\lambda\}|\{z\})\Bigr]^2}
  {\prod\limits_{j=1}^N{\cal Y}_\kappa(z_j|\{z\})\cdot
 \det_N\frac{\partial{\cal Y}(\lambda_j|\{\lambda\})}{\partial\lambda_k}}
 \, .
\label{1-GME}
 \end{align}
%
The various quantities entering this representation are all rather universal objects in the context of the Bethe ansatz.
Let us explain their meaning.

The parameters $\lambda_1,\dots,\lambda_N$ describe the specific
eigenstate $|\psi(\{\lambda\})\rangle$ in which the average value is computed. From now on, this state will be the ground state of the XXZ chain \eqref{0-HamXXZ} in a given magnetic field $h$ and for some anisotropy $\Delta=\cos\zeta$, $0<\zeta<\pi$. In this case, these parameters are real numbers \cite{YanY66} which satisfy the system of Bethe equations
 \begin{equation}\label{1-BE_Y}
 {\cal Y}(\lambda_j|\{\lambda\})=0, \qquad
 j=1,\dots,N,
 \end{equation}
with
 \begin{equation}\label{1-Y-def}
 {\cal Y}(\mu|\{\lambda\}) =
 a(\mu)\prod_{k=1}^{N}\sinh(\lambda_k-\mu-i\zeta) + d(\mu)
 \prod_{k=1}^{N}\sinh(\lambda_k-\mu+i\zeta) \, .
 \end{equation}
Here $a(\mu)=a_0^M(\mu)$ and $d(\mu)=d_0^M(\mu)$ are the vacuum eigenvalues of the transfer matrix, with
 \begin{equation}\label{1-ad}
 a_0(\mu)=\sinh(\mu-\textstyle{\frac{i\zeta}2}),\qquad
 d_0(\mu)=\sinh(\mu+\textstyle{\frac{i\zeta}2}).
 \end{equation}

The whole explicit dependence of \eqref{1-GME} on the distance $m$ is contained in the function $l(z)$, which can be expressed in terms of $a_0$ and $d_0$ as
 \begin{equation}\label{1-l}
 l(\mu)=\left(\frac{a_0(\mu)}{d_0(\mu)}\right)^m=\left(\frac{\sinh(\mu-\textstyle{\frac{i\zeta}2})}
 {\sinh(\mu+\textstyle{\frac{i\zeta}2})}\right)^m\equiv
 e^{im(p_0(\mu)+\pi)},
 \end{equation}
where
 \begin{equation}\label{1-p0}
 p_0(\mu)
 =-i\log\left(\frac{\sinh(\textstyle{\frac{i\zeta}2}-\mu)}
 {\sinh(\textstyle{\frac{i\zeta}2}+\mu)}\right), \qquad p_0(0)=0,
 \end{equation}
plays the role of the bare momentum of the pseudo particles (see
also \eqref{0-K-XXZ}, \eqref{0-rho}).

Besides the function ${\cal Y}(\lambda_j|\{\lambda\})$,
the master equation
\eq{1-GME} contains also a function ${\cal Y}_\kappa(z_j|\{z\})$, $\kappa = e^{\beta}$,
which naturally appears in the definition of the $\kappa$-twisted Bethe equations associated to the spectrum of the $\kappa$-twisted monodromy matrix (see \cite{KitMST05me}). It is given as
 \begin{equation}\label{1-TY-def}
 {\cal Y}_\kappa(\mu|\{z\}) =
 a(\mu)\prod_{k=1}^{N}\sinh(z_k-\mu-i\zeta) + \kappa\,d(\mu)
 \prod_{k=1}^{N}\sinh(z_k-\mu+i\zeta)\, .
 \end{equation}

One more object entering  \eq{1-GME} is the
determinant of the matrix $\Omega_\kappa$, with matrix elements
\begin{multline} \label{1-matH}
 \left(\Omega_\kappa\right)_{jk}(\{z\},\{\lambda\}|\{z\})=
   a(\lambda_j)\,t(z_k,\lambda_j)\,\prod_{a=1}^{N} \sinh(z_a-\lambda_j-i\zeta)\\
   -\kappa\, d(\lambda_j)\,t(\lambda_j,z_k)\,\prod_{a=1}^{N}
   \sinh(z_a-\lambda_j+i\zeta)\, ,
\end{multline}
where
 \begin{equation}\label{1-t}
 t(z,\lambda)=\frac{-i\sin\zeta}{\sinh(z-\lambda)\sinh(z-\lambda-i\zeta)}.
 \end{equation}
The determinant of this matrix describes the scalar product between
any eigenstate of the twisted transfer-matrix and some arbitrary state (see
\cite{KitMST05me,KitKMST07}).

It remains to describe the integration contour $\Gamma(\{\lambda\})$ appearing
in equation \eq{1-GME}. For every variable $z_j$ the integral is
taken over a closed contour  surrounding the points
$\lambda_1,\dots,\lambda_N$ and such that  any other
singularities of the integrand (i.e. the roots of the system ${\cal
Y}_\kappa(z_j|\{z\})=0$) are located outside this contour.
Since here the parameters $\lambda_1,\dots,\lambda_N$ decribe the ground state of the XXZ chain in a non-zero magnetic field $h$, they are real numbers and, in the thermodynamic limit, they fill the interval $[-q,q]$ (with $q$ finite, see \eqref{epsilon}) with a density $\rho(\la)$ satisfying the linear integral equation~\eqref{0-rho}. Therefore, the integration contour will be chosen such as to surround the interval $[-q,q]$.

It is worth mentioning that the integral representation \eq{1-GME}
only holds for $|\kappa|$ small enough. This restriction does not
cause any problems: the expectation value of $e^{\beta \mc{Q}_m}$ is
a polynomial in $\kappa$; thus it is enough to know this polynomial
in a vicinity of $\kappa=0$ and then consider its analytic
continuation. One should however remember that it is not always possible to analytically
continue directly the integrand of
\eq{1-GME}. For instance, at $\kappa\to1$, one of the solutions of
the system ${\cal Y}_\kappa(z_j|\{z\})=0$ goes to $\{\lambda\}$, and an integration contour $\Gamma(\{\lambda\})$ satisfying the above desired properties does not exist anymore.


\subsection{Transformation of the determinants\label{PME}}

The integrand of \eqref{1-GME} has poles at $z_k=\la_j$ and, formally,
the multiple contour integral in \eq{1-GME} can be computed by
residues at these points. These poles are contained in the
determinants $\det\Omega_\kappa$. More precisely, each
of these determinants has simple poles at $z_k=\la_j$, which means that the integrand has double poles at these points.

In order to be able to analyze the contribution of these poles, it is convenient to extract them explicitly from $\det\Omega_\kappa$.
This extraction can be done in several ways.
The method we present here consists in extracting a Cauchy determinant from each $\det\Omega_\kappa$ using the fact that the parameters $\la$ are solution of the Bethe equations. It allows us in particular to show
that  $\det \Omega_\kappa$
is proportional to $\kappa-1$.

\begin{prop}\label{Extract-Cauchy}
If the parameters $\lambda_1,\dots,\lambda_N$ satisfy the system of
Bethe equations, then
 \begin{multline}\label{2-l-rep}
 \det_N\Omega_\kappa(\{z\},\{\lambda\}|\{z\})=\det_N\left(\frac1{\sinh(\lambda_j-z_k)}\right)\cdot
 \prod_{j=1}^N\left\{a(\lambda_j)\left[\kappa \frac{V_+(\lambda_j)}{V_-(\lambda_j)}-1\right]\right\}
  \num
 \times \frac{1-\kappa}{V_+^{-1}(\theta)-\kappa
 V_-^{-1}(\theta)}\cdot\prod_{a,b=1}^N\sinh(z_a-\lambda_b-i\zeta)\cdot
 \det_N\left(\delta_{jk}+ U^{(\lambda)}_{jk}(\theta)\right),
 \end{multline}
or
 \begin{multline}\label{2-z-rep}
 \det_N\Omega_\kappa(\{z\},\{\lambda\}|\{z\})=\det_N\left(\frac1{\sinh(z_k-\lambda_j)}\right)\cdot
 \prod_{j=1}^N\left\{d(\lambda_j)\left[\frac{V_-(z_j)}{V_+(z_j)}-\kappa\right]\right\}
  \num
 \times \frac{1-\kappa}{V_-(\theta)-\kappa
 V_+(\theta)}\cdot\prod_{a,b=1}^N\sinh(\lambda_a-z_b-i\zeta)\cdot
 \det_N\left(\delta_{jk}+ U^{(z)}_{jk}(\theta)\right).
 \end{multline}
In these expressions, $\theta$ is an arbitrary complex number, and
 \begin{equation}\label{1-Vpm}
 V_{\pm}(\mu)\equiv
 V_{\pm}\left(\mu \mid
        \begin{matrix}\{\lambda\} \\ \{z\} \end{matrix}
        \right)
 = \prod_{a=1}^{N}
 \frac {\sinh(\mu-\lambda_a\pm i\zeta)}{\sinh(\mu-z_a\pm i\zeta)}.
 \end{equation}
The entries of the matrices $U^{(\lambda)}(\theta)$ and
$U^{(z)}(\theta)$ are given by
 \begin{equation}\label{2-Tl}
 U^{(\lambda)}_{jk}(\theta)=%
 \frac{\prod\limits_{a=1}^N\sinh(z_a-\lambda_j)}
 {\prod\limits_{\substack{a=1\\ a\ne j}}^N\sinh(\lambda_a-\lambda_j)}\cdot
 \frac{K_\kappa(\lambda_{j}-\lambda_{k})-K_\kappa(\theta-\lambda_{k})}{V_+^{-1}(\lambda_j)-\kappa
 V_-^{-1}(\lambda_j)}\, ,
 \end{equation}
 \begin{equation}\label{2-Tz}
 U^{(z)}_{jk}(\theta)=
 %
 \frac{\prod\limits_{a=1}^N\sinh(z_k-\lambda_a)}
 {\prod\limits_{\substack{a=1\\ a\ne k}}^N\sinh(z_k-z_a)}\cdot
 \frac{K_\kappa(z_{j}-z_k)-K_\kappa(z_j-\theta)}{V_-(z_k)-\kappa
 V_+(z_k)} \, ,
 \end{equation}
with
 \begin{equation}\label{2-Kk}
  K_\kappa(\lambda)=\coth(\lambda+i\zeta)-\kappa\coth(\lambda-i\zeta) \, .
 \end{equation}
\end{prop}
The proof of this proposition is given in Appendix \ref{ap-DT}.

\begin{rem}
 One can observe that $\det\Omega_\kappa$ is proportional to
$1-\kappa$. However, the second line of \eqref{2-l-rep} or of \eqref{2-z-rep} does not necessarily vanish at
$\kappa=1$ ($\beta=0$). Indeed, if $z_j=\lambda_j$ (modulo the
permutation group) for all $j=1,\dots,N$, then $V_\pm(\theta)=1$ and
the factor $1-\kappa$ cancels. Such situation can arise in the process of evaluating the multiple integral \eq{1-GME}.
\end{rem}

In the representations \eq{2-l-rep}, \eq{2-z-rep}, the poles at
$z_k=\lambda_j$ are gathered explicitly in the two Cauchy determinants.
The entries of the matrices $U^{(\lambda,z)}$ contain
singularities at $\lambda_j=\lambda_k$ (respectively $z_j=z_k$), but the corresponding determinants are obviously not singular,
since the original $\det\Omega_\kappa$ vanishes at $\lambda_j=\lambda_k$
or $z_j=z_k$. To make this fact explicit, we present the
determinants of the finite size matrices
$\delta_{jk}+U_{jk}^{(\lambda,z)}$ as Fredholm determinants of
integral operators acting on a contour $\Gamma$ surrounding the
points $\{\lambda\}$ and $\{z\}$ (see Fig.~1):
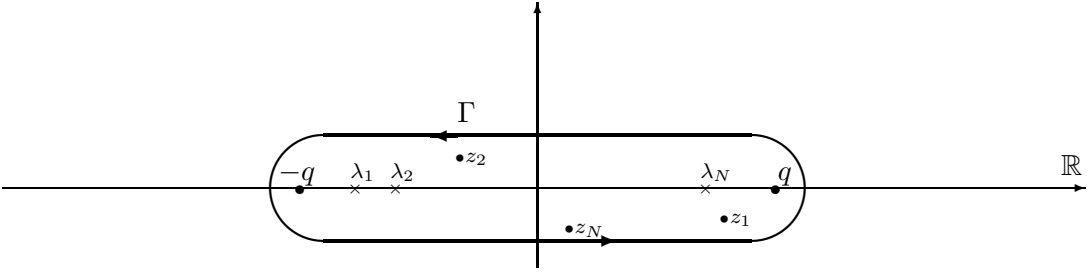
\begin{figure}[h]
\begin{picture}(440,140)
 \put(400,50){\vector(1,0){5}}
 \put(0,50){\line(1,0){400}}
 \put(396,55){$\mathbb{R}$}
 \put(200,20){\vector(0,1){100}}
 \put(290,53){$q$}
 \put(103,53){$-q$}
 \put(109,47.5){$\scriptstyle{\bullet}$}
 \put(287,47.5){$\scriptstyle{\bullet}$}
 \put(129,48){$\scriptscriptstyle{\times}$}
 \put(144,48){$\scriptscriptstyle{\times}$}
 \put(260,48){$\scriptscriptstyle{\times}$}
 \put(130,54){$\scriptstyle{\lambda_1}$}
 \put(145,54){$\scriptstyle{\lambda_2}$}
 \put(261,54){$\scriptstyle{\lambda_N}$}
 \put(169,60){$\scriptscriptstyle{\bullet}$}
 \put(210,33){$\scriptscriptstyle{\bullet}$}
 \put(268,37){$\scriptscriptstyle{\bullet}$}
 \put(173,60){$\scriptstyle{z_2}$}
 \put(214,33){$\scriptstyle{z_N}$}
 \put(272,37){$\scriptstyle{z_1}$}
 %
 %
 \thicklines
  \put(200,50){\oval(200,40)}
 \put(170,69.5){\vector(-1,0){10}}
 \put(220,30){\vector(1,0){10}}
 \put(170,75){$\Gamma$}
\end{picture}
\caption{The contour $\Gamma$. \label{1}}
\end{figure}
 \begin{equation}\label{2-det-lz}
 \det_N\left[\delta_{jk}+U^{(\lambda,z)}_{jk}(\theta) \right]=
 \det\left[I+\frac1{2\pi i}\hat U_\theta^{(\lambda,z)}(w,w')\right],
 \end{equation}
where
 \begin{equation}\label{2-U1}
 \hat U_\theta^{(\lambda)}(w,w')=
 -\prod\limits_{a=1}^N\frac{\sinh(w-z_a)}{\sinh(w-\lambda_a)}\cdot
 \frac{K_\kappa(w-w')-K_\kappa(\theta-w')}{V_+^{-1}(w)-\kappa
 V_-^{-1}(w)},
 \end{equation}
and
 \begin{equation}\label{2-U2}
 \hat U_\theta^{(z)}(w,w')=
 \prod\limits_{a=1}^N\frac{\sinh(w'-\lambda_a)}{\sinh(w'-z_a)}\cdot
 \frac{K_\kappa(w-w')-K_\kappa(w-\theta)}{V_-(w')-\kappa
 V_+(w')}.
 \end{equation}
The equivalence of the two representations \eq{2-det-lz} is proven in
Appendix \ref{ap-FDR}.

Using these results,  we can rewrite the master equation
\eq{1-GME} in the following form:
 \begin{multline}\label{2-ME}
 \langle e^{\beta \mc{Q}_m}\rangle
=\frac1{N!}\oint\limits_{\Gamma(\{\lambda\})}
\prod_{j=1}^N\left\{
 \frac{\dd z_j}{2\pi i}\;
 e^{im(p_0(z_j)-p_0(\lambda_j))}\,
 \frac{
 \left(\kappa\frac{V_+(\lambda_j)}{V_-(\lambda_j)}-1\right)
 \left(\kappa-\frac{V_-(z_j)}{V_+(z_j)}\right)
 }
 {\kappa+(-1)^N\frac{a(z_j)}{d(z_j)}\prod\limits_{a=1}^N
 \frac{\sinh(z_{a}-z_j-i\zeta)}{\sinh(z_{j}-z_a-i\zeta)}}
 \right\}\num
 \times \frac{1}{\det_N\Theta_{jk}}\cdot 
 \widetilde
 W_N\left(\begin{matrix}\lambda_1,\dots,\lambda_N\\z_1,\dots,z_N\end{matrix}\right)
 \cdot \left(\det_N\frac1{\sinh(z_k-\lambda_j)}\right)^2,
 \end{multline}
where
 \begin{multline}\label{2-W}
 \widetilde
 W_N\left(\begin{matrix}\lambda_1,\,\dots,\,\lambda_N\\
                        z_1,\,\dots,\,z_N\end{matrix}\right)
 \equiv
 \widetilde{W}_N\left(\begin{array}{c} \{\la\}\\ \{z\}\end{array}\right)
 =
 \prod_{a,b=1}^N\frac{\sinh(z_a-\lambda_b-i\zeta)\sinh(\lambda_b-z_a-i\zeta)}
 {\sinh(z_a-z_b-i\zeta)\sinh(\lambda_a-\lambda_b-i\zeta)}\numa{35}
 \times\frac{(\kappa-1)^2
 \det\left(I+\frac1{2\pi i}\hat U_{\theta_1}^{(\lambda)}(w,w')\right)
 \det\left(I+\frac1{2\pi i}\hat U_{\theta_2}^{(z)}(w,w')\right)}
 {(V_+^{-1}(\theta_1)-\kappa V_-^{-1}(\theta_1) ) (V_-(\theta_2)-\kappa
 V_+(\theta_2))},
 \end{multline}
and
 \begin{equation}\label{2-G}
 \Theta_{jk}=\left(a(\lambda_j)\prod_{a=1}^N\sinh(\lambda_{a}-\lambda_j-i\zeta)\right)^{-1}
 \cdot\frac{\partial{\cal
 Y}(\lambda_j|\{\lambda\})}{\partial\lambda_k}.
\end{equation}
It is important to note that the function $\widetilde W$ \eqref{2-W} is a symmetric function of the variables
$\{z\}$ and also of the variables $\{\lambda\}$.
Moreover, it possesses the
recursive reduction property
 \begin{equation}\label{2-recW}
 \left.\widetilde W_N\left(\begin{matrix}\lambda_1,\,\dots,\,\lambda_N\\
                        z_1,\,\dots,\,z_N\end{matrix}\right)\right|_{z_N=\lambda_N}=
 \widetilde W_{N-1}\left(\begin{matrix}\lambda_1,\,\dots,\,\lambda_{N-1}\\
                        z_1,\,\dots,\,z_{N-1}\end{matrix}\right).
 \end{equation}
The entries of the matrix $\Theta$ contain the Lieb kernel
$K(\lambda)$ defined in  \eq{0-K-XXZ},
 \begin{equation}\label{2-Gjk}
 \Theta_{jk}=\delta_{jk} \left(\log'\frac{a(\lambda_j)}{d(\lambda_j)}-i\sum_{a=1}^NK(\lambda_{j}-\lambda_{a})\right)+
 iK(\lambda_j-\lambda_k)\,\, .
 \end{equation}
%


\subsection{Expansion of the master equation\label{EME}}

We now express the master equation as  series
appropriate for taking the thermodynamic limit. Replacing one of
the Cauchy determinant by the product of its diagonal elements  we
re-write equation \eq{2-ME} as
 \begin{multline}\label{3-ME}
 \langle e^{\beta \mc{Q}_m}\rangle
 =\frac1{\det_N\Theta}
  \oint\limits_{\Gamma(\{\lambda\})}
  \prod_{j=1}^N\left\{
  \frac{\dd z_j}{2\pi i}\cdot
  e^{im(p_0(z_j)-p_0(\lambda_j))}\cdot
  \mathcal{V}_N
  \left(\la_j\mid\begin{matrix}\{\la\}\\ \{z\}\end{matrix}\right)
  \right\}\cdot
  \widetilde W_N\left(\begin{array}{c}\{\lambda\}\\  \{z\}\end{array}\right)
  \num
 \times
 \prod_{k=1}^N\frac1{\sinh(z_k-\lambda_k)}\cdot\det_N\left[\frac{ h_N(z_k|\{z\}|\{\lambda\})-1}{\sinh(z_k-\lambda_j)}
 +\frac1{\sinh(z_k-\lambda_j)}\right],
 \end{multline}
where
\begin{equation}\label{5-mcV}
 {\cal V}_N\left(\mu \mid\begin{matrix} \{\lambda\} \\
                                        \{z\} \end{matrix} \right)
 =
\kappa\, { V_+\left(\mu\mid\begin{matrix} \{\lambda\} \\
                           \{z\} \end{matrix} \right)}\,
       {V_-^{-1}\left(\mu\mid\begin{matrix} \{\lambda\} \\
                           \{z\} \end{matrix} \right)}
 -1 ,
\end{equation}
and
 \begin{equation}\label{3-h}
 h_N(z_k|\{z\}|\{\lambda\})
 =\frac{\kappa-
  V_-\left(z_k\mid\begin{matrix}\{\la\}\\ \{z\}\end{matrix}\right)\,
  V_+^{-1}\left(z_k\mid\begin{matrix}\{\la\}\\ \{z\}\end{matrix}\right)         }
 {\kappa+(-1)^N\frac{a(z_k)}{d(z_k)}\prod\limits_{a=1}^N\frac{\sinh(z_{a}-z_k-i\zeta)}{\sinh(z_{k}-z_a-i\zeta)}}.
 \end{equation}
Note that the function $\mathcal{V}$ defined by \eqref{5-mcV} satisfy similar properties to $\widetilde{W}$: it is also a symmetric function of the variables $\{\la\}$ and of the variables $\{z\}$ separately, and possesses the same reduction property:
\begin{equation}\label{5-redV}
 \left.\mathcal{V}_N\left(\mu\mid
       \begin{matrix}\lambda_1,\,\dots,\,\lambda_N\\
                        z_1,\,\dots,\,z_N\end{matrix}\right)
       \right|_{z_N=\lambda_N}=
 \mathcal{V}_{N-1}\left(\mu\mid
       \begin{matrix}\lambda_1,\,\dots,\,\lambda_{N-1}\\
                        z_1,\,\dots,\,z_{N-1}\end{matrix}\right).
\end{equation}
As for $h_N$ \eqref{3-h}, it possesses also interesting reduction properties. Indeed, the Bethe equations \eq{1-BE_Y} for $\{\la\}$ ensure that, for any $k$ and $s$ in $1, \dots, N$,
 \begin{equation}\label{3-prop-h1}
 \left.h_N(z_k|\{z\}|\{\lambda\})\right|_{z_k=\lambda_s}=1 \, ,
 \end{equation}
and
 \begin{equation}\label{3-prop-h2}
 \left.\frac\partial{\partial z_k}h_N(z_k|\{z\}|\{\lambda\})\right|_{z_k=\lambda_s}
 \hspace{-2mm}=
 \mathcal{V}_{N-1}^{-1}\left(\la_s\mid
       \begin{matrix}\{\la\}\setminus \la_s \\
          \{z\}\setminus z_k \end{matrix}\right)
 \left(
 \log'\frac{a(\lambda_s)}{d(\lambda_s)}-i\sum_{a=1}^NK(\lambda_{s}-\la_a)\right) .
 \end{equation}
It follows in particular from \eqref{3-prop-h1} that  the first part of the determinant in \eqref{3-ME}
is holomorphic at $z_k=\la_j$.

Let us now expand this determinant via the Laplace formula. We obtain
 \begin{multline}\label{3-ME-Lapl}
 \langle e^{\beta \mc{Q}_m}\rangle
 =\frac1{\det_N\Theta}
  \oint\limits_{\Gamma(\{\lambda\})}
  \prod_{j=1}^N\left\{
  \frac{\dd z_j}{2\pi i}\cdot
  \frac{e^{im(p_0(z_j)-p_0(\lambda_j))}}{\sinh(z_j-\la_j)}\cdot
  \mathcal{V}_N
  \left(\la_j\mid\begin{matrix}\{\la\}\\ \{z\}\end{matrix}\right)
  \right\}
  \cdot
  \widetilde W_N\left(\begin{array}{c}\{\lambda\}\\  \{z\}\end{array}\right)
  \num
 \hspace{-5mm}\times
 \sum_{n=0}^N
 \sum_{\substack{
   \{z\}=\{z\}_{\gamma}\cup\{z\}_{\bar\gamma} \\
   \{\lambda\}=\{\lambda\}_{\alpha}\cup\{\lambda\}_{\bar\alpha} \\
   \#\bar\gamma=\#\bar\alpha=n
       }}
 \hspace{-4mm}
 (-1)^{[P(\alpha)]+[P(\gamma)]}\,
 {\det_{\substack{k\in\gamma\\ j\in\alpha}}}\left[\frac{ h_N(z_k|\{z\}|\{\lambda\})-1}{\sinh(z_k-\lambda_j)}\right]
 {\det_{\substack{k\in\bar\gamma\\ j\in\bar\alpha}}}
 \left[\frac1{\sinh(z_k-\lambda_j)}\right].
 \hspace{-2mm}
 \end{multline}
The above sum runs through all possible partitions $\alpha \cup {\bar\alpha}$ and $\gamma \cup {\bar\gamma}$ of $1, \dots, N$ such that ${\#\bar\gamma=\#\bar\alpha=n}$. The sets of parameters $\{\lambda\}$ and $\{z\}$ are divided accordingly  into
disjoint subsets
$\{\lambda\}=\{\lambda\}_{\alpha}\cup\{\lambda\}_{\bar\alpha}$,
and similarly $\{z\}=\{z\}_{\gamma}\cup\{z\}_{\bar\gamma}$.
Moreover, we specify that the parameters in
each subset are ordered in the canonical way, namely
$\{\lambda\}_{\alpha} = \{\lambda_{\alpha_1},\lambda_{\alpha_2},\dots\}$  with
$\alpha_1<\alpha_2<\dots$ etc.  Finally,
$[P(\alpha)]$ (respectively $[P(\gamma)]$) stands for the signature the permutation $P$ such that
$P(\alpha,\bar\alpha)=1,\dots,N$ (respectively
$P(\gamma,\bar\gamma)=1,\dots,N$).
\bigskip
\begin{lemma}
In the multiple integral \eq{3-ME-Lapl},
\begin{enumerate}
\item the only non-zero contributions come from partitions 
such that
$\gamma=\alpha$;
\item
the first
determinant in \eq{3-ME-Lapl} contributes only through the product of its diagonal elements.
\end{enumerate}
\end{lemma}

\Proof Let us consider a partition such that $\gamma\ne\alpha$. Then there
exists $\ell\in\gamma$ such that $\ell\notin\alpha$. Let us consider the integrand of
\eq{3-ME-Lapl} as a function of $z_\ell$. At first sight it has a simple pole at
$z_\ell=\lambda_\ell$ and also simple poles at the points $z_\ell=\lambda_j$ for all
$j\in \alpha$.
Yet, due to the property \eq{3-prop-h1}, all the corresponding
residues are equal to zero.  Hence, the integral over $z_\ell$
vanishes. Thus we conclude that only partitions such that  $\gamma = \alpha$ yield non-vanishing contributions to the integral.

It remains to prove the second part of the lemma. Let us consider an
off-diagonal contribution coming from the first determinant. Hence
there is a variable $z_\ell\in\{z\}_{\gamma}$ such that the
integrand, considered  as a function of this $z_\ell$, has at most
simple poles. However, due to the presence of the factor
$h_N(z_\ell|\{z\}|\{\lambda\})-1$, their residues vanish. \qed

Thus,
 \begin{multline}\label{3-ME-Lapl-1}
 \langle e^{\beta \mc{Q}_m}\rangle=\frac1{\det_N\Theta}\oint\limits_{\Gamma(\{\lambda\})}
\prod_{j=1}^N\left\{
  \frac{\dd z_j}{2\pi i}\cdot
  e^{im(p_0(z_j)-p_0(\lambda_j))}\cdot
  \mathcal{V}_N
  \left(\la_j\mid\begin{matrix}\{\la\}\\ \{z\}\end{matrix}\right)
  \right\}\cdot \widetilde W_N\left(\begin{array}{c}\{\lambda\}\\  \{z\}\end{array}\right)
  \num
 \times \sum_{n=0}^N
 \sum_{\substack{\alpha\cup\bar\alpha
 \\ \#\bar\alpha=n}}
 \prod_{k\in\bar\alpha}\frac1{\sinh(z_k-\lambda_k)}
 \prod_{k\in\alpha}\left[\frac{ h_N(z_k|\{z\}|\{\lambda\})-1}{\sinh^2(z_k-\lambda_k)}\right]
 {\det_{j,k\in\bar\alpha}}~\frac1{\sinh(z_k-\lambda_j)}  .
 \end{multline}

The integrals over $\{z\}_{\alpha}$ can now be easily computed using \eq{3-prop-h2}. We obtain
 \begin{multline}\label{3-partit}
 \langle e^{\beta \mc{Q}_m}\rangle=\frac1{\det_N\Theta}
 \sum_{n=0}^N
 \sum_{\substack{\alpha\cup\bar\alpha
 \\ \#\bar\alpha=n}}
 \oint\limits_{\;\Gamma(\{\lambda\})}
 \prod_{j\in\bar\alpha}\left\{
  \frac{\dd z_j}{2\pi i}\cdot
  e^{im(p_0(z_j)-p_0(\lambda_j))}\cdot
  \mathcal{V}_{n}
  \left(\la_j\mid\begin{matrix}\{\la\}_{\bar\alpha}\\ \{z\}_{\bar\alpha}\end{matrix}\right)
  \right\}\num
 \hspace{-3mm}
 \times
 \widetilde W_{n}\left(\begin{array}{c}
 \{\lambda\}_{\bar\alpha}\\ \{z\}_{\bar\alpha}\end{array}\right)\cdot
 \prod_{j\in\alpha}\left(2\pi iM\tilde\rho(\lambda_j)\right) \cdot
 \prod_{k\in\bar\alpha}\frac1{\sinh(z_k-\lambda_k)}
 \cdot {\det_{j,k\in\bar\alpha}}~\frac1{\sinh(z_k-\lambda_j)}  ,
 \end{multline}
where
 \begin{equation}\label{3-trho}
 2\pi iM\tilde\rho(\lambda_j)=
 \log'\frac{a(\lambda_j)}{d(\lambda_j)}-i\sum_{a=1}^NK(\lambda_{j}-\lambda_a) \, .
 \end{equation}
The notation $\tilde\rho(\la)$ is motivated by the fact that
$\tilde\rho(\lambda)$ goes
to the ground state density $\rho(\lambda)$ \eq{0-rho}
in the thermodynamic limit.

It remains to replace the summation over partitions by a sum over
individual $\lambda$'s. Let us denote by $\Lambda$ the original set of
spectral parameters $\{\lambda_1,\dots,\lambda_N\}$ describing the
ground state. We then have
 \begin{multline}\label{3-fin-answ}
 \langle e^{\beta \mc{Q}_m}\rangle
 =\frac{\prod_{j=1}^N\left(2\pi iM\tilde\rho(\lambda_j)\right)}
       {\det_N\Theta}
  \sum_{n=0}^N\frac{1}{n!}
  \sum_{\lambda_1,\dots,\lambda_n\in\Lambda}
  \prod_{j=1}^n \left(\frac1{2\pi iM\tilde\rho(\lambda_j)}\right)
 \num
 \times
\oint\limits_{\Gamma(\{\la\})}\prod_{j=1}^n \frac{\dd z_j}{2\pi i}
 \cdot
 \prod_{j=1}^n
 \frac{e^{im(p_0(z_j)-p_0(\lambda_j))}}{\sinh(z_j-\lambda_j)}\cdot
 \mc{F}_n\tbl{\la}{z}
 \cdot\det_n\frac1{\sinh(z_k-\lambda_j)}.
 \end{multline}
with
 \begin{equation}\label{6-F-VW}
 \mc{F}_n\tbl{\la}{z}
 =\widetilde{W}_n\left(\barr{c} \{\lambda\} \\ \{z\} \ea
 \right)
 \prod_{j=1}^n {\cal V}_n
 \left(\lambda_j\mid\barr{c} \{\lambda\} \\ \{z\} \ea \right).
 \end{equation}
%


\subsection{Thermodynamic limit\label{TD}}

Recall that  the thermodynamic limit  corresponds to
$N,M\to\infty$ in such a way that the average density $D=N/M$
remains fixed. In this limit, discrete sums over  the ground state Bethe parameters $\lambda_k$  turn into integrals over
$[-q,q]$ weighted by a density function $\rho(\lambda)$,
 \begin{equation}\label{4-sum-int}
 \frac1M\sum_{\lambda_j\in\Lambda}f(\lambda_j)\to
 \int\limits_{-q}^qf(\lambda)\rho(\lambda)\,d\lambda.
 \end{equation}

It is easy to compute the limit of the pre-factor in
\eq{3-fin-answ}. Using the notation \eqref{3-trho}, we can express the determinant of the matrix $\Theta$ \eq{2-Gjk} as
 \begin{equation}\label{4-pre-lim}
 \det_N\Theta
 =
 \prod_{j=1}^N\left(2\pi iM\tilde\rho(\lambda_j)\right)
 \cdot
 \det_N\left[\delta_{jk}+\frac{K(\lambda_j-\lambda_k)}
 {2\pi M\tilde\rho(\lambda_j)}\right]  .
 \end{equation}
The equation \eq{3-trho} for
$\tilde\rho(\lambda)$ can be written in the form
 \begin{equation}\label{4-trho}
 2\pi \tilde\rho(\lambda_j)=p'_0(\lambda_j)-\frac1M\sum_{a=1}^NK(\lambda_{j}-\lambda_a)\, .
 \end{equation}
Replacing the  sum by the integral and using the integral equation
for the ground state density \eq{0-rho}, we find that in the
thermodynamic limit $\tilde\rho(\lambda)\to\rho(\lambda)$.
Therefore the determinant in the r.h.s. of \eq{4-pre-lim} goes to the Fredholm
determinant of the integral operator $I+\frac1{2\pi}K$ acting on
$[-q,q]$, namely,
 \begin{equation}\label{4-lim-prefact}
 \det_N\Theta\cdot\prod_{j=1}^N\left(2\pi
 iM\tilde\rho(\lambda_j)\right)^{-1}\to
 \det\left[I+\frac1{2\pi}K\right]  ,\qquad
 N,M\to\infty,
 \end{equation}
where the integral operator acts on the interval $[-q,q]$.

Hence, in the thermodynamic limit, the generating function
$\langle e^{\beta \mc{Q}_m}\rangle$ is given by the following series of
multiple integrals:
 \begin{multline}\label{4-fin-answ}
 \langle e^{\beta \mc{Q}_m}\rangle
 =\frac1{{\det}\left(I+\frac1{2\pi}K\right)}
 \sum_{n=0}^\infty\frac{1}{n!}
 \oint\limits_{\Gamma([-q,q])}
 \prod_{j=1}^n \frac{\dd z_j}{2\pi i}
 \int\limits_{-q}^q
 \prod_{j=1}^n \frac{\dd\lambda_j}{2\pi i}
 \num
 \times
 \mc{F}_n\tbl{\la}{z}\cdot
 \prod_{j=1}^n\frac{e^{im(p_0(z_j)-p_0(\lambda_j))}}{\sinh(z_j-\lambda_j)}
 \cdot\det_n\frac1{\sinh(z_k-\lambda_j)} \, .
 \end{multline}

The question of convergence of this series is quite complicated. To
understand this topic a little better, 
let us consider the free fermions point $\Delta=0$
($\zeta=\frac\pi2$).

In this point the terms of the series can be significantly simplified.
Indeed, it is obvious from \eqref{1-Vpm} that $\mathcal{V}_n\equiv \kappa- 1$ for
$\zeta=\frac\pi2$. It is less trivial, but true (see Appendix \ref{ap-FF}), that $\widetilde W_n\equiv1$.
Therefore $\mc{F}_n\equiv (\kappa-1)^n$, and we have
 \begin{multline}
 \langle e^{\beta \mc{Q}_m}\rangle_{{}_{\zeta=\frac\pi2}}
 =
 \sum_{n=0}^\infty\frac{(\kappa-1)^n}{n!}
 \oint\limits_{\Gamma([-q,q])}
 \prod_{j=1}^n \frac{\dd z_j}{2\pi i}
 \int\limits_{-q}^q\prod_{j=1}^n \frac{\dd\lambda_j}{2\pi i}
 \\
 \times
 \prod_{j=1}^n\frac{e^{im(p_0(z_j)-p_0(\lambda_j))}}{\sinh(z_j-\lambda_j)}
 \cdot
 \det_n\frac1{\sinh(z_k-\lambda_j)} \, .
 \label{4-FD-exp0}
 \end{multline}
The integrals over $z_j$ are now factorized and can be taken
explicitly. We obtain 
 \begin{equation}\label{4-FD-exp}
 \langle e^{\beta
 \mc{Q}_m}\rangle_{{}_{\zeta=\frac\pi2}}=\sum_{n=0}^\infty\frac{(\kappa-1)^n}{n!}
 \int\limits_{-q}^q\prod_{j=1}^n \dd\lambda_j
 \cdot\det_n\left[\frac{\sin\left(\frac{m}2(p_0(\lambda_j)-p_0(\lambda_k))\right)}
 {\pi\sinh(\lambda_j-\lambda_k)} \right] .
 \end{equation}
The series \eq{4-FD-exp} is an expansion of the Fredholm determinant
of the integral operator $I+V_{0}$ with kernel
 \begin{equation}\label{4-kernel}
 V_{0}(\lambda,\mu)
 =(\kappa-1)
 \frac{\sin\left\{\frac{m}2[p_0(\lambda)-p_0(\mu)]\right\}}
 {\pi\sinh(\lambda-\mu)}\, .
 \end{equation}
The general theory of Fredholm determinants ensures that the series
\eq{4-FD-exp} is absolutely convergent and  define an
entire function of $\kappa$. It  means that the expansion coefficients decay faster  than exponentially.

In the case of general $\Delta$, the series \eq{4-fin-answ} cannot
be reduced to a simple form as in \eq{4-FD-exp}.
As mentioned above, the analysis of its convergence is therefore rather complicated.
Nevertheless, taking into account that the XXZ chain
with a general anisotropy parameter can be considered as a smooth
deformation of the $\Delta=0$ case, we shall assume in the following that the
expansion \eq{4-fin-answ} also is absolutely convergent.

As a conclusion to this section, we would like to stress the role of
the non-zero magnetic field. As we already  mentioned, the support
of the ground state density at non-zero magnetic field is a finite
interval $[-q,q]$. Therefore, all the multiple integrals in the
series \eq{4-fin-answ}  are convergent. The pre-factor
$\det[I+\frac1{2\pi} K]^{-1}$ is also finite. However, at zero
magnetic field, $q\to\infty$, and it is easy to see that each
separate integral in the series \eq{4-fin-answ} becomes divergent.
This divergence should be of course compensated by the divergence of
the Fredholm determinant $\det[I+\frac1{2\pi} K]$. The latter
becomes ill-defined at $q\to\infty$, because the kernel of the
integral operator depends on the difference of the parameters, i.e.
$K=K(\lambda-\mu)$. It is not clear, however, how one can extract
the finite part from the representation \eq{4-fin-answ} directly at
$h=0$. Therefore one can say that the non-zero magnetic field plays
the role of a regularization in the framework of our approach. One
can check however that, in the end of the computation, the obtained
results have smooth limits at $h \to 0$, although the corresponding
proof is highly non-trivial.


\section{Cycle integrals}
\label{sec-cycles}

Even in the simplest case of free fermions where the multiple
integrals are decoupled, the calculation of the large $m$ behavior
of $\moy{\ex{\beta \mc{Q}_m}}$ remains non-trivial. Indeed, one has
to analyze the large $m$ behavior of the Fredholm determinant of the
integral operator $I+V_0$ whose kernel is given in \eq{4-kernel}. In the general case, the situation is even worse as the multiple integrals are highly coupled and hence the corresponding series  cannot be identified with the Fredholm series for the determinant of
some integral operator.

However, the series \eq{4-fin-answ} can be
rearranged in a specific way, making possible the asymptotic analysis
of its separate terms. The key observation towards such a rearrangement comes from the free fermion point.
There the leading asymptotic effect is produced by
traces of powers of $V_0$ : $\tr(V_0^s)$ behaves as
$\e{O}(m)$ when $m\to\infty$ for $s=1,2,\dots$.
Analogs of such traces of powers of $V_0$ will be given, in the general case, by the following type of integrals:
 \begin{equation}\label{5-def-CI}
 \mathcal{I}_s\pac{\mc{G}_s}
 =\oint\limits_{\Gamma([-q,q])}
 \hspace{-3mm} \f{\dd^s z}{(2\pi i)^s}
 \Int{-q}{q}\f{\dd^s \la}{(2\pi i)^s} \;
 \mc{G}_{s}\tbl{\la}{z}
 \pl{j=1}{s}
 \f{\ex{i m (p_0(z_j)-p_0(\la_j))  }}
 {\s{z_j-\la_j}\s{z_j-\la_{j+1}}}\, ,
 \end{equation}
in which the $2s$ integrals are coupled via some holomorphic function $\mc{G}_s$ of $2s$ variables $\la_1,\ldots,\la_s$, $z_1,\ldots,z_s$, symmetric separately in the $s$ variables $\la$ and in the $s$ variables $z$. Note that, in \eqref{5-def-CI}, we identify $\la_{s+1}$ with $\la_1$.

In this section, we explain how the representation \eq{4-fin-answ} can be decomposed into such integrals, and how the asymptotic behavior of the latter can be analyzed in the large $m$ limit.

\subsection{The cycle expansion}

The decomposition of the series \eqref{4-fin-answ} into  multiple integrals of the form \eqref{5-def-CI}, which plays a key role in its asymptotic analysis, comes directly from the {\em cycle expansion} of the Cauchy determinant. In this article
we use the following definition for a {\em cycle}:

\begin{Def} Let $A=\pa{a_{i,j}}$ be an arbitrary $n\times n$ matrix. A
 cycle of length $\ell$  is any product of entries
of A of the form
\begin{equation*}
 a_{j_1,j_2}\; a_{j_2,j_3}\dots  a_{j_{\ell-1},j_\ell}\;
 a_{j_\ell,j_1}\, .
\end{equation*}
With this terminology, integrals of the form \eqref{5-def-CI} will be called {\em cycle integrals}.
\end{Def}

It is well known (see e.g. \cite{Bon04L}) that  the determinant  of
any matrix $A$ can be presented as a sum of products of cycles of
different lengths. We call such a representation of a determinant  its {\em cycle expansion}.
In our case, the determinant we need to decompose into cycles (namely, the Cauchy determinant in \eqref{4-fin-answ}) is part, together with some symmetric function, of the integrand of a $n$-fold integral.
For such kind of representations we have the following result:
\begin{prop}\label{5-mult-cycl}
Let $\msc{C}_{\la}$ and $\msc{C}_z$ be two curves in $\Cx$,
$g$ a continuous function in $(z,\la)\in\msc{C}_{z}\times
\msc{C}_\la$, and $\mc{G}_n$ a continuous
function in $n$ variables $\la$ and $n$ variables $z$ on
$\msc{C}^n_{\la}\times \msc{C}^n_z$.
Assume moreover that, for any permutation $\sg\in\mf{S}_n$, $\mc{G}_n$ is
invariant under replacements of pairs $\pa{\la_i,z_i}\mapsto
\pa{\la_{\sg\pa{i}},z_{\sg\pa{i}}}$. Then,
\begin{multline}
\frac{1}{n!}\Int{\msc{C}_{\la}}{} \! \dd^n \la  \Int{\msc{C}_z}{} \! \dd^n z \  \mc{G}_n\pab{\paa{\la}}{\paa{z}} \, \det_n \big[g(z_{i},\la_{j})\big] \\
=
\sul{\substack{\ell_1,\dots,\ell_n =0 \\ \sul{k=1}{n}k\ell_k=n }
}{n}
C(n|\{\ell\})
\Int{\msc{C}_{\la}}{} \! \dd^n \la  \Int{\msc{C}_z}{} \! \dd^n z \
\mc{G}_n\pab{\paa{\la}}{\paa{z}}
\pl{t\in J_{\{\ell\}}  }{}
g(z_t ,\la_{t+1} )
,
\label{expansion of n-fold integral in cycles}
\end{multline}
where
\begin{equation}
 C(n|\{\ell\})= \pl{s=1}{n} \f{1}{\ell_{s}!}
 \paf{(-1)^{s+1}}{s}^{\ell_s}.
\end{equation}
Here $J_{\{\ell\}}$ denotes the following set of triplets,
associated to the configuration given by $\ell_1,\ldots,\ell_n$,
\begin{equation}\label{5-set-Jell}
J_{\{\ell\}}=\{(s,p,j)  : 1\le s \le n ,\,  1\le p \le \ell_s ,
\,  1\le j \le s \},
\end{equation}
and, for $t=(s,p,j)$, $t+1$ denotes the triplet $(s,p,j+1)$, with $(s,p,s+1)\equiv(s,p,1)$.
\end{prop}

\begin{rem}
The notations introduced above should be understood as follows:
$s$ labels the length of a cycle, $\ell_s$ stands for the number of
cycles of length $s$ and $j$ marks the position of the variable in
the corresponding cycle. Hence, for $t=(s,p,j)$, $\mu_t\equiv \mu_{s,p,j}$ stands for the
$j^{\e{th}}$ variable of the $p^{\e{th}}$ cycle of length $s$. In
this notation we assume that $\mu_{s,p,s+1}\equiv \mu_{s,p,1}$.
\end{rem}

\Proof Let $g$ be a  $n\times n$ matrix. Then,
 \begin{align}
 \det_n[g]
 &=\frac1{n!}\left.\frac{\partial^n}{\partial\gamma^n}
          \det_n[I+\gamma g]\right|_{\gamma=0}
  =\frac1{n!}\left.\frac{\partial^n}{\partial\gamma^n}
 \exp\bigl(\tr\log[I+\gamma g]\bigr)\right|_{\gamma=0}
        \nonumber\\
 &=\frac1{n!}\left.\frac{\partial^n}{\partial\gamma^n}
   \prod_{s=1}^n\exp\left\{\frac{(-1)^{s+1}\gamma^s}s
                           \tr(g^s)\right\}\right|_{\gamma=0}
         \nonumber\\
 &=\frac1{n!}\left.
 \frac{\partial^n}{\partial\gamma^n}
 \sul{\ell_1,\dots,\ell_n =0}{\infty}\hspace{-2mm}
 C(n|\{\ell\})
 \prod_{s=1}^n \bigl[\gamma^s\tr(g^s)\bigr]^{\ell_s}\right|_{\gamma=0}
          \nonumber\\
 &=\sul{\substack{\ell_1,\dots,\ell_n =0\\ \Sg k\ell_k=n }}{n}
  \hspace{-2mm}
  C(n|\{\ell\})\,
  \prod_{s=1}^n \bigl[\tr(g^s)\bigr]^{\ell_s}.
\label{5-Proof-CE-1}
 \end{align}
Substituting here $g_{ij}=g(z_i,\lambda_j)$ and writing explicitly
all $\bigl[\tr(g^s)\bigr]^{\ell_s}$, we obtain
 \begin{equation}
 \det_n\big[g(z_i,\lambda_j)\big]
 =
 \sul{\substack{\ell_1,\dots,\ell_n =0 \\ \Sg k\ell_k=n } }{n}
 \hspace{-2mm}
 C(n|\{\ell\})\,
 \prod_{s=1}^n\pl{p=1}{\ell_s}\hspace{2mm}
 \sum_{i_{(s,p,1)},\ldots,i_{(s,p,s)}=1}^n
 \pl{j=1}{s} g\big(z_{i_{(s,p,j)}},\la_{i_{(s,p,j+1)}}\big)\; ,
 \label{5-Proof-CE-2}
 \end{equation}
It remains to observe that, in the product of the functions
$g\big(z_{i_{(s,p,j)}},\la_{i_{(s,p,j+1)}}\big)$, all $i_{(s,p,j)}$ should be
different, otherwise the corresponding term does not contribute to
the determinant (all other contributions eventually cancel out). Hence, the multiple sum over $i_{(s,p,j)}$ is in
fact the sum over permutations:
 \begin{equation}\label{5-Proof-CE-3}
 \det_n\big[g(z_i,\lambda_j)\big]=
 \sul{\substack{\ell_1,\dots,\ell_n =0 \\ \Sg k\ell_k=n } }{n}
 \hspace{-2mm}
 C(n|\{\ell\})
 \sum_{\sigma\in\mf{G}_n} \,
 \prod_{t\in J_{\{\ell\}}}
 g\big(z_{\sg(t)},\la_{\sg(t)+1}\big).
 \end{equation}
Due to the symmetry properties of the function $\mc{G}_n$, all the
terms of the sum over permutations give the same contribution to the integral.
\qed

Let us apply this result to the Cauchy determinant under the
multiple integrals in the series \eq{4-fin-answ} for $\moy{e^{\beta \mc{Q}_m}}$. Denoting
 \begin{equation}\label{5-Doubleangle}
 \moy{\ex{\beta \mc{Q}_m}}
 =\frac{\Mmoy{\ex{\beta \mc{Q}_m}}}
       {\det\left[I+{\textstyle\frac1{2\pi}}K\right]}\,,
 \end{equation}
we obtain
 \begin{multline}\label{exp beta Q en termes de cycles}
 \Mmoy{\ex{\beta \mc{Q}_m}}=
 \sul{n=0}{\infty}
 \sul{\substack{\ell_1, \dots, \ell_n=0 \\
 \Sg k\ell_k=n}}{\infty}
 \hspace{-2mm}
 C(n|\{\ell\})
 \hspace{-1mm}
 \oint\limits_{\Gamma([-q,q])}\hspace{-3mm} \f{\dd^n z}{(2\pi i)^{n}}
 \int\limits_{-q}^{q} \f{\dd^n \la}{(2\pi i)^{n}}\;
 \mc{F}_n\left(\barr{c} \{\lambda\} \\ \{z\} \ea \right)\num
 \times\pl{t\in J_{\{l\}}}{} \f{\ex{im(p_0(z_t)-p_0(\la_t)
 )}}{\s{z_t-\la_t}\s{z_t-\la_{t+1}}}\,.
 \end{multline}
It is actually convenient to remove the constraint $\Sg k\ell_k=n$ on the summation variables $\ell_k$ in the sum \eqref{exp beta Q en termes de cycles}. This can be done by introducing, in the $n^{\e{th}}$ term of the series, an $n^{\e{th}}$-derivative over some auxiliary parameter $\gamma$, similarly as in the proof of Proposition~\ref{5-mult-cycl}:
\begin{multline}
 \Mmoy{\ex{\beta {\mc{Q}_m}}}
 = \sul{n=0}{\infty}
 \f{1}{n!}
 \f{\Dp{}^n}{\Dp{}\ga^n }
 \sul{\ell_1, \dots,\, \ell_n=0 }{\infty}\,
 \pl{k=1}{n} \frac1{\ell_k!}
   \left(\f{(-1)^{k+1} \ga^k}{k}\right)^{\!\ell_k}
 \hspace{-2mm}
 \oint\limits_{\Gamma([-q,q])}\hspace{-1mm}
 \prod_{t\in J_{\{\ell\}}}\!\!\f{\dd z_t}{2\pi i}
 \int\limits_{-q}^{q}
 \prod_{t\in J_{\{\ell\}}}\!\!\f{\dd \la_t}{2\pi i}
          \\
 \times
 \mc{F}_{|J_{\{\ell\} }|}
 \left(\barr{c} \{\lambda\} \\ \{z\} \ea \right)
 \left.\pl{t\in J_{\{\ell\}}}{}\!
 \f{\ex{im\left(p_0(z_t)-p_0(\la_t) \right)}}
   {\s{z_t-\la_t}\s{z_t-\la_{t+1}}}
 \,\right|_{\gamma=0},
\label{exp beta Q sans contrainte}
\end{multline}
where $|J_{\{\ell\} }|=\Sg k \ell_k$ denotes the cardinality of the set $J_{\{\ell\}}$ (this cardinality is no longer equal to $n$ since we have removed the constraint).

The $n^{\e{th}}$ term in the series for $\Mmoy{\ex{\beta
\mc{Q}_m}}$ is thus expressed as some sum of multiple cycle integrals of the form \eqref{5-def-CI}. Note however that these cycle integrals are coupled through the function $\mc{F}_{|J_{\{\ell\} }|}$ which involves the total number $2|J_{\{\ell\} }|$ of variables, and not only the $2s$ variables associated to a given cycle.
In order to formalize this fact, we need to introduce some notations.

Let us first recall the standard order on $\mathbb{N}^3$:
\begin{equation*}
 (t_1,t_2,t_3)< (u_1,u_2,u_3)
  \Leftrightarrow
 \{ t_1 < u_1 \; \e{or} \;
 \{ t_1=u_1\; \e{and}\;
 \{ t_2 <u_2 \; \e{or} \;  \{ t_2=u_2 \; \e{and} \;  t_3< u_3 \} \}\}\} ,
\end{equation*}
and let $\mf{id}$ be the identity operator on functions of one variable $\la$ and one variable $z$:
\begin{equation}
\mf{id}\pac{\mc{G}\pab{\la}{z}}=\mc{G}\pab{\la}{z} \;.
\end{equation}
For some cycle configuration $\{\ell\}$, we define $\mc{I}_{(s,p)}$ as the operator acting on functions $\mc{G}_{|J_{\{\ell\}}|}$ of $2|J_{\{\ell\}}|$ variables $\la_t$ and $z_t$, $t\in J_{\{\ell\}}$, as the cycle integral $\mathcal{I}_s$ \eqref{5-def-CI} over the variables $\la_{s,p,j}$, $z_{s,p,j}$, $1\le j \le s$, i.e.
\begin{equation}
\mc{I}_{(s,p)}
=\bigotimes\limits_{t<(s,m,1)}
 \underbrace{\id_{t}}_{\substack{ \e{variables} \\
                            \la_{t} ,\, z_{t}}}
   \bigotimes\
                            \mc{I}_{s}
   \hspace{-1mm}
 \bigotimes\limits_{{t}>(s,m,s)}
         \underbrace{\id_{{t}}}_{\substack{ \e{variables} \\
                            \la_{t} ,\, z_{t}}}     .
\end{equation}
Note that the action on $\mc{G}_{|J_{\{\ell\}}|}$ of a single $\mc{I}_{(s,p)}$ (associated to a given cycle $(s,p)$) still produces a
function (with respect to the other variables). However, the action on $\mc{G}_{|J_{\{\ell\}}|}$ (denoted here with the symbol *) of the product of such operators $\mc{I}_{(s,p)}$ with respect to all cycles,
\begin{multline}\label{act-Isp}
\left(\pl{s=1}{n}\pl{p=1}{\ell_s} \mc{I}_{(s,p)}\right)
*
\mc{G}_{|J_{\{\ell\}}|} 
 = \pl{s=1}{n} \pl{p=1}{\ell_s} \pl{j=1}{s}
\left( \oint \f{d z_{s,p,j}}{2\pi i}
 \int\limits_{-q}^{q}\f{d\la_{s,p,j}}{2\pi i}\right)
 \,\cdot\mc{G}_{  \abs{J_{ \paa{\ell}} }  } \pab{  \paa{\la_{s,p,j}}  }{  \paa{z_{s,p,j}}
 }\\
%
 %
 \times\prod_{s=1}^{n}\pl{p=1}{\ell_s}\pl{j=1}{s}
 \f{\ex{im(p_0(z_{s,p,j})-p_0(\la_{s,p,j}))  }}
 {\s{z_{s,p,j}-\la_{s,p,j}}\s{z_{s,p,j}-\la_{s,p,j+1}}}\;,
\end{multline}
produces a number, since all variables $\la$
and $z$ are integrated.

This notation enables us
to recast the generating function in a quite compact form:
%
%
%
%
 \begin{equation}\label{6-Q1-Gn}
 \Mmoy{\ex{\beta {\mc{Q}_m}}}= \sul{n=0}{\infty}
 \f{1}{n!}\left.\f{\Dp{}^n}{\Dp{}\ga^n }\,
  G_{1\dots n}(\gamma)\right|_{\gamma=0},
 \end{equation}
where
 \begin{equation}\label{6-Gn}
 G_{1\dots n}(\gamma)=\sul{\ell_1, \dots,\, \ell_n=0 }{\infty} \pl{k=1}{n}
 \frac1{\ell_k!}
   \left(\f{(-1)^{k+1} \ga^k}{k}\right)^{\!\ell_k}\cdot\left(
  \pl{s=1}{n}\pl{p=1}{\ell_s} \mc{I}_{(s,p)}\right)
  *
  \mc{F}_{|J_{ \{\ell\} } | } 
   \, .
 \end{equation}
%

Therefore, the whole problem of the asymptotic analysis of the series
\eqref{4-fin-answ} for $\moy{e^{\beta \mc{Q}_m}}$ reduces to the derivation of an asymptotic formula for $G_{1 \dots n}\pa{\ga}$.
The advantage of such a representation is the following: in spite of the fact that all cycle integrals are coupled with each others through the function $\mc{F}_{ | J_{ \paa{\ell} } | }$,
the asymptotic behavior for large $m$ of every set  of cycle integrals can
nevertheless be computed separately, independently of the other integrals; then
the asymptotic formula for multiple cycle integrals can be obtained
by applying consecutively the results of the analysis of each pure cycle
integral $\mc{I}_{(s,p)}$ to the corresponding group of cycle variables
$ \paa{\la_{s,p,j}}_{1\le j\le s}$ and $ \paa{z_{s,p,j}}_{1\le j\le s}$, while considering the
remaining variables fixed. Passing from the exact formula to the
asymptotic one produces a very powerful effect: due to peculiar
properties of the function $\mc{F}_{  \abs{ J_{ \paa{\ell} } } }$,
the cycles become ``quasi-decoupled'' in the $m\tend +\infty$ limit.
Such a quasi-decoupling is enough to perform the sum over all the
possible numbers $\ell_k$ of cycles of length $k$.

The summation procedure  of the asymptotics of multi-cycle integrals will be
presented in  Section~\ref{sec-sum}. However, first we need to remind the results of \cite{KitKMST08a} concerning the
asymptotic analysis of pure cycle integrals.

\subsection{Asymptotic analysis of cycle integrals\label{AACI}}

The asymptotic study of cycle integrals of the form \eqref{5-def-CI} was performed in \cite{KitKMST08a} from the one of the Fredholm determinant of a generalized sine kernel. It was shown there that such integrals admit, up to any arbitrary order $n_0$, an asymptotic expansion of the form
\begin{multline}\label{5-asympt-Is}
 \mc{I}_s[\mc{G}_s]=I_s^{(0)}[\mc{G}_s]
 + \sum_{n=1}^{n_0}\frac{1}{m^n} I_s^{(n;\,\e{nosc})}[\mc{G}_s]
                 \\
 + \sum_{\substack{r\in\mathbb{Z}^*\\ |r|\le \frac{n_0}{2} }}
   e^{irm[p_0(q)-p_0(-q)]}
   \sum_{n=2|r|}^{n_0} \frac{1}{m^n} I_s^{(n;\, r)}[\mc{G}_s]
 + \e{O}\bigg(\frac{\log^{s} m}{m^{n_0+1}}\bigg).
\end{multline}
The leading behavior (up to o(1) corrections) is given by $I_s^{(0)}[\mc{G}_s]$, which contains a linear term in $m$, a term proportional to $\log m$, and a constant term. Its explicit form, which was computed in \cite{KitKMST08a}, will be specified later. This leading term admits two kinds of corrections: {\em oscillating}  (containing an oscillating factor $e^{irm[p_0(q)-p_0(-q)]}$), and {\em non-oscillating} ones\footnote{$I_s^{(n;\, r)}[\mc{G}_s]$ and $I_s^{(n;\,\e{nosc})}[\mc{G}_s]$ still depend on $m$, but simply as a polynomial (of degree at most $s$) of $\log m$.}. Note that the leading term $I_s^{(0)}[\mc{G}_s]$ itself is {\em non-oscillating}.

In order to derive the asymptotic behavior of $\moy{e^{\beta\mc{Q}_m}}$, one will have to sum up \eqref{5-asympt-Is} according to \eqref{6-Gn}. If we suppose that subleading corrections in \eqref{5-asympt-Is} remain subleading through this process of summation (which is not completely true, as it will be explained later), then it is in principle enough, so as to obtain the leading asymptotic behavior of $\moy{e^{\beta\mc{Q}_m}}$, to consider only the leading term $I_s^{(0)}[\mc{G}_s]$ in \eqref{5-asympt-Is}. Recall however that our main goal is not $\moy{e^{\beta\mc{Q}_m}}$ itself, but the spin-spin correlation function $\moy{\sg_1^z\,\sg_{m+1}^z}$ obtained from $\moy{e^{\beta\mc{Q}_m}}$ by second derivative with respect to $\beta$ and by {\em second lattice derivative} (see \eqref{0-corr-funct}).
This process of taking the second lattice derivative will decrease the order of the non-oscillating terms, but not of the oscillating ones, as we may merely differentiate the exponential.
Therefore, to be sure to get the leading behavior of $\moy{\sg_1^z\,\sg_{m+1}^z}$, we will have to consider not only the leading (non-oscillating) term $I_s^{(0)}[\mc{G}_s]$, but also the {\em leading oscillating correction}, given by
\begin{equation}\label{5-def-O}
O_s[\mc{G}_s]=O_s^+[\mc{G}_s] + O_s^-[\mc{G}_s],
\quad \text{with}\quad
O_s^\pm= \frac{1}{m^2} e^{\pm i m [p_0(q)-p_0(-q)]} I_s^{(2;\,\pm 1)}.
\end{equation}
%
It will indeed appear that the non-oscillating term gives rise to an $m^{-2}$ behavior for $\moy{\sg_1^z\,\sg_{m+1}^z}$, whereas the oscillating one produces an $m^{-\theta}$ decrease, $\theta$ being lower or higher than $2$ according to the value of $\Delta$.

To specify the functional action of $I_s^{(0)}$ and $O_s$ on the function $\mc{G}_s$ and to apply it to our particular case, let us first briefly remind the main idea of the analysis performed in \cite{KitKMST08a}.

\subsubsection{From generalized sine kernel to cycle integrals}

Suppose that the function $\mc{G}_s$ appearing in \eqref{5-def-CI}
is realized as a product of functions of one variable, i.e.
\begin{equation}\label{5-pure-prod}
\mc{G}_s\pab{\paa{\la}}{\paa{z}}
=\pl{i=1}{s} \varphi\pa{\la_i}\,\ex{g\pa{z_i}},
\end{equation}
with $g$ and $\varphi$ holomorphic in some vicinity of $[-q,q]$.
Then the cycle integral $\mc{I}_{s}[\mc{G}_s]$ \eqref{5-def-CI} can
be explicitly computed in terms of the $s^{\e{th}}$ $\ga$-derivative
of the  Fredholm determinant of the operator $I+\ga V$ on $[-q,q]$,
with kernel
 \begin{equation}\label{5-kernel}
 V(\lambda,\mu)= F(\lambda)\frac{\sin\left\{\frac{m}2\pac{p_0(\lambda)-p_0(\mu)}-
 \frac{i}2\pac{g(\lambda)-g(\mu)}\right\}}
 {\pi\sinh(\lambda-\mu)} ,
 \qquad F\pa{\la}=\ex{g(\la)}\varphi(\la).
 \end{equation}
Indeed, from the series representation of $\log \ddet{}{I+V}$, one has,
\begin{equation}\label{5-s-derivFD}
\f{\pa{-1}^{s-1}}{\pa{s-1}!}\Bigl. \Dp{\ga}^s\log \ddet{}{I+ \ga
V}\Bigr|_{\gamma=0} =\Int{-q}{q} \dd^s\la
\pl{k=1}{s}V\pa{\la_{k},\la_{k+1}}, \quad\text{with}\quad
\lambda_{s+1}\equiv\lambda_1,
\end{equation}
which is exactly the cycle integral
$\mc{I}_{s}\pac{\mc{G}_s}$ once the residues in $z$ have all been computed.

In order to obtain the asymptotic expansion for cycle integrals
involving more general holomorphic  functions
$\mc{G}_s$ that are symmetric with respect
to the variables $\la_1,\ldots,\la_s$ and $z_1,\ldots,z_s$ separately,
one uses the fact that such functions
admit an expansion of the form
 \begin{equation}\label{5-factorization}
 \mc{G}_{s}\tbl{\la}{z}
 =\sum_{k=1}^\infty\prod_{j=1}^s\varphi_k(\la_j)\,\phi_k(z_j),
 \end{equation}
where $\varphi_k(\lambda)$ and $\phi_k(z)$ are holomorphic in a
neighborhood of the interval $[-q,q]$.
Setting then
\begin{equation}
F_k(\la)=\varphi_k(\la)\phi_k(\la) \qquad
\ex{g_k(\lambda)}=\phi_k(\lambda)  ,
\label{choix de F et g}
\end{equation}
and denoting the corresponding kernel \eqref{5-kernel} as $V_k$, we obtain
\begin{equation}\label{5-Is-Fredholm}
\mc{I}_{s}\pac{\mc{G}_s}=\sul{k=1}{\infty}
\f{(-1)^{s+1}}{(s-1)!}\left. \Dp{\ga}^s \log\det\pac{I+ \ga
V_k}\right|_{\gamma=0}  .
\end{equation}
Hence, the asymptotics of cycle integrals for such $\mc{G}_s$ can be
inferred from those of the Fredholm determinant of $I+\ga V$, $V$
being a generalized sine kernel of the form \eqref{5-kernel} (see
\cite{KitKMST08a} for a more rigorous proof of all this procedure).

The large $m$ asymptotic behavior of $\log\det[I+\ga V]$ was
obtained in \cite{KitKMST08a} by Riemann-Hilbert approach. There it
was proved that, in the $m\tend+\infty$ limit,
\begin{equation}\label{5-asympt-V}
 \log\det[I+\ga
V]= \mc{W}_0(m)+\e{o}(1),
\end{equation}
and the leading asymptotic term $\mc{W}_0$ was computed explicitly
in terms of the function
 \begin{equation}\label{5-nu-g}
 \nu(\lambda)=-\frac1{2\pi i}\log\bigl(1+\gamma F(\lambda)\bigr),
 \qquad \nu_\pm=\nu(\pm q).
 \end{equation}
It reads
 \begin{multline}\label{5-asy-n-osc}
 {\cal W}_0(m,[\nu])
 =-\hspace{-1mm}
  \int\limits_{-q}^q \! \bigl[imp'_0(\lambda)+g'(\lambda)\bigr]
 \nu(\lambda)\,\dd\lambda
  -\hspace{-1mm}
 \sum_{\sg=\pm}\!
 \big\{\nu_\sigma^2 \log[m\sinh(2q)p'_0(\sg q)]
        - \log G(1,\nu_\sg)\big\}\\
 +\frac12
  \int\limits_{-q}^{q}\!
  \frac{\nu'(\lambda)\nu(\mu)-\nu(\lambda)\nu'(\mu)}
       {\tanh(\lambda-\mu)}\,
  \dd\lambda\,\dd\mu
 +\sum_{\sg=\pm} \sg\nu_\sg
  \int\limits_{-q}^{q}\!
  \frac{\nu_\sg-\nu(\lambda)}{\tanh(\sg q-\lambda)}\,\dd\lambda\, ,
 \end{multline}
where $G(1,z)=G(1+z)G(1-z)$, and $G(z)$ is the Barnes function
defined as the unique solution of the equation
\begin{equation}\label{5-Barnes}
 G(1+z)=\Gamma(z)G(z),\quad\mbox{with}\quad G(1)=1\quad\mbox{and}\quad
 \frac{\dd^3}{\dd z^3}\log G(z)\ge 0, \quad z>0.
\end{equation}
It was proved in \cite{KitKMST08a} that this leading term $\mc{W}_0$ gives rise to the leading term $I_s^{(0)}[\mc{G}_s]$ of \eqref{5-asympt-Is} through the procedure described above.

In \cite{KitKMST08a} were also computed the first oscillating and non-oscillating corrections to \eqref{5-asympt-V}.
Non-oscillating corrections to \eqref{5-asympt-V} lead to non-oscillating
corrections in \eqref{5-asympt-Is}, and we {\em a priori} do not need
them for our purpose. We have seen however that, even if the first
oscillating correction \eqref{5-def-O} in \eqref{5-asympt-Is} gives rise, through the
process of summation that we will see in the next section, to a
subleading term in the expansion for $\moy{e^{\beta\mc{Q}_m}}$, it may
become leading once the difference-differential operator \eqref{0-corr-funct} is applied.
Therefore we also recall the explicit expression of the first oscillating correction which was computed in \cite{KitKMST08a}. It reads,
\begin{equation}\label{5-leading-osc}
  \mc{W}_1(m,[\nu])\, e^{im[p_0(q)-p_0(-q)]}
                    +\mc{W}_{-1}(m,[\nu])\, e^{-im[p_0(q)-p_0(-q)]},
\end{equation}
with
 \begin{equation}\label{5-Wosc}
 {\cal W}_{\pm1}(m,[\nu])
 =\frac{\nu_+\nu_-\, u(q)^{\pm 1}}{\sinh^2(2q)\,p'_0(q)\,p'_0(-q)}\;
   m^{-2\pm 2(\nu_++\nu_-)}
 \end{equation}
and
 \begin{equation}\label{5-h}
 u(q)=\prod_{\sg=\pm}
      \left\{
      e^{\sg g(\sg q)}
      \left[\sinh(2q)\, p'_0(\sg q)\right]^{2\nu_\sg}
      \frac{\Gamma(-\nu_\sg)}{\Gamma(\nu_\sg)}\;
      \exp{\left[
        -2\sg\!\!\int\limits_{-q}^{q}\!\!
        \frac{\nu_\sg-\nu(\lambda)}{\tanh(q-\sg\lambda)}\dd\lambda
        \right]
        }
     \right\}
 .
 \end{equation}
The asymptotic estimates given above are uniform over $\ga$ for $\ga$ small enough: for all $\varepsilon>0$, there exists $\gamma_0$ such that, $\forall\ga<\ga_0$, $2|\nu_++\nu_-|<\varepsilon$ and the corrections to \eqref{5-leading-osc} are of order less than $\e{O}(m^{-3+\varepsilon})$.
Therefore \eqref{5-leading-osc} indeed gives rise, through the process described above (see \eqref{5-Is-Fredholm}), to the first oscillating correction $O_s$ \eqref{5-def-O} of \eqref{5-asympt-Is}.

\begin{rem}
 We would like to draw the reader's attention to a remarkable
relation between the terms ${\cal W}_0$  and ${\cal W}_{\pm1}$:
 \begin{equation}\label{5-relation}
 {\cal W}_{\pm 1}(m,[\nu])\, \ex{\pm i m  [p_0(q)-p_0(-q)]}
=\ex{{\cal W}_0(m,[\nu\mp 1])-{\cal W}_0(m,[\nu])}\,.
 \end{equation}
Hence, the terms $\mc{W}_{\pm 1}$ partly restore the original
periodicity $\nu\rightarrow \nu+n$, $n\in\mathbb{Z}$, of $\det[I+\ga
V]$, that is broken if one considers the leading term ${\cal W}_0$
alone.
\end{rem}

\subsubsection{Leading asymptotic behavior of cycle integrals}

We have seen how to generate the leading term $I_s^{(0)}[\mc{G}_s]$ and its first oscillating correction $O_s[\mc{G}_s]$, in the asymptotic expansion \eqref{5-asympt-Is} of the cycle integral \eqref{5-def-CI}:
$I_s^{(0)}[\mc{G}_s]$ is generated through \eqref{5-Is-Fredholm} by the leading asymptotic part $\mc{W}_0$ \eqref{5-asy-n-osc}, whereas $O_s[\mc{G}_s]$ comes from \eqref{5-leading-osc}. It remains, in order to be able to sum up their respective contributions in \eqref{6-Gn}, to describe more precisely their action as a functional of $\mc{G}_s$.

For the purpose of the summation of the series \eq{6-Gn}, it will be convenient to distinguish the $g$-dependent part of $\mc{W}_0$, presenting the latter as a sum of
two terms ${\cal W}_{0}={\cal W}_{0}^{(0)}+{\cal W}_{0}^{(g)}$,
where
 \begin{equation}\label{5-W0g}
 {\cal W}_{0}^{(0)}(m,[\nu])={\cal W}_{0}(m,[\nu])\Bigl.\Bigr|_{g=0}
 \qquad
 \text{and}\qquad
 {\cal W}_{0}^{(g)}[\nu]=-\int\limits_{-q}^q g'(\lambda)\,
 \nu(\lambda)\,\dd\lambda.
 \end{equation}
It will split accordingly the functional $I_s^{(0)}[\mc{G}_s]$ into two parts:
\begin{equation}\label{6-Funct-form}
 I_s^{(0)}[\mc{G}_s]
 =\pa{H_s+D_s}[\mc{G}_s],
 \end{equation}
where   $H_s[\mc{G}_s]$ is the part of $I_s^{(0)}[\mc{G}_s]$ issued from ${\cal W}_{0}^{(0)}$, whereas $D_s[\mc{G}_s]$ comes from $\mc{W}_{0}^{(g)}$.

To obtain the explicit expression of $D_s$ as a functional of $\mc{G}_s$, one should proceed as follows: compute the $s^{\e{th}}$
$\gamma$-derivative of $\mc{W}_0^{(g)}[\nu]$ at $\gamma=0$,
\begin{equation}\label{5-s-derivWg}
\f{\pa{-1}^{s-1}}{\pa{s-1}!}\;
\Bigl. \Dp{\ga}^s\,\mc{W}_0^{(g)}[\nu]\Bigr|_{\gamma=0}
=\Int{-q}{q}\frac{\dd\la}{2\pi i}\;
 g'(\lambda)\, F^s(\lambda)\,,
\end{equation}
substitute in this expression $F(\lambda)$ and $g(\lambda)$ by \eq{choix de F et g},
\begin{equation}\label{5-s-derivWg1}
\f{\pa{-1}^{s-1}}{\pa{s-1}!}\;
\Bigl. \Dp{\ga}^s\,\mc{W}_0^{(g)}[\nu]\Bigr|_{\gamma=0}
 =\Int{-q}{q}\frac{\dd\la}{2\pi i}\;
 g'_{k}(\lambda)\,\big[\varphi_k(\lambda)\,e^{g_k(\lambda)}\big]^s\, ,
\end{equation}
and take finally the sum over $k$ as in \eqref{5-Is-Fredholm}. This gives
\begin{equation}\label{6-actionD}
D_s\pac{\mc{G}_s}
=\Int{-q}{q}\f{\dd \la}{2\pi i}\;
\biggl.
\Dp{\eps}\mc{G}_s\pab{\paa{\la}^s}{\la+\eps,\paa{\la}^{s-1}}\biggr|_{\eps=0}
\,.
\end{equation}
The notation $\{\lambda\}^s$ means here that all the original cycle
variables $\lambda_1,\dots,\lambda_s$ are now set equal to the same value
$\lambda$. The variables $z_2,\dots,z_s$ are likewise equal to $\lambda$. Notwithstanding, one should set here $z_1=\lambda$
only after taking the derivative with respect to this variable.

One can derive similarly explicit formulas for the action of
$H_s$ and $O_s$. However, for general functions $\mc{G}_s$,
the corresponding expressions are rather cumbersome (see
\cite{KitKMST08a}). Therefore we restrict our analysis to a specific
class of functions. Let us suppose from now on that $\mc{G}_s$ can be presented as
 \begin{equation}\label{8-formG}
 \mc{G}_s\left(\barr{c} \{\lambda\} \\ \{z\} \ea \right)
 =\prod_{j=1}^s
  \Phi_1\left(\lambda_j\mid\barr{c} \{\lambda\} \\ \{z\} \ea \right)
  \cdot
  \Phi_2\left(\barr{c} \{\lambda\} \\ \{z\} \ea \right),
 \end{equation}
where the functions
\begin{equation}\label{phi1-phi2}
 \Phi_1\left(\omega\mid\!\barr{c} \{\lambda\} \\ \{z\} \ea \!\right),
 \qquad
  \Phi_2\left(\!\barr{c} \{\lambda\} \\ \{z\} \ea \!\right),
\end{equation}
are symmetric functions of the $s$ variables
$\{\lambda\}$ and of the $s$ variables $\{z\}$ separately, and satisfy the reduction property
\begin{equation}\label{8-recurs}
\left.
\Phi_1\left(\omega\mid\!\barr{c} \{\lambda\} \\ \{z\} \ea\! \right)
\right|_{\lambda_j=z_j }
\hspace{-4mm}
=
 \Phi_1\left(\omega\mid\!\barr{c} \{\lambda\}\setminus \lambda_j
              \\ \{z\}\setminus z_j   \ea\! \right),
\quad
\left.
\Phi_2\left(\!\barr{c} \{\lambda\} \\ \{z\} \ea \!\right)
\right|_{\lambda_j=z_j }
\hspace{-4mm}
=
 \Phi_2\left(\!\barr{c} \{\lambda\}\setminus \lambda_j
             \\ \{z\}\setminus z_j   \ea \!\right).
\end{equation}
Note that, if the set $\{\lambda\}$ coincides with the set $\{z\}$, then the function
$\Phi_2$ becomes a constant whereas $\Phi_1$ becomes a one-variable function:
 \begin{equation}\label{8-full-red2}
 \Phi_2\left(\!\barr{c} \{\lambda\}
              \\ \{\lambda\} \ea \!\right)=
 \Phi_2\left(\!\barr{c} \emptyset\\ \emptyset \ea \!\right)
 \equiv\Phi_2=\text{const},
 \quad
 \Phi_1\left(\omega\mid\!\barr{c} \{\lambda\}
             \\ \{\lambda\} \ea \!\right)
 =\Phi_1\left(\omega\mid\!\barr{c}
              \emptyset\\ \emptyset \ea \!\right)
 \equiv\Phi_1(\omega).
\end{equation}
Observe that the functions $\mc{V}$ and $\wt{W}$ in \eq{6-F-VW}
are precisely of this type and, hence,
$\mc{F}_{|J_{\{\ell\}}|}$ is a particular case of \eq{8-formG}.
Moreover, in the process of consecutive
summation of the series \eq{6-Gn}, we will permanently deal with
functions of the form \eq{8-formG}. A remarkable property of such functions is that, regardless of their multi-variable nature, everything happens in the
asymptotic regime as if they were a
pure product of one variable functions.

Let us give now the explicit form of the action of the functionals
$H_s$ and $O_s$ on such functions.

\begin{prop}\label{8-actHs}
Let $\mc{G}_s$ be a function of the form \eq{8-formG}. Then
 \begin{equation}\label{8-funct}
 H_s[\mc{G}_s]
 =\frac{(-1)^{s-1}}{(s-1)!}\left.\partial^s_\gamma\,
 \mc{W}_0^{(0)}[\hat\nu]\right|_{\gamma=0}\hspace{-1mm}\cdot\,\Phi_2\;,
\end{equation}
 where
 \begin{equation}\label{8-hatnu1}
 \hat\nu(\omega)=-\frac1{2\pi i}\log\left[1+\gamma\Phi_1(\omega)\right],
\end{equation}
and with  $\Phi_1(\omega)$ and $\Phi_2$ defined
as in \eq{8-full-red2}.
\end{prop}

\Proof
The explicit form of the  functional $\mc{W}_0^{(0)}[\nu]$ follows from \eqref{5-asy-n-osc}. However, to prove Proposition~\ref{8-actHs}, it is enough that it can be written in the following
quite general form:
 \begin{equation}\label{D1-Funct}
 \mc{W}_0^{(0)}[\nu]
 =\sum_{n=0}^\infty\frac1{n!}
  \int\limits_{-q}^q T_n(\xi_1,\dots,\xi_n)\,
  \prod_{j=1}^n \nu(\xi_j)\,\dd^n\xi\, ,
 \end{equation}
where $\nu(\xi)$ is given by \eq{5-nu-g} and $T_n$ are some functions or distributions.
In order to obtain
the corresponding part of the asymptotics of the cycle integral of
length $s$, one should first take the $s^{\e{th}}$ $\gamma$-derivative
 \begin{equation}\label{D1-der-Funct}
 \partial^s_\gamma \mc{W}_0^{(0)}[\nu]\Bigl.\Bigr|_{\gamma=0}
 =\sum_{n=0}^\infty\,
  {\sum_{k_1,\dots,k_n=1}^s}
  \hspace{-4mm}{\vphantom\sum}'\hspace{3mm}
  \prod_{j=1}^n\frac{\partial_\gamma^{k_j}\nu_0}{k_j!}
  \Bigl.\Bigr|_{\gamma=0}
  \int\limits_{-q}^q T_n(\xi_1,\dots,\xi_n)\,
  \prod_{j=1}^n F^{k_j}(\xi_j) \;\dd^n\xi\,,
 \end{equation}
where $\nu_0$ is given by
 \begin{equation}\label{5-bncn}
 \nu_0=\frac{-1}{2\pi i}\log(1+\gamma)\, ,
 \end{equation}
and prime means that the summation over $k_1,\ldots,k_n$ is taken under the
constraint $\sum_{j=1}^nk_j=s$.
Then, using \eq{5-factorization} and \eq{choix de F et g}, we obtain
 \begin{equation*}
 H_s[\mc{G}_s]
 =\frac{(-1)^{s-1}}{(s-1)!}
 \sum_{n=0}^\infty\,
 {\sum_{k_1,\dots,k_n=1}^s}\hspace{-4mm}{\vphantom\sum}'
 \hspace{3mm}
 \prod_{j=1}^n\!\frac{\partial_\gamma^{k_j}\nu_0}{k_j!}
 \Bigl.\Bigr|_{\gamma=0}
 %
 \int\limits_{-q}^q \! T_n(\xi_1,\dots,\xi_n)\,
 \mc{G}_s\!\left(\! \begin{array}{c}
 \{\xi_1\}^{k_1},\dots, \{\xi_n\}^{k_n}\\
 \{\xi_1\}^{k_1},\dots, \{\xi_n\}^{k_n}\end{array}\!\!\right)
 \dd^n\xi.
 \end{equation*}
It remains to observe that, due to the reduction property \eq{8-recurs},
 \begin{equation}\label{8-observe}
 \mc{G}_s\left(\begin{array}{c}
 \{\xi_1\}^{k_1},\dots ,\{\xi_n\}^{k_n}\\
 \{\xi_1\}^{k_1},\dots ,
 \{\xi_n\}^{k_n}\end{array}\!\right)=\Phi_2\prod_{j=1}^n
 \Phi_1^{k_j}(\xi_j)\, .
 \end{equation}
Comparing with \eq{D1-der-Funct} we see that it gives
\eq{8-funct} with $\hat\nu$ defined in \eq{8-hatnu1}.
\qed

\begin{prop}\label{8-act-O}
Let  $\mc{G}_s$ be as in \eq{8-formG}. Then
 \begin{equation}\label{8-functO}
 O_s^\pm[\mc{G}_s]
 =\frac{(-1)^{s-1}}{(s-1)!}\left.\partial^s_\gamma\,
  e^{\mc{W}^{(0)}_{0}[\hat\nu^{(\pm)}\mp1]
     -\mc{W}^{(0)}_{0}[\hat\nu^{(\pm)}]}\right|_{\gamma=0}
  \hspace{-1mm}\cdot\,
  \Phi_2\left(\barr{c}\mp q \\ \pm q  \ea \right),
\end{equation}
 where
 \begin{equation}\label{8-hatnu1O}
 \hat\nu^{(\pm)}(\omega)
 =-\frac1{2\pi i}\log\left[1+\gamma
   \Phi_1\left(\omega\mid\barr{c} \mp q \\ \pm q \ea \right)\right].
\end{equation}
\end{prop}

\Proof
The proof is very similar to the one of Proposition~\ref{8-actHs}.
Consider, for example, the oscillating functional $O_s^+$.
The action of this functional is generated by the oscillating term
${\cal W}_{+1}[\nu]\, \ex{i m [p_0(q)-p_0(-q)]}$, which in its turn is expressed in terms of $\mc{W}_0^{(0)}[\nu]$ by \eq{5-relation}:
 \begin{equation}\label{8-relation}
 {\cal W}_{+ 1}[\nu]\; \ex{i m  [p_0(q)-p_0(-q)]}
 =\ex{{\cal W}_0^{(0)}[\nu- 1]-{\cal W}_0^{(0)}[\nu]+ g(q)-g(-q)}.
 \end{equation}
Once again, we do not need the complete explicit expression of ${\cal W}_0^{(0)}$.
Let simply
 \begin{equation}\label{8-defT}
 \ex{{\cal W}_0^{(0)}[\nu- 1]-{\cal W}_0^{(0)}[\nu]}
 =\nu(q)\,\nu(-q)\;T[\nu]\,,
 \end{equation}
in which we have stressed that the oscillating term of the asymptotics
of the Fredholm determinant is proportional to $\nu(q)\,\nu(-q)$.
For the remaining part $T[\nu]$, we can use a representation similar to \eq{D1-Funct}:
 \begin{equation}\label{8-Funct-O}
 T[\nu]=\sum_{n=0}^\infty
 \frac1{n!}\int\limits_{-q}^q T_n(\xi_1,\dots,\xi_n)\,
 \prod_{j=1}^n \nu(\xi_j)\;\dd^n\xi\,,
 \end{equation}
where $T_n$ are some functions or
distributions. The $s^{\e{th}}$ $\gamma$-derivative then reads
 \begin{multline}\label{8-der-Funct-O}
 \partial^s_\gamma
 \Bigl(\nu(q)\,\nu(-q)\; T[\nu]\Bigr) \Bigl.\Bigr|_{\gamma=0}
 =\sum_{n=0}^\infty\,
  {\sum_{k_+,\, k_-,\, k_i=1}^s}\hspace{-5mm}{\vphantom\sum}'
  \hspace{3mm}
  \frac{\partial_\gamma^{k_+}\nu_0}{k_+!}\,
  \frac{\partial_\gamma^{k_-}\nu_0}{k_-!}\,
  \prod_{j=1}^n\frac{\partial_\gamma^{k_j}\nu_0}{k_j!}
  \Bigl.\biggr|_{\gamma=0}\\
 \times \int\limits_{-q}^q
 T_n(\xi_1,\dots,\xi_n)\,F^{k_+}(q)\,F^{k_-}(-q)\,
 \prod_{j=1}^n F^{k_j}(\xi_j)\;\dd^n\xi\, ,
 \end{multline}
where $\nu_0$ is defined in \eq{5-bncn}, and prime means that the
sum over $k_+$, $k_-$, $k_1,\ldots,k_n$ is taken under the constraint $k_++k_-+\sum_{j=1}^nk_j=s$. Then,
using \eq{5-factorization}, \eq{choix de F et g}, and multiplying
\eq{8-der-Funct-O} by $e^{g(q)-g(-q)}$, we obtain
 \begin{multline}\label{D1-equiv-FunctO}
 O_s^+[\mc{G}_s]
 =\sum_{n=0}^\infty\,
 {\sum_{k_+,\, k_-,\, k_i=1}^s}
 \hspace{-5mm}{\vphantom\sum}'  \hspace{3mm}
 \frac{\partial_\gamma^{k_+}\nu_0}{k_+!}\,
 \frac{\partial_\gamma^{k_-}\nu_0}{k_-!}\,
 \prod_{j=1}^n\frac{\partial_\gamma^{k_j}\nu_0}{k_j!}
 \Bigl.\biggr|_{\gamma=0}\\
 \times
 \int\limits_{-q}^q T_n(\xi_1,\dots,\xi_n)\;
 \mc{G}_s\!\left(\!\!\begin{array}{rc}
  -q,&\hspace{-2mm}\{q\}^{k_+},\{-q\}^{k_--1},
                     \{\xi_1\}^{k_1},\ldots,\{\xi_n\}^{k_n}\\
   q,&\hspace{-2mm}\{q\}^{k_+},\{-q\}^{k_--1},
                     \{\xi_1\}^{k_1},\ldots,\{\xi_n\}^{k_n}
  \end{array}\!\!\right)
 \, \dd^n\xi\,.
 \end{multline}
It remains to observe that, due to the reduction property \eq{8-recurs},
 \begin{multline}\label{8-observeO}
 \mc{G}_s\!\left(\!\!\begin{array}{rc}
  -q,&\hspace{-2mm}\{q\}^{k_+},\{-q\}^{k_--1},
                     \{\xi_1\}^{k_1},\ldots,\{\xi_n\}^{k_n}\\
   q,&\hspace{-2mm}\{q\}^{k_+},\{-q\}^{k_--1},
                     \{\xi_1\}^{k_1},\ldots,\{\xi_n\}^{k_n}
  \end{array}\!\!\right)
                 \\
 =
 \Phi_1^{k_+}\left( q\mid\barr{r} \hspace{-2mm}-q
 \\ \hspace{-2mm} q \ea \right)
  \Phi_1^{k_-}\left(- q\mid\barr{r} \hspace{-2mm}-q
 \\ \hspace{-2mm} q \ea \right)
  \prod_{j=1}^n
 \Phi_1^{k_j}\left(\xi_j\mid\barr{r} \hspace{-2mm}-q
 \\ \hspace{-2mm}q \ea \right)\cdot\Phi_2\left(\barr{r}\hspace{-2mm} -q
 \\ \hspace{-2mm} q \ea \right)\,.
 \end{multline}
Comparing with \eq{8-der-Funct-O} we  arrive at the action
\eq{8-functO} with $\hat\nu^{(\pm)}$ defined in \eq{8-hatnu1O}.\qed

Thus, the action of the functionals $D_s$, $H_s$ and $O_s$ is
explicitly defined at least on the class of functions $\mc{G}_s$
admitting the representation \eq{8-formG}. In that way we have found
the leading non-oscillating and oscillating asymptotics of  cycle
integrals. However, this is not yet enough to obtain a correct
estimate of the infinite series of multi-cycle integrals \eq{6-Gn}.
The matter is that some a priori sub-leading corrections after their
summation may give non-vanishing leading contribution. We will
consider this question in more details in Section~\ref{RAD}, where
the mechanism of summation will become clear. Here we merely would
like to note that  some part of the corrections could easily be
included into the action of the functionals considered above. For
example, we did not use the explicit form of the functional
$\mc{W}_0^{(0)}[\nu]$ in the proof of Proposition~\ref{8-actHs}.
Hence, instead of $\mc{W}_0^{(0)}[\nu]$ alone, we could as well have
considered all the terms in the asymptotic expansion of
$\log\det[I+\ga V]$ which can be written in the form
\eqref{D1-Funct}, and the statement of this proposition would still
be valid.


\section{Asymptotic summation of cycle integrals}
\label{sec-sum}

In this section we sum up the series \eq{6-Gn} in the
asymptotic regime $m\to\infty$. We already labeled each pure cycle
integral $\mc{I}_{(s,p)}$ with respect to the position of the
variables on which  it acts. We define similarly $D_{(s,p)}$ (resp.
$H_{(s,p)}$ and $O_{(s,p)}$)  as the operator that acts on the variables
of the $p^{\e{th}}$ cycle of length $s$ as $D_s$ (resp.
$H_s$ and $O_s$), and by doing nothing to all the other
variables. Then, if we also define $R_s$ to be the functional given by the remaining part of \eqref{5-asympt-Is} (which is in principle of order o(1) for non-oscillating corrections, and of order $\e{o}(\frac{1}{m^2})$ for oscillating ones), and $R_{(s,p)}$ to be its conterpart acting on the cycle $(s,p)$, we can reexpress \eqref{6-Gn} as
\begin{equation}\label{G-DHO}
%
{G}_{1\dots n}(\gamma) =  \sul{\ell_1,\ldots,\ell_n=0}{\infty}\,
\pl{s=1}{n}\f{(u_s)^{\ell_s}}{\ell_s!}
\pl{p=1}{\ell_s} \big[ D_{(s,p)}+H_{(s,p)}+O_{(s,p)}+R_{(s,p)} \big] *
\mc{F}_{|J_{\{\ell\}}|}
\, ,
\end{equation}
where $u_s=\frac{(-1)^{s-1}\gamma^s}s$.
There, just as in the case of multi-cycle  integrals, the order of the
different
operators is not important due to  the symmetry of the function $\mc{F}_{|J_{\{\ell\}}|}$ in its two sets of arguments
$\{\la\}$ and $\{z\}$.

Before starting to sum up explicitely the series \eqref{G-DHO}, let us make a usefull remark:
since we only need the $n^{\e{th}}$ $\ga$-derivative of $G_{1\ldots n}$ at $\ga=0$ (see \eqref{6-Q1-Gn}), it is enough for our purpose to obtain a {\it $\gamma$-equivalent} form of the result.

\begin{Def}
Two functions $\varphi_1(\gamma)$ and $\varphi_2(\gamma)$ are said to be
{\em $\gamma$-equivalent of order $n$} if
\begin{equation}
 \partial^k_\gamma\varphi_1(\gamma)\Bigl.\Bigr|_{\gamma=0}=
 \partial^k_\gamma\varphi_2(\gamma)\Bigl.\Bigr|_{\gamma=0},
 \qquad k=0,1,\dots,n\;.
\end{equation}
\end{Def}
Therefore, at any stage of the computation, we may replace $G_{1\ldots n}$ by some $\gamma$-equivalent function of order $n$ which, by abuse of notations, will still be called $G_{1\ldots n}$.

\subsection{Summation of the action of $H_{(s,p)}$\label{RS-Gen}}

Let us start by summing up the action of the operators $H_{(s,p)}$. We first expand the products of the operators in \eq{G-DHO}
according to the binomial formula:
\begin{multline}
\f{1}{\ell_s!}\pl{p=1}{\ell_s}
 \big[D_{(s,p)}+H_{(s,p)}+O_{(s,p)}+R_{(s,p)}  \big]
         \\
=\sul{r_s=0}{\ell_s} \f{1}{r_s! \, (\ell_s-r_s)!}
 \pl{p=1}{r_s}\big[D_{(s,p)}+O_{(s,p)}+R_{(s,p)} \big]
 \pl{p=r_s+1}{\ell_s} H_{(s,p)}\,.
\label{developement produit d'operateurs}
\end{multline}
As we already mentionned, such an operation is possible due to the symmetry property of the function $\mc{F}_{| J_{\{\ell\}} |}$
on which the operators in \eqref{developement produit
d'operateurs} act. Substituting this into \eq{G-DHO} and changing
the order of summation, we arrive at
\begin{multline}\label{8-G-DHO-H}
{G}_{1\dots n}(\gamma)
=\sul{r_1,\dots,\, r_n =0 }{\infty}
%
\pl{s=1}{n}\f{(u_s)^{r_s}}{r_s!}
\pl{s=1}{n}\pl{p=1}{r_s}\big[ D_{(s,p)}+O_{(s,p)}+R_{(s,p)}\big] \\
\times
 \sul{\ell_1, \dots,\, \ell_n =0}{\infty}
\pl{s=1}{n}\f{(u_s)^{\ell_s}}{\ell_s!}
\pl{s=1}{n}\pl{p=1}{\ell_s} H_{(s,p)} *
{\mc{F}}_{|J_{\{r\}}|+|J_{\{\ell\}}|}
\left(\barr{c}
\{\lambda\}_{J_{\{r\}}} \cup \{\mu\}_{J_{\{\ell\}}}\\
\{z\}_{J_{\{r\}}} \cup \{y\}_{J_{\{\ell\}}} \ea  \right).
\end{multline}
Here, for convenience, we have divided the set of arguments of ${\mc{F}}_{\abs{J_{\{\ell\}}}+\abs{J_{\{r\}}}}$ into two
subsets: the operators $D_{(s,p)}$, $O_{(s,p)}$ and $R_{(s,p)}$ act on the variables
$\{\lambda\}_{J_{\{r\}}}$ and $\{z\}_{J_{\{r\}}}$, while
the operators $H_{(s,p)}$ act on $\{\mu\}_{J_{\{\ell\}}}$ and
$\{y\}_{J_{\{\ell\}}}$.

Recall that each individual $H_{(s,p)}$
acts non trivially (as the functional $H_s$) only on variables $\mu_{s,p,j}$ and
$y_{s,p,j}$, with $j=1,\dots,s$. Therefore one can write
\begin{equation}\label{8-actH-1}
 H_{(s,p)}*{\mc{F}}_{|J_{\{r\}}|+|J_{\{\ell\}}|}=
 H_s\pac{{\mc{F}}_{|J_{\{r\}}|+|J_{\{\ell\}}|}
 \left(\barr{c}\{\mu_{s,p,j}\}_{1\le j\le s}\cup
 \{\lambda\}_{J_{\{r\}}}
 \cup \{\mu\}_{\! J^{\widehat{s,p}}_{\{\ell\}}}\\
 \{y_{s,p,j}\}_{1\le j\le s}\cup\{z\}_{J_{\{r\}} }
 \cup\{y\}_{\! J^{\widehat{s,p}}_{\{\ell\}}}   \ea  \right)},
\end{equation}
where $H_s$ acts  on the first sets of variables
$\{\mu_{s,p,j}\}_{1\le j\le s}$ and $\{y_{s,p,j}\}_{1\le j\le s}$ according to
\eq{8-funct}, while all other variables can be considered as
auxiliary parameters. The set $J_{\{\ell\}}^{\widehat{s,p}}$ is defined as
\begin{equation}
J_{\{\ell\}}^{\widehat{s,p}}  =J_{\{\ell\}} \setminus \{(s,p,j) \, :\,
 1\le j \le s \}\, .
\end{equation}
%
It is clear
that $\mc{F}_{\abs{J_{\{r\}}}+\abs{J_{\{\ell\}}}}$ has the form
\eq{8-formG}, with
 \begin{equation}\label{8-Phi1}
 \Phi_1(\om )
=\mc{V}_{|J_{\{r\}}|+|J_{\{\ell\}}|}
   \left(\om \mid\barr{c}
   \{\mu_{s,p,j}\}_{1\le j\le s}
 \\
  \{y_{s,p,j}\}_{1\le j\le s}
     \ea  \right),
 \end{equation}
 \begin{equation}\label{8-Phi2}
 \Phi_2
=\hspace{-1.9mm}
 \prod_{\omega\in\{\lambda\}_{J_{\!\{r\}}} \!
 \cup \{\mu\}_{\! J^{\widehat{s,p}}_{\{\ell\}} }}
 \hspace{-5mm}
 \mc{V}_{|J_{\{r\}}|+|J_{\{\ell\}}|}
 \left(\omega \mid\!\! \barr{c}
 \{\mu_{s,p,j}\}_{1\le j\le s} 
 \\ \{y_{s,p,j}\}_{1\le j\le s} 
 \ea  \! \right)
 \cdot \wt{W}_{|J_{\{r\}}|+|J_{\{\ell\}}|}
 \left( \!\!
   \barr{c}\{\mu_{s,p,j}\}_{1\le j\le s}
 \\ \{y_{s,p,j}\}_{1\le j\le s}
 \ea \!\! \right),
 \end{equation}
where, for simplicity,  we have omitted  the additional sets
$\{\lambda\}_{J_{\{r\}}} \cup \{\mu\}_{J^{\widehat{s,p}}_{\{\ell\}}}$
\vspace{-1mm} and
$\{z\}_{J_{\{r\}}} \cup\{y\}_{J^{\widehat{s,p}}_{\{\ell\}}}$ in all  the
arguments in \eq{8-Phi1}, \eq{8-Phi2}.
Applying Proposition~\ref{8-actHs}, we obtain
\begin{equation}\label{8-actH-2}
 H_{(s,p)}* {\mc{F}}_{\abs{J_{\{r\}}}+\abs{J_{\{\ell\}}}}=
  \frac{(-1)^{s-1}}{(s-1)!}\left.\partial^s_\gamma
 \mc{W}_0^{(0)}[\nu^{\widehat{s,p} }]\right|_{\gamma=0}
 \cdot
 {\mc{F}}_{\abs{J_{\{r\}}}+\abs{J^{\widehat{s,p} }_{\{\ell\}}}}\;,
 \end{equation}
where
 \begin{equation}\label{8-nu-sp}
 \nu^{\widehat{s,p} }(\omega)
=\frac {-1}{2\pi i}\log\left[1+\gamma
 \mc{V}_{\abs{J_{\{r\}}}+\abs{J^{\widehat{s,p} }_{\{\ell\}}}}
 \left(\omega\mid
 \barr{c}\{\lambda\}_{J_{\{r\}}}
 \cup \{\mu\}_{\! J^{ \widehat{s,p}}_{\{\ell\}}} \\
 \{z\}_{J_{\{r\}}}
 \cup \{y\}_{\! J^{\widehat{s,p}}_{\{\ell\}}}\ea \right)\right] .
 \end{equation}

The obtained result still admitting the representation \eq{8-formG}, we can act with all operators $H_{(s,p)}$ consecutively.
It gives
 \begin{equation}\label{8-actH-all}
 \pl{s=1}{n}\pl{p=1}{\ell_s} H_{(s,p)}*
 {\mc{F}}_{\abs{J_{\{r\}}}+\abs{J_{\{\ell\}}}}
=
 \pl{s=1}{n}\left(\frac{(-1)^{s-1}}{(s-1)!}
 \left.\partial^s_\gamma
 \mc{W}_0^{(0)}[\hat\nu]\right|_{\gamma=0}\right)^{\ell_s}
 {\mc{F}}_{\abs{J_{\{r\}}}}\left(\!\barr{c}
 \{\lambda\}_{J_{\{r\}}} \\ \{z\}_{J_{\{r\}}} \ea \! \right),
\end{equation}
where
 \begin{equation}\label{8-hatnu}
 \hat\nu(\omega)=\frac {-1}{2\pi i}\log\left[1+\gamma
 \mc{V}_{|J_{\{r\}}|}\left(\omega\mid
 \barr{c}\{\lambda\}_{J_{\{r\}}} \\
 \{z\}_{J_{\{r\}}}\ea \right)\right].
 \end{equation}
Substituting this into \eq{8-G-DHO-H} and summing up over all
$\ell_s$ we arrive at
\begin{equation}\label{Ghat1an apres resomation Hs}
G_{1\dots n}(\ga)
= \!\!\sul{r_1, \dots , \, r_n =0}{\infty}
\prod_{s=1}^n\f{(u_s)^{r_s}}{r_s!}
\prod_{s=1}^n\pl{p=1}{r_s}\big[ D_{(s,p)}+O_{(s,p)}+R_{(s,p)}\big] *
{\mc{F}}^{(H)}_{\abs{J_{\{r\}}}}\left(\!\barr{c}
\{\lambda\}_{J_{\{r\}}} \\ \{z\}_{J_{\{r\}}} \ea \!\! \right),
\end{equation}
where
\begin{equation}
\label{8-dressedF}
{\mc{F}}^{(H)}_{\abs{J_{\{r\}}}}
\left(\barr{c}\!
 \{\lambda\}_{J_{\{r\}}} \\
 \{z\}_{J_{\{r\}}} \ea \!\!\right)=
 {\mc{F}}_{\abs{J_{\{r\}}}}\left(\!\barr{c}
 \{\lambda\}_{J_{\{r\}}} \\ \{z\}_{J_{\{r\}}} \ea \!\! \right)
%
%
 \cdot
 \exp\left\{\sul{s=1}{n}
\frac{\gamma^{s}}{s!}
\left.\partial^s_\gamma
 \mc{W}_0^{(0)}(m,[\hat\nu])\right|_{\gamma=0}
\right\}.
\end{equation}
Since it will turn out to be rather important for the summation of the remaing terms, we have here explicitly indicated that the functional
$\mc{W}_0^{(0)}(m,[\hat\nu])$ depends on the distance $m$.

We can now perform the summation over $s$ in \eq{8-dressedF} by extending it up to infinity, which means that we replace ${\mc{F}}^{(H)}$ by the following
$\gamma$-equivalent function, still denoted by ${\mc{F}}^{(H)}$:
 \begin{equation}\label{8-dressedF1}
 {\mc{F}}^{(H)}_{\abs{J_{\{r\}}}}\left(\!\barr{c}
 \{\lambda\}_{J_{\{r\}}} \\
 \{z\}_{J_{\{r\}}} \ea \!\!\right)=
 {\mc{F}}_{\abs{J_{\{r\}}}}\left(\!\barr{c}
 \{\lambda\}_{J_{\{r\}}} \\ \{z\}_{J_{\{r\}}} \ea  \!\!\right)
\cdot
\exp\left\{ \mc{W}_0^{(0)}(m,[\hat\nu]) \right\}.
\end{equation}
%


Thus, the successive action of operators $H_{(s,p)}$ produces
a complete decoupling of parts of the variables, and  eventually restores some of the Fredholm determinant asymptotics \eq{5-asy-n-osc}.

\begin{rem}
 Note that the summation performed here is {\em exact}, and that we did not use the explicit expression of $\mc{W}_0^{(0)}(m,[\hat\nu])$.
We could as well have considered the action of some operator $\widetilde{H}_s$ including not only the leading contribution $H_{s}$, but also subleading (non-oscillating) ones, provided they originate from the same kind of terms \eqref{D1-Funct}.
In fact, the very same process will enable us, in the next subsection, to sum up some part of $R_s$ as well.
\end{rem}


\subsection{Contributions from $R_{(s,p)}$ \label{RAD}}

Unlike the original function ${\mc{F}}_{\abs{J_{\{r\}}}}$
\eq{6-F-VW}, the obtained  function
${\mc{F}}^{(H)}_{\abs{J_{\{r\}}}}$ depends on the distance $m$.
Let us specify this dependence. We have
 \begin{equation}\label{8-W0}
 {\cal W}_0^{(0)}(m,[\hat\nu])
 =-im\int\limits_{-q}^q p'_0(\lambda)\,
 \hat\nu(\lambda)\,\dd\lambda
-
 \sum_{\sg=\pm}\hat\nu_\sigma^2\,\log\big[m\, \sinh(2q)\, p'_0(\sg q)\big]
+\tilde C[\hat\nu]\,,
 \end{equation}
with
 \begin{equation}\label{8-C1}
 \tilde C[\hat\nu]
=\frac12\!\int\limits_{-q}^{q}\!
\frac{\hat\nu'(\lambda) \hat\nu(\mu)
      -\hat\nu(\lambda) \hat\nu'(\mu)}
     {\tanh(\lambda-\mu)}
 \dd\lambda\dd\mu
+\! \sum_{\sg=\pm} \! \Bigg[\sg\hat\nu_\sg \!\! \int\limits_{-q}^{q}\!\!
 \frac{\hat\nu_\sg-\hat\nu(\lambda)}{\tanh(\sg q-\lambda)}\dd\lambda
 +\log G(1,\hat\nu_\sg)\Bigg] ,\hspace{-1mm}
 \end{equation}
in which $\hat\nu(\omega)$ is given by \eq{8-hatnu}. Then
the function ${\mc{F}}^{(H)}_{\abs{J_{\{r\}}}}$ reads
\begin{multline}\label{8-calF}
{\mc{F}}^{(H)}_{\abs{J_{\{r\}}}}
=\exp\left\{-im\int\limits_{-q}^q p'_0(\om)\,
 \hat\nu(\om)\,\dd\om \right\}
 \cdot
 \prod_{\sg=\pm}
 \left[m\,\sinh(2q)\,p'_0(\sg q)\right]^{-\hat\nu^2(\sigma q)}
 \cdot e^{\tilde C[\hat\nu]}  \\
 \times \widetilde{W}_{\abs{ J_{ \paa{r} } }}
 \left(\barr{c} \{\lambda\}_{J_{\{r\}}} \\ \{z\}_{J_{\{r\}}} \ea
 \right)
 \prod_{t\in J_{\{r\}}}
 {\cal V}_{\abs{ J_{ \paa{r} } }}\left(\lambda_t\mid\barr{c} \{\lambda\}_{J_{\{r\}}}
  \\ \{z\}_{J_{\{r\}}} \ea
 \right).
 \end{multline}
In order to complete the calculations, we should in principle sum up the action
of the operators $D_{(s,p)}$ and $O_{(s,p)}$ on 
${\mc{F}}^{(H)}_{\abs{J_{\{r\}}}}$, {\em provided that $R_{(s,p)}$ produces only subleading corrections}.
This was the case when acting on $m$-independent functions of the form \eqref{8-formG} but, due to the fact that ${\mc{F}}^{(H)}_{\abs{J_{\{r\}}}}$ \vspace{-0.8mm}
now depends on $m$, this is no longer true. Let us explain why.

The new function ${\mc{F}}^{(H)}_{\abs{J_{\{r\}}}}$ is still of the form \eqref{8-formG}, but now $\Phi_2$ depends on $m$ as
\begin{equation}\label{tilde-G}
 \Phi_2\!\left(\hspace{-1mm}\barr{c} \{\lambda\} \\ \{z\} \ea \hspace{-1mm}\right)
  =\exp\left\{m \,
   \Phi\!\left(\hspace{-1mm}\begin{array}{c}\{\lambda\}\\
   \{z\}\end{array} \hspace{-1mm}\right)
 +\log m \;
  \widetilde\Phi\!\left(\hspace{-1mm}\begin{array}{c}\{\lambda\}\\
                    \{z\}\end{array} \hspace{-1mm}\right)
  \right\}
  \cdot
  \widetilde{\Phi}_2\!\left(\hspace{-1mm}\barr{c} \{\lambda\} \\ \{z\} \ea \hspace{-1mm}\right),
\end{equation}
with $\Phi$, $\widetilde\Phi$ and $\widetilde{\Phi}_2$ satisfying the reduction property \eqref{8-recurs}.
We see that the
differential operator in $D_{(s,p)}$, when acting on such $m$-dependent functions, may produce terms that are linear over the distance $m$: hence, although this operator originates from a term of order $\e{O}(1)$  in \eqref{5-asympt-Is}, it behaves as a term of order $m$ when acting on
\eqref{8-calF}. The same phenomenon may happen for the action of  $R_{(s,p)}$: if the latter contains some derivative operators, then these operators may produce
non vanishing contributions to the final answer although they are effectively subleading when acting on each cycle; for instance, a
differential operator $\frac1m\partial$ gives a term of order $\e{O}(\frac1m)$ when
acting on the original function \vspace{-0.8mm} ${\mc{F}}_{\abs{J_{\{r\}}}}$, but it
gives a contribution of order $\e{O}(1)$ when acting on
the new function ${\mc{F}}^{(H)}_{\abs{J_{\{r\}}}}$ which is the result of summation over all cycles of the action of $H_{(s,p)}$ on ${\mc{F}}_{\abs{J_{\{r\}}}}$.
Therefore, we need  more informations on the structure of the subleading terms in \eqref{5-asympt-Is} in order to be able to extract from $R_s$ the part of the corrections that eventually contributes to the leading order.

The structure of the series \eqref{5-asympt-Is} for $\mc{I}_s[\mc{G}_s]$ was studied from  the asymptotic expansion of the Fredholm determinant \eq{5-kernel} and
related Riemann--Hilbert problem in \cite{KitKMST08a}. This
analysis being extremely cumbersome, we do not have the  explicit expression of the series \eqref{5-asympt-Is}. However, we obtained in \cite{KitKMST08a} some information on the generic form of the corrections: they are given in terms of the function $\mc{G}_s$ or of its derivatives, the latter being evaluated at the endpoints $\pm q$ or integrated, together with some weight function, on the interval $[-q,q]$.
Therefore the functional $R_s$ indeed contains differential operators. However, it was proved in \cite{KitKMST08a} that the non-oscillating correction $I_s^{(n;\,\e{nosc})}[\mc{G}_s]$ of order $n$ in \eqref{5-asympt-Is} contains {\em at most derivatives of order $n$} of $\mc{G}_s$, whereas the oscillating correction $I_s^{(n;\,\e{osc})}[\mc{G}_s]$  of order $n$ contains {\em at most derivatives of order $n-2$}.
They correspond respectively to the contribution of some differential operator $\frac{1}{m^n}\mf{D}_s^{(n;\, \e{nosc})}$ and   $\frac{1}{m^n}\mf{D}_s^{(n;\, \e{osc})}$ in $R_s$, $\mf{D}_s^{(n;\, \e{nosc})}$ being of order at most $n$ and $\mf{D}_s^{(n;\, \e{osc})}$ of order at most $n-2$.
Therefore, only the part of {\em maximal order} of these operators may produce, when acting on functions of the form \eqref{8-formG}, \eqref{tilde-G} (and more precisely on the exponent), contributions on the same order as $D_s$ or $O_s$. These are the contributions that we need to determine.

For convenience, in order to distinguish the effect of non-oscillating and oscillating corrections, we define two functionals $R_s^{\e{nosc}}$ and $R_s^{\e{osc}}$:
$R_s^{\e{nosc}}$ contains all the non-oscillating corrections with respect to $I_s^{(0)}$, and $R_s^{\e{osc}}$ contains all the oscillating corrections with respect to $O_s$, in the sense of the asymptotic series \eqref{5-asympt-Is}\footnote{It means that we choose some arbitrary integer $n_0$ and consider the asymptotic development for $R_s^{\e{nosc}}$ and $R_s^{\e{osc}}$ up to $\e{O}(\log^s m/m^{n_0+1})$.}.

\subsubsection{Non-oscillating contribution}

Let us first discuss the effect of the non-oscillating part $R_s^{\e{nosc}}$ of the corrections. Clearly, as we already noticed,  at each order $n$, only the differential operators
$\frac{1}{m^n}\mf{D}_s^{(n;\, \e{nosc})}$, and more precisely {\em the part $\frac{1}{m^n}\widetilde{{\mf{D}}}^{(n;\,\e{nosc})}_s$ of $\frac{1}{m^n}\mf{D}_s^{(n;\, \e{nosc})}$ which results in the maximal order $n$ of derivatives}, may eventually produce, when acting on $m$-dependent functions \eqref{tilde-G}, contributions of order $\e{O}(1)$.
For $n\ge 1$, the explicit expression of the differential operator $\widetilde{{\mf{D}}}^{(n;\,\e{nosc})}_s$ of maximal order $n$ is not known, but it can be shown from \cite{KitKMST08a} that it has the form
 \begin{equation}\label{struct}
 \widetilde{{\mf{D}}}^{(n;\,\e{nosc})}_s
 =\sum_{\sg=\pm}\sum_{\{k\},\{\ell\}}
  C_\sg(\{k\},\{\ell\})
\prod_{j=1}^s\partial_{z_j}^{k_j}\partial_{\lambda_j}^{\ell_j}
\bigg|_{\la_j=z_j=\sg q}
 ,
\qquad \sum_{j=1}^s(k_j+\ell_j) = n,
 \end{equation}
where $C_\sg(\{k\},\{\ell\})$ are some computable coefficients.
Here the action of the derivatives (on some function ${\cal G}_s$) is followed by the evaluation of all the variables $z$ and $\la$ at the same point $\sg q$, $\sigma=\pm$.

Obviously, when acting on some $m$-dependent function $\mc{G}_s$ of the type \eqref{8-formG}, \eqref{tilde-G}, this operator can produce
$\e{O}(1)$ contribution only if it acts on the part of the exponent which is linear in $m$, i.e. on
 \begin{equation}\label{8-eff-exp}
 \exp\left\{m\, \Phi\left(\begin{array}{c}\{\lambda\}\\
 \{z\}\end{array} \right)\right\}  .
 \end{equation}
Using the reduction property \eqref{8-recurs} for $\Phi_1$ and $\Phi_2$ and the fact that the contributions of order O(1) can be obtained only in case each derivative hits exactly once the exponent \eqref{8-eff-exp}, we obtain
 \begin{multline}\label{8-eff-act}
 \frac{1}{m^n}
 \widetilde{{\mf{D}}}^{(n;\,\e{nosc})}_s[\mc{G}_s]
%
 = \sum_{\sg=\pm}\Phi_1^s\pa{\sg q}\cdot \Phi_2 \\
 \times
 \Bigg\{
 \sum_{
    \{k\},\{\ell\} 
      }\hspace{-2mm}\vphantom{\sum}'\hspace{1mm}
 C_\sg(\{k\},\{\ell\})
 \prod_{j=1}^s\bigl[\tilde g'(z_j)\bigr]^{k_j}
 \bigl[-\tilde g'(\lambda_j)\bigr]^{\ell_j}\Bigg\}\Bigg|_{\la_j=z_j=\sg q}
 \cdot
 \left[ 1
 +\e{O}\bigg({\frac{\log m}m}\bigg)\right],
%
 \end{multline}
where the prime means that the sum is taken under the constraint $\sum(k_j+\ell_j)=n$.
Here we have used the conventions \eqref{8-full-red2} for the reduced functions $\Phi_1$ and $\Phi_2$, and we have defined the one-variable function $\tilde g'$ to be given in terms of the reduced function $\Phi$ as
 \begin{equation}\label{8-tg}
 \tilde g'(\lambda)=
  \partial_\eps\Phi\bigg(\begin{array}{c}
 \la \\ \lambda+\eps\end{array}
 \bigg)\bigg|_{\eps=0}\,.
 \end{equation}
Hence, everything happens (at least as far as the leading term is concerned) as if we would be acting on functions
$\wt{\mc{G}}_s$ of the type
\begin{equation}
\wt{\mc{G}}_s\pab{ \paa{\la} }{ \paa{z} }
 = \Phi_2\cdot \pl{j=1}{s}\Phi_1(\la_j)
   \cdot
 \exp\Bigg\{m \sul{j=1}{s}\big[\wt{g}(z_j)-\wt{g}(\la_j)\big]\Bigg\}\, .
\label{7-partcase}
\end{equation}

Thus, the problem of finding the structure of the $\e{O}(1)$ contributions coming from the action of $R_{s}^{\e{nosc}}$
on some $m$-dependent function $\mc{G}_s$ of the form \eqref{tilde-G} reduces to the computation of the leading behavior of the functional action of
${\cal I}_s$ on $\wt{\mc{G}}_s$ \eqref{7-partcase}.
Strictly speaking, the functional argument of the cycle integral
${\cal I}_s[\wt{\mc{G}}_s]$ should not depend on the distance $m$ but, in
the case \eq{7-partcase}, we can simply re-define $p_0\to \hat
p=p_0-i \tilde g$ and  $\wt{\mc{G}}_s\to \hat{\cal G}_s = \prod_{j=1}^{s}\Phi_1\pa{\la_j}$ (we recall that $\Phi_2$ is just a constant that will be in factor of the result).
This is then a special case of \eqref{5-pure-prod}, and therefore the leading contribution
of the corresponding cycle integral can be obtained from the $s^{\e{th}}$ $\ga$-derivative at $\ga=0$ of
\begin{equation}\label{8-asy-n-osc}
 {\cal W}_0^{(0)}(m,[\hat\nu])\Big|_{p_0\to\hat p}\hspace{-2mm}
 =-im \int\limits_{-q}^q  \hat p'(\lambda)\,
 \hat{\nu}(\la)\,\dd\lambda
 -
 \sum_{\sg=\pm}\!
 \left\{
 \hat{\nu}_{\sg}^2 \, \log\big[m\sinh(2q)\, \hat p'(\sg q)\big]
 \right\} + C[\hat{\nu}] \, ,
 \end{equation}
where $\hat{\nu}$ is given in terms of $\Phi_1$ as in \eq{8-hatnu1}.
The fisrt term of the r.h.s. of \eqref{8-asy-n-osc}, which is linear in $m$, corresponds to the action on \eqref{7-partcase} of the operator $D_{s}$ that we already know.
It is the other (constant in $m$) $\hat{p}'$-dependent part that eventually reproduces
the contribution we want to compute, namely the $\e{O}(1)$ contribution coming from the action of the whole series $\sum_{n\geq 1}{} \frac{1}{m^n} \widetilde{\mf{D}}^{(n;\,\e{nosc})}_s$ on $m$-dependent functions of the form \eqref{7-partcase}. It reads
 \begin{equation}\label{8-asy-tg}
 \frac{(-1)^{s-1}}{(s-1)!}\;
 \partial_\gamma^s
 \sum_{\sg=\pm}
 \left\{ -\hat{\nu}^2(\sigma q)\,
 \log \frac{ \hat{p}'(\sigma q)}{{p}'_0(\sigma q)}
 \right\}\bigg|_{\gamma=0} \!\!
 \cdot \; \Phi_2\, ,
 \end{equation}
where we recall that $\hat{\nu}$ is given in terms of $\Phi_1$ as in \eq{8-hatnu1}, and that
 \begin{equation}\label{p-hat-q}
\hat{p}'(\la )=p'_0(\la )
  -i \, \partial_\eps
  \Phi\bigg(\begin{array}{c}
       \la \\ \la +\eps \end{array}
       \bigg)\bigg|_{\eps=0}\,.
 \end{equation}

We have thus determined the explicit contribution of order $\e{O}(1)$ coming from the action of $R_s^{\e{nosc}}$
on  $m$-dependent function $\mc{G}_s$ of the type \eqref{8-formG}, \eqref{tilde-G}. Let us call this quantity $R_s^{\e{(1;\,nosc)}}[\mc{G}_s]$: explicitely,  $R_s^{\e{(1;\,nosc)}}[\mc{G}_s]$ is given by the expression \eqref{8-asy-tg}. If moreover we denote by ${R}_s^{\e{(sub;\,nosc)}}[\mc{G}_s]$ the remaining part of $R_s^{\e{nosc}}[\mc{G}_s]$, we have
\begin{equation}
 R_s^{\e{nosc}}[\mc{G}_s]=R_s^{\e{(1;\,nosc)}}[\mc{G}_s]
      +{R}_s^{\e{(sub;\,nosc)}}[\mc{G}_s].
\end{equation}

\begin{rem}\label{rem-reorg-serie}
What we have done here is to act formally with the asymptotic series \eqref{5-asympt-Is} on $m$-dependent functions ${\mc{G}}_s$ of the form \eqref{8-formG}, \eqref{tilde-G}, and to  identify, at each formal order $n$ in $m$, the part of $\frac{1}{m^n}I_s^{(n;\, \e{nosc})}[{\mc{G}}_s]$ contributing to the order $\e{O}(1)$. We know that the summed contribution 
of these leading parts, that we have called $R_s^{\e{(1;\,nosc)}}[\mc{G}_s]$, is given by \eqref{8-asy-tg}.
We also know that the  remaining terms in each $\frac{1}{m^n}I_s^{(n;\, \e{nosc})}[{\mc{G}}_s]$ are of order $\e{O}(\frac{\log m}m)$.
However, it is not obvious to prove rigorously that the sum ${R}_s^{\e{(sub;\,nosc)}}[\mc{G}_s]$ of these remaining terms stay subdominant in the final answer. It should follow from the analysis presented in \cite{KitKMST08a} but here we just assume that it is so.
\end{rem}

It is now possible to sum up the successive actions of the corresponding operators \vspace{-1mm} $R_{(s,p)}^{\e{(1;\,nosc)}}$  on the function  ${\mc{F}}^{(H)}$ \eqref{8-calF}.
The action of $R_s^{\e{(1;\,nosc)}}$ being similar to the one of $H_s$ (compare \eqref{8-asy-tg} to \eqref{8-funct}), we can apply exactly the same procedure as in the previous subsection, and the action of $R_{(s,p)}^{\e{(1;\,nosc)}}$ exponentiates similarly as in \eqref{8-dressedF}. Therefore, a $\gamma$-equivalent form of the series \eqref{Ghat1an apres resomation Hs} is
\begin{equation}\label{Ghat1an apres resomation Rs-osc}
G_{1\dots n}(\ga) = \!\!\!\sul{r_1, \dots , \, r_n =0}{\infty}
\!\f{u_s^{r_s}}{r_s!}
\pl{p=1}{r_s}\Big[D_{(s,p)}+O_{(s,p)}
+{R}_{(s,p)}^{\e{(sub;\,nosc)}}+{R}_{(s,p)}^{\e{osc}}\Big]*
{\mc{F}}^{(R)}_{\abs{J_{\{r\}}}} \!
\left(\! \barr{c}
 \{\lambda\}_{J_{\{r\}}} \! \\
 \{z\}_{J_{\{r\}}} \! \ea \! \right),
\end{equation}
where
\begin{equation}
\label{8-dressed-calF}
{\mc{F}}^{(R)}_{\abs{J_{\{r\}}}}
\left(\barr{c}
 \{\lambda\}_{J_{\{r\}}} \\
 \{z\}_{J_{\{r\}}} \ea \right)=
 {\mc{F}}^{(H)}_{\abs{J_{\{r\}}}}\left(\barr{c}
 \{\lambda\}_{J_{\{r\}}} \\ \{z\}_{J_{\{r\}}} \ea  \right)
%
%
 \cdot
\prod_{\sigma=\pm}
\left[ \frac{ \hat{p}'(\sigma q)}{{p}'_0(\sigma q)}
       \right]^{-\hat{\nu}^2(\sigma q)}.
\end{equation}
Using the explicit expression \eqref{8-calF} for ${\mc{F}}^{(H)}$, we obtain for the new function ${\mc{F}}^{(R)}$:
\begin{multline}\label{8-hatF}
{\mc{F}}^{(R)}_{\abs{J_{\{r\}}}}
=\exp\left\{-im\int\limits_{-q}^q p'_0(\om)\,
 \hat\nu(\om)\,\dd\om \right\}
 \cdot
 \prod_{\sg=\pm}
 \left[m\,\sinh(2q)\,\hat{p}'(\sg q)\right]^{-\hat\nu^2(\sigma q)}
 \cdot e^{\tilde C[\hat\nu]}  \\
 \times \widetilde{W}_{\abs{ J_{ \paa{r} } }}
 \left(\barr{c} \{\lambda\}_{J_{\{r\}}} \\ \{z\}_{J_{\{r\}}} \ea
 \right)
 \prod_{t\in J_{\{r\}}}
 {\cal V}_{\abs{ J_{ \paa{r} } }}\left(\lambda_t\mid\barr{c} \{\lambda\}_{J_{\{r\}}}
  \\ \{z\}_{J_{\{r\}}} \ea
 \right),
 \end{multline}
in which $\hat{\nu}$ is still given by \eqref{8-hatnu} and
\begin{equation}\label{p-hat-q-F}
\hat{p}'(\mu)=p'_0(\mu)
  -\int\limits_{-q}^q p'_0(\omega)\;
 \partial_\eps \hat\nu\bigg(\omega \mid \begin{array}{cc}
       \mu, &\{\lambda\}_{J_{\{r\}}} \\ \mu +\eps, &\{z\}_{J_{\{r\}}} \end{array}
       \bigg)\bigg|_{\eps=0}\,\dd\omega\,.
 \end{equation}

\subsubsection{Oscillating contribution}\label{sec-R-osc}

Let us also compute the leading action of the oscillating part $R_s^{\e{osc}}$ on $m$-dependent function $\mc{G}_s$ of the type \eqref{8-formG}, \eqref{tilde-G}. Similar considerations based on the explicit structure of the oscillating corrections in the asymptotic series \eqref{5-asympt-Is} \cite{KitKMST08a} allow us to decompose $R_s^{\e{osc}}[\mc{G}_s]$ as
\begin{equation}
 {R}_s^{\e{osc}}[\mc{G}_s]
 =R_s^{\e{(1;\,osc)}}[\mc{G}_s]+{R}_s^{\e{(sub;\,osc)}}[\mc{G}_s].
\end{equation}
Here $R_s^{\e{(1;\,osc)}}[\mc{G}_s]$  takes into account the contributions to the leading oscillating order of the action of each subleading oscillating functional in \eqref{5-asympt-Is}, and therefore is an oscillating contribution of the same order as $O_s[\mc{G}_s]$, whereas the remainder ${R}_s^{\e{(sub;\,osc)}}[\mc{G}_s]$ is subleading with respect to $O_s[\mc{G}_s]$.
Like in the previous case, the
modification at the leading order reduces to the extension of the term $\log p'_0$ to
$\log \hat p'= \log\left( p'_0-\frac im\partial_\eps\right)$.
Namely, let us define $\widetilde{O}_s=O_s+R_s^{\e{(1;\,osc)}}\equiv
\widetilde{O}_s^+ +\widetilde{O}_s^-$. Then, for  $\mc{G}_s$ being
given by \eq{8-formG}, \eqref{tilde-G}, we have
 \begin{multline}\label{7-functO}
 \widetilde{O}^\pm_s[\mc{G}_s]
=\frac{(-1)^{s-1}}{(s-1)!}\, \partial^s_\gamma
 \exp\left\{\pm im\bigl[p'_0(q)-p'_0(-q)\bigr]+
 \tilde C[\hat\nu^{(\pm)}\mp1]-\tilde C[\hat\nu^{(\pm)}]\right\}\\
 \times\prod_{\sg=\pm}
 \left[m\sinh(2q)\,\hat p'_\pm(\sigma q) \right]^{\pm2\hat\nu^{(\pm)}(\sg q)-1}
 \Bigl.\Bigr|_{\gamma=0}
  \cdot\Phi_2\left(\barr{c}
 \mp q \\ \pm q  \ea \right),
\end{multline}
 where $\hat\nu^{(\pm)}(\omega)$ is still given by \eq{8-hatnu1O}
 and
 \begin{equation}\label{8-pOsc}
 \hat p'_\pm(\sg q)=p'_0(\sg q)
 \mp i\sg
 \left. \partial_{\eps_\sg}\Phi\left(\barr{c}
 \mp q +\eps_\mp\\ \pm q+\eps_\pm \ea \right)\right|_{\eps=0}.
\end{equation}
%

The successive action of the oscillating operators $\widetilde{O}_{(s,p)}$ will be summed up perturbatively in Section~\ref{sec-osc}.

\subsection{The action of ${D}_{(s,p)}$ as a continuous generalization of multiple Lagrange series\label{AD-CLS}}


Let us now sum up the successive action of operators ${D}_{(s,p)}$.
Applying again the binomial formula, we obtain
\begin{multline}\label{8-G-DHO}
{G}_{1\dots n}(\gamma)
=\sul{r_1,\dots,\, r_n =0 }{\infty}
%
\pl{s=1}{n}\f{(u_s)^{r_s}}{ r_s!}
\pl{s=1}{n}\pl{p=1}{r_s}
\left[ \widetilde{O}_{(s,p)}+\widetilde{R}_{(s,p)}\right]\\
\times
 \sul{\ell_1, \dots,\, \ell_n =0}{\infty}
\pl{s=1}{n}\f{(u_s)^{\ell_s}}{\ell_s!}
\pl{s=1}{n}\pl{p=1}{\ell_s}{D}_{(s,p)}*
{\mc{F}}^{(R)}_{\abs{J_{\{r\}}}+\abs{J_{\{\ell\}}}}
\left(\barr{c}
\{\mu\}_{J_{\{r\}}} \cup \{\lambda\}_{J_{\{\ell\}}} \\
\{y\}_{J_{\{r\}}} \cup \{z\}_{J_{\{\ell\}}}  \ea  \right),
\end{multline}
in which the operators ${D}_{(s,p)}$ act on the variables $\{\lambda\}$ and
$\{z\}$, whereas $\widetilde{O}_{(s,p)}$ and \vspace{-0.5mm} $\widetilde{R}_{(s,p)}\equiv {R}_{(s,p)}^{\e{(sub;\,nosc)}}+{R}_{(s,p)}^{\e{(sub;\,osc)}}$ act on $\{\mu\}$ and $\{y\}$.
We therefore need to compute
\begin{multline}\label{8-defS}
{\mc{F}}^{(D)}_{\abs{J_{\{r\}}}}\left(\barr{c}
\{\mu\}_{J_{\{r\}} }\\
\{y\}_{J_{\{r\}} } \ea \right)
=
 \sul{\ell_1, \dots,\, \ell_n =0}{\infty}
%
\pl{s=1}{n}\f{(u_s)^{\ell_s}}{\ell_s!}
   \\
\times\pl{s=1}{n}\pl{p=1}{r_s}
{D}_{(s,p)}
*{\mc{F}}^{(R)}_{\abs{J_{\{r\}}}+\abs{J_{\{\ell\}}}}\left(\barr{c}
\{\mu\}_{J_{\{r\}}} \cup \{\lambda\}_{J_{\{\ell\}}} \\
\{y\}_{J_{\{r\}}} \cup \{z\}_{J_{\{\ell\}}}
 \ea  \right).
\end{multline}
This series is actually a {\em continuous generalization of
the multiple Lagrange series}.

Recall that the standard Lagrange series
has the form (see e.g. \cite{WhiW})
 \begin{equation}\label{6-A4-SCser}
 G_0=\sum_{n=0}^\infty\frac{1}{n!}
\left.\frac{\dd^n}{\dd\epsilon^n}
\bigl(F(\epsilon)
 \phi^n(\epsilon)\bigr)\right|_{\epsilon=0},
 \end{equation}
where $F(z)$ and $\phi(z)$ are some functions holomorphic in a
vicinity of the origin. If the series \eq{6-A4-SCser} is convergent,
then it can be summed up in terms of the solution of the equation
 \begin{equation}\label{6-A4-eq}
 z-\phi(z)=0,
 \end{equation}
and the sum is given by
 \begin{equation}\label{6-A4-SC-res}
 G_0=\frac{ F(z)}{1-\phi'(z)}\, .
 \end{equation}

It is possible to generalize the Lagrange series \eq{6-A4-SCser} to the case
of functions $\phi$ and $F$ depending on several variables,
and even to consider the corresponding continuous limit.
The sum of such generalized series is then expressed in terms of
a {\em solution of an
integral equation}. All the details about these
generalizations are given in Appendix \ref{ap-LSD}.
We just apply here the result of this appendix  to our particular case.

Substituting the action \eq{6-actionD} into \eq{8-defS} and setting $k=|J_{\{r\}}|$, we obtain
\begin{multline}
 {\mc{F}}^{(D)}_k\left(\barr{c}\{\mu\} \\ \{y\} \ea \right)
= \sul{\ell_1,\dots,\, \ell_n =0 }{\infty}\,
  \pl{s=1}{n} \frac{u_s^{\ell_s}}{\ell_s!}
 \int\limits_{-q}^q
 \left.\left\{\prod_{s=1}^n\prod_{p=1}^{\ell_s}
 \frac{\dd\lambda_{s,p}}{2\pi i}
 \cdot \partial_{\epsilon_{s,p}}
 \right\}\right|_{\epsilon_{s,p}=0} \\
%
%
%
 \times
  {\mc{F}}^{(R)}_{ \abs{ J_{\paa{ \ell }}  }+k }
  \pab{ \{\mu\}\; , \ \underset{s,p}{\cup} \paa{\la_{s,p}}^{s} }
{ \{y\}\; ,\  \underset{s,p}{\cup} \big( \paa{\la_{s,p}+\eps_{s,p}} \cup \paa{ \la_{s,p} }^{s-1}\big) }  .
\end{multline}
%
%
Recall that the notation $\paa{\la_{s,p}}^{s}$ means that the
variable $\la_{s,p}$ is repeated $s$ times.
Moreover, here and in the following, we use simplified notations for
\begin{equation}\label{notation-sp}
 \underset{s,p}{\bigcup}\equiv
   \underset{\substack{1\le s\le n\\ 1\le p\le \ell_s}}{\bigcup},
 \qquad
 \sul{s,p}{}\equiv\sum_{s=1}^n\sum_{p=1}^{\ell_s}\, .
\end{equation}
Using now the explicit expression \eqref{8-hatF} of
${\mc{F}}^{(R)}_{ \abs{ J_{\paa{ \ell }} } }$,
we have
 \begin{align}
 {\mc{F}}^{(D)}_k &= \sul{\ell_1,\dots,\, \ell_n =0 }{\infty}\,
 \pl{s=1}{n} \frac{u_s^{\ell_s}}{\ell_s!}
 \int\limits_{-q}^q
 \left.\left\{\prod_{s=1}^n\prod_{p=1}^{\ell_s}
 \frac{\dd\lambda_{s,p}}{2\pi i}
 \cdot \partial_{\epsilon_{s,p}}
 \right\}\right|_{\epsilon_{s,p}=0}
 \exp\left\{-im\int\limits_{-q}^q p'_0(\omega)\,
 \hat\nu(\omega)\,\dd\omega \right\} \nonumber\\
 &\ \ \times
 \prod_{\sg=\pm}
      \big[m\sinh(2q)\,\hat{p}'(\sg q)\big]^{-\hat\nu^2(\sigma q)}
 \cdot e^{\tilde C[\hat\nu]}
 \cdot
\widetilde{W}_{\abs{\ell}+k}\!
  \left(\hspace{-1mm}\barr{c}
 \underset{u,v}{\cup} \{\lambda_{u,v}\} \\
 \underset{u,v}{\cup} \{\lambda_{u,v}+\epsilon_{u,v}\}
  \ea
 \hspace{-1mm}\right) \nonumber\\
 &\ \ \times
 \prod_{t=1}^k
 {\cal V}_{\abs{\ell}+k}\!
     \left(\mu_t\mid\hspace{-1mm}\barr{c}
 \underset{u,v}{\cup} \{\lambda_{u,v}\} \\
 \underset{u,v}{\cup} \{\lambda_{u,v}+\epsilon_{u,v}\} \ea
 \hspace{-1mm}\right)
 \prod_{s=1}^n\prod_{p=1}^{\ell_s}
 {\cal V}^s_{\abs{\ell}+k}\!
    \left(\lambda_{s,p}\mid\hspace{-1mm}\barr{c}
 \underset{u,v}{\cup} \{\lambda_{u,v}\} \\
 \underset{u,v}{\cup} \{\lambda_{u,v}+\epsilon_{u,v}\} \ea
 \hspace{-1mm}\right),
 \label{9-actDF0}
\end{align}
where $|\ell|=\sum_{s=1}^n\ell_s$. In order to lighten notations, we
did not write explicitly the arguments $\{\mu\}$ and $\{y\}$ of the
functions  $\mc{V}$ and $\wt{W}$. We also have omitted most
arguments of the functions $\hat\nu$ and $p'$.
For instance,
 \begin{equation}\label{p-hatnu}
 \hat\nu(\omega)\equiv \hat\nu\left(\omega\mid
 \barr{cc}
  \{\mu\},& \hspace{-2mm} \underset{u,v}{\cup} \{\lambda_{u,v}\} \\
 \{y\}, & \hspace{-2mm}  \underset{u,v}{\cup} \{\lambda_{u,v}+\epsilon_{u,v}\}
 \ea \right)
 =\frac {-1}{2\pi i}\log\left[1+\gamma
 \mc{V}\left(\omega\mid
 \barr{cc}
  \{\mu\}, &\hspace{-2mm} \underset{u,v}{\cup} \{\lambda_{u,v}\} \\
 \{y\} , & \hspace{-2mm} \underset{u,v}{\cup} \{\lambda_{u,v}+\epsilon_{u,v}\}
 \ea \right)\right].
 \end{equation}
and one should understand $p'(\sigma q)$ in the similar way.

Observe that we have, at most, to compute the first order derivative
with respect to each
variable $\eps_{s,p}$. It is thus enough to linearize  the arguments
of the functions $\mc{V}$ and $\wt{W}$ in the vicinities of
$\eps_{s,p}=0$. The linearized version of the function $\mc{V}$ reads
\begin{equation}\label{fonction V linearisee}
\mc{V}_{\abs{\ell}+k}\pa{\om \mid \barr{c}
\underset{s,p}{\cup}\paa{\la_{s,p}} \\
\underset{s,p}{\cup}\paa{\la_{s,p}+\eps_{s,p}}  \ea }\to
\exp\paa{\beta -i \sul{s,p}{} \eps_{s,p}\, K\pa{\om- \la_{s,p}}
+\Psi_k(\omega)}-1\, ,
 \end{equation}
where $K(\omega-\lambda)$ is given by \eq{0-K-XXZ}, and the function
$\Psi_k(\omega)$ by
 \begin{align}
 \Psi_k(\omega)
 &\equiv
 \Psi_k \bigg(\omega\mid \begin{array}{c} \{\mu\}\\ \{ y\} \end{array}\bigg)
     \nonumber\\
 &=\sum_{t=1}^k
\log\frac{\sinh(\omega-\mu_t+i\zeta)\,
                        \sinh(\omega-y_t-i\zeta)}
    {\sinh(\omega-y_t+i\zeta)\,\sinh(\omega-\mu_t-i\zeta)}
 =i\sum_{t=1}^k\int\limits_{y_t}^{\mu_t}
 K(\omega-\lambda)\,\dd\lambda\, .
 \label{8-Psi}
 \end{align}
The linearized form of $\hat\nu(\omega)$ follows immediately from
\eq{fonction V linearisee}:
 \begin{equation}\label{8-hatnu-lin}
 \hat\nu(\omega)
 \rightarrow
 -\frac1{2\pi i}\log
 \left\{1+\gamma\left[
 \exp\left(\beta
   -i \sul{s,p}{} \eps_{s,p}\, K\pa{\om- \la_{s,p}}
 +\Psi_k(\omega)\right)-1\right]\right\}\, .
 \end{equation}
As for the function $\wt{W}$, its linearized form is
rather cumbersome. On the other hand this function will not play
a significant role in this section, therefore we give here the
linearization of $\wt{W}$ in a rather formal way, namely
\begin{equation}\label{fonction W linearisee}
  \wt{W}_{\abs{\ell}+k}\!
 \pab{\hspace{-1mm}    \underset{s,p}{\cup}\paa{\la_{s,p}}
}{\hspace{-1mm}   \underset{s,p}{\cup}\paa{\la_{s,p}+\eps_{s,p}}  \hspace{-1mm}}
 \to
 F_k\! \pa{ \sul{s,p}{}\eps_{s,p}\, g^{\pa{1}}(\la_{s,p}) ;
\sul{s,p}{} \hspace{-.2mm}\sul{t,r}{} \eps_{s,p}\, \eps_{t,r}\,
g^{\pa{2}}(\la_{s,p},\la_{t,r} ) ; \dots  } ,
\end{equation}
where $g^{(1)}$, $g^{(2)},\dots$ are some functions of
$\lambda_{s,p}$. The explicit formulas for them will be given in
Section~\ref{AEebQ}. Then we have
 \begin{multline}\label{6-Gn-3}
 {\mc{F}}^{(D)}_k=\sul{\ell_1, \dots,\, \ell_n=0 }{\infty} \pl{s=1}{n}
 \f{1}{\ell_s!}\Int{-q}{q}
 \left.\left\{\prod_{s=1}^n\prod_{p=1}^{\ell_s}
 \dd\lambda_{s,p}
 \cdot \partial_{\epsilon_{s,p}}
 \right\}\right|_{\epsilon_{s,p}=0}
 \exp\left\{-im\int\limits_{-q}^q p'_0(\omega)\,
 \hat\nu(\omega)\,\dd\omega \right\}
 \\
\times
 \prod_{\sg=\pm}
      \big[m\sinh(2q)\,\hat{p}'(\sg q)\big]^{-\hat\nu^2(\sigma q)}
 \cdot e^{\tilde C[\hat\nu]}
 \cdot F_k\pa{\{g^{\pa{i}}\}}
 \\
 \times
 \pl{t=1}{k}f\!\pa{\mu_t\mid -i\sul{u,v}{}\eps_{u,v}\,
 K(\mu_t-\la_{u,v}) }
 \,
 \pl{s=1}{n}\pl{p=1}{\ell_s} \frac{u_s}{2\pi i}
 f^{s}\!\pa{\lambda_{s,p}\mid -i\sul{u,v}{}\eps_{u,v}\,
 K(\la_{s,p}-\la_{u,v}) }
 ,
\end{multline}
where the notation $F_k\pa{\paa{g^{\pa{i}}} }$ is a compact form of
\eq{fonction W linearisee} and where we have set
\begin{equation}
 \label{7-fs}
f\pa{\lambda \mid x}= \ex{\beta+x+\Psi_k(\lambda)} -1  \, .
\end{equation}

The representation \eq{6-Gn-3} is exactly the continuous
generalization of the multiple Lagrange series considered in
Appendix~\ref{ap-LSD} (see \eq{A4-MI-mult}). Thus, we can simply
apply the results derived in this appendix to the concrete case of
functions $f$, $F_k$.

\bigskip

Let us give the final result: a $\gamma$-equivalent
form of the new function ${\mc{F}}^{(D)}_k$  \eq{6-Gn-3} reads
 \begin{multline}\label{8-S-result}
 {\mc{F}}^{(D)}_k=
 \exp\left\{
 im\int\limits_{-q}^q p'_0(\omega)\,
 z(\omega)\,\dd\omega\right\}
 \prod_{\sg=\pm}\bigl[m\sinh(2q)\, p'(\sg q)\bigr]^{-z^2(\sigma q)}
 \cdot  e^{\tilde C[-z]}\\
 \times
 \f{F_k \!\pa{\Int{-q}{q} g^{\pa{1}}(\om)\, z(\om) \,\dd \om ;\quad \dots}
 \pl{t=1}{k}f\!\pa{\mu_t\mid -i\int\limits_{-q}^q
 K(\om-\mu_t)\, z(\om)\,\dd\om }}
 { \ddet{}{ I+iK(\xi-\la)\,
   f'_{\Sg}\!\pa{\xi\mid  -i\Int{-q}{q} \,
  K(\om-\xi)\, z(\om) \,\dd \om}
 }} \, .
\end{multline}
Here the function $z(\la)$ solves a non-linear integral equation,
 \begin{equation}\label{8-Int-eq0}
 z(\la)=f_{\Sg}\!\left(\lambda\mid -i\int\limits_{-q}^q
 K(\om-\la)\,z(\om)\,\dd\om\right) ,
 \end{equation}
with $f_{\Sg}$ given by
 \begin{equation}\label{8-fS}
 f_{\Sg}(\la\, |\, x)=\frac1{2\pi i}\log\left[1+\gamma\left(
 e^{\beta +x+\Psi_k(\la)}-1\right) \right].
 \end{equation}
The symbol $f'_{\Sg}$ means the derivative of $f_{\Sg}(\la \, |\, x)$
over $x$. The Fredholm determinant in the denominator of
\eq{8-S-result} is the Jacobian of the equation \eq{8-Int-eq0}
(compare with \eq{6-A4-eq} and \eq{6-A4-SC-res}).
The function\footnote[1]{%
We use the notation  $p'(\lambda)$ for this function, since
eventually  $p(\lambda)$ will have the sense of the dressed momentum
\eq{0-Dmom}.}
$p'(\lambda)$, which appears in  \eq{8-S-result} taken in the points $\lambda=\pm q$,
satisfies a linear integral equation,
 \begin{equation}\label{8-Int-eq-lin}
 p'(\la)+\int\limits_{-q}^q
 K_\Sg(\la,\om)\, p'(\om) \,\dd\om=p'_0(\la)\, ,
 \end{equation}
 where
 \begin{equation}\label{8-K-Sg}
 K_\Sg(\la,\om)=i\, K(\la-\om)\;
 f_{\Sg}'\!\pa{\om\mid  -i \Int{-q}{q}  K(\xi-\om)\,
 z(\xi)\,\dd \xi}.
 \end{equation}
We would like finally to mention that the
functions $z(\lambda)$ and $p'(\lambda)$ depend on the parameters
$\{\mu\}$ and $\{y\}$ through $\Psi_k$ (see \eqref{8-fS}),
 \begin{equation}\label{8-zp}
 z(\la)= z\left(\la\mid
 \barr{c}\{\mu\} \\ \{y\}\ea \right),\qquad
  p'(\la)= p'\left(\la\mid
 \barr{c}\{\mu\} \\ \{y\} \ea \right),
 \end{equation}
and still satisfy the reduction property \eqref{8-recurs}.

\bigskip

Let us now explain how to obtain this result from the results of Appendix~\ref{ap-LSD}.

According to Appendix~\ref{ap-LSD-4}, the final answer for
${\mc{F}}^{(D)}_k$ is given in terms of a function $z^{(n)}(\la)$ solving an
integral equation (see \eq{A4-RE-Int-eq2})
\begin{equation}\label{6-int-eq-zn}
z^{(n)}(\la)-f_{\Sg_n}\!\!\pa{\la\mid -i\Int{-q}{q} \!
K(\om-\la)\, z^{(n)}(\om)\,\dd \om } = 0 ,
 \qquad
f_{\Sg_n}(\la\,|\,x)=\sul{s=1}{n} \frac{u_s}{2\pi i}\,f^{s}(\la\,|\, x) .
\end{equation}
Following our usual strategy, we can replace this equation with a
$\gamma$-equivalent one by sending $n\tend\infty$ in $f_{\Sg_n}$ as
$u_s \propto \ga^s$. Then the function $z^{(n)}(\mu)$ is replaced
by $z(\mu)$ satisfying the integral equation \eq{8-Int-eq0}, since
\begin{align}
\lim_{n\tend\infty} f_{\Sg_n}(\la\,|\, x)
&=\sul{s=1}{\infty} \f{ \pa{-1}^{s-1}\ga^s }{2\pi is}
  \pa{ \ex{\beta+x+\Psi_k(\la) }-1 }^s
\nonumber\\
&=\frac1{2\pi i}\log\big[1+\gamma
f(\la\,|\, x) \big]
=f_{\Sg}(\la\,|\, x)\,.
\label{6-limit-z}
\end{align}

Comparing \eq{8-S-result} and \eq{6-Gn-3} one can observe that the
action of the operators ${D}_{(s,p)}$ leads to the replacement of the
functional argument $\hat\nu(\la)$ by $-z(\la)$. This happens due to
the fact that the linearized form of $\hat\nu$ \eq{8-hatnu-lin}
coincides with the r.h.s. of the equation \eq{8-Int-eq0} up to the
sign
 \begin{equation}\label{8-hnu-z}
 \hat\nu\left(\omega \mid \barr{c}
 \underset{s,p}{\cup}\{\la_{s,p}\} \\
 \underset{s,p}{\cup}\{\la_{s,p}+\eps_{s,p}\}
 \ea
 \right)\to-f_{\Sg}\left(\om\mid -i \sul{s,p}{}
 \eps_{s,p} \, K\pa{\omega- \la_{s,p}}\right).
 \end{equation}
Since the summation of the continuous generalization of the Lagrange
series reduces to the replacement of the sum over $\eps_{s,p}$ by
the integral with $z(\lambda)$ (see \eq{A4-2-RE-res-1}), we obtain
 \begin{equation}\label{8-hnu-z1}
 \hat\nu(\omega)\to -f_{\Sg}\left(\om\mid -i\int\limits_{-q}^q
 K(\omega-\la)\, z(\lambda)\,\dd\lambda\right)=-z(\omega).
 \end{equation}

Finally, let us explain how the function $p'(\lambda)$ and the
equation  \eq{8-Int-eq-lin} arise. From \eqref{p-hat-q-F} and
\eqref{8-hnu-z1},  we have\footnote{Here also, we do not specify
that $z$ depends on the extra sets of parameters $\{\mu\}$ and
$\{y\}$.}
 \begin{equation}\label{8-phat-p}
 \hat p'(\la)\to
 p'(\la)=p'_0(\la)
 +\int\limits_{-q}^q p'_0(\omega)\;
  \partial_\epsilon z\bigg(\omega \mid \begin{array}{c}
       \la \\ \la +\eps \end{array}
       \bigg)\bigg|_{\eps=0}.
 \end{equation}
Differentiating the integral equation \eq{8-Int-eq0} we find
 \begin{equation}\label{10-Ie-der}
 \partial_\epsilon z\bigg(\omega \mid \begin{array}{c}
       \la \\ \la +\eps \end{array}
       \bigg)\bigg|_{\eps=0}
 +
 \int\limits_{-q}^q K_\Sg(\xi,\om)\;
 \partial_\epsilon z\bigg(\xi \mid \begin{array}{c}
       \la \\ \la +\eps \end{array}
       \bigg)\bigg|_{\eps=0}
 \,\dd\xi
 =
 -K_\Sg(\la,\om)\, .
 \end{equation}
Here $-K_\Sg(\la,\om)$ in the r.h.s. appears due to the derivative
of the function $\Psi_k$ \eq{8-Psi}. Hence the derivative
$\partial_\epsilon z$ can be expressed in terms of the resolvent
$R_\Sg(\la,\om)$  of the integral operator $K_\Sg$:
 \begin{equation}\label{8-derivat1}
 \partial_\epsilon z\bigg(\omega \mid \begin{array}{c}
       \la \\ \la +\eps \end{array}
       \bigg)\bigg|_{\eps=0}
 = - R_\Sg(\la,\om)\, .
 \end{equation}
Substituting this into \eq{8-phat-p} we arrive at
 \begin{equation}\label{10-resolv}
 p'(\la)=p'_0(\la)-\int\limits_{-q}^q R_\Sg(\la,\om)\, p'_0(\om)
 \,\dd\om\, ,
 \end{equation}
which means that $p'(\la)$ solves the equation \eq{8-Int-eq-lin}.


\subsection{Oscillating corrections}\label{sec-osc}

The series \eq{8-G-DHO} takes now the form
\begin{equation}\label{8-G-DHO-1}
{G}_{1\dots n}(\gamma) =\sul{r_1,\dots,\, r_n =0 }{\infty} \;
 %
\pl{s=1}{n}\f{(u_s)^{r_s}}{r_s!} \pl{s=1}{n}\pl{p=1}{r_s}
\left[ \widetilde{O}_{(s,p)}+\widetilde{R}_{(s,p)}\right]*
{\mc{F}}^{(D)}_{\abs{J_{\{r\}}}}\left(\barr{c}
\{\mu\}_{J_{\{r\}}}\\
\{y\}_{J_{\{r\}}} \ea \right),
\end{equation}
with ${\mc{F}}^{(D)}_{\abs{J_{\{r\}}}}$ given by \eqref{8-S-result}.
Note that ${\mc{F}}^{(D)}_{\abs{J_{\{r\}}}}$ is still an $m$-dependent function of the type \eq{8-formG}, \eq{tilde-G},
and therefore we can use the results of Section~\ref{sec-R-osc}.

At this stage, the functionals which remain to be summed up produce only subleading contributions. Hence, if we were only interested in the leading term of $\moy{e^{\beta\mc{Q}_m}}$, it would be enough to consider only the leading term of these series, namely the term $r_1=\dots=r_n=0$. Recall however that we need to keep the leading oscillating contribution since, after second lattice derivative, it may produce a term competing with the leading non-oscillating one. We can nevertheless forget completely the remaining terms $\widetilde{R}_{(s,p)}$, since they merely produce corrections to one of these two main contributions. Moreover,
since the oscillating operators $\widetilde{O}_{(s,p)}$ are themselves of sub-leading order, it
is enough to restrict our analysis to the terms of \eqref{8-G-DHO-1} that are linear over $\widetilde{O}_{(s,p)}$, namely the terms for which one $r_k$ takes the value
$0$ or $1$ while  all other are equal to $0$. We thus obtain
 \begin{equation}\label{8-Lin-O}
 {G}_{1\dots n}(\gamma)
 ={\mc{F}}^{(D)}_{0}\left(\barr{c}\emptyset \\
 \emptyset\ea \right)+
 \sul{s=1}{n} u_s\, \widetilde{O}_s\!
\left[ {\mc{F}}^{(D)}_{s}\left(\barr{c}\{\mu\} \\
                  \{y\} \ea \right)\right]+ \text{corrections},
\end{equation}
in which the {\em corrections} are either non-oscillating corrections to the first term, or oscillating corrections to the second term (note that the non-oscillating corrections may be of order higher than this second term, but they are nevertheless not important for the final result).
Observe once again that, due to the fact that $u_s\propto\gamma^s$, the summation
over $s$ in \eq{8-Lin-O} can be extended to infinity. This will be
done in the end of our calculations.

To complete the summation of the series, we should now
calculate the functionals $\widetilde{O}_s\big[{\mc{F}}^{(D)}_s\big]$. Let us
present ${G}_{1\dots n}(\gamma)$  in \eq{8-Lin-O} as the sum of
three parts
 \begin{equation}\label{8-G-G0+-}
 {G}_{1\dots n}(\gamma)
 =\sum_{\sigma=0,\pm 1} {G}^{(\sigma)}_{1\dots n}(\gamma)
 + \text{corrections}\,,
 \end{equation}
where
 \begin{equation}\label{8-def-G0+-}
 {G}^{(0)}_{1\dots n}(\gamma)
 ={\mc{F}}^{(D)}_{0}\left(\barr{c}\emptyset \\
 \emptyset\ea \right),
 \qquad
 {G}^{(\pm)}_{1\dots n}(\gamma)=
 \sul{k=1}{n}u_k\,\widetilde{O}_k^{\pm}\!
\left[ {\mc{F}}^{(D)}_{k}\left(\barr{c}\{\mu\} \\
 \{y\} \ea \right)\right].
\end{equation}

The non-oscillating part ${G}^{(0)}_{1\dots n}(\gamma)$
has in fact already been computed. It is given by \eq{8-S-result} in which
one should set $\{\mu\}=\{y\}=\emptyset$, which is easy since we almost did not specify this dependence:
basically, the parameters $\{\mu\}$ and $\{y\}$ enter the
functions $z(\lambda)$ and $p'(\lambda)$ \eq{8-zp} through the function $\Psi_k$ \eq{8-Psi}.
Thus, in order to obtain ${G}^{(0)}_{1\dots n}(\gamma)$, it is enough
to set $\Psi_k=0$ in \eq{8-Psi} and $k=0$ in \eq{8-S-result}.

Let us now compute the oscillating corrections ${G}^{(\pm)}_{1\dots n}(\gamma)$. The new function ${\mc{F}}^{(D)}_k$ \eq{8-S-result} is of the type \eq{8-formG}, \eq{tilde-G},
with
 \begin{equation}\label{10-Phi1}
 \Phi_1\left(\xi \mid\barr{c}\{\mu\} \\
 \{y\} \ea \right)
 = f\pa{\xi \mid -i\int\limits_{-q}^q
 K(\om-\xi)\, z(\om)\,\dd\om }\, ,
 \end{equation}
where $f(\la | x)$ is given by \eq{7-fs} and
 \begin{equation}\label{10-Phi}
 \Phi\left(\barr{c}\{\mu\}\\
 \{y\} \ea \right)
 =i\int\limits_{-q}^q p'_0(\om)\,
 z(\om)\,\dd\om\,.
 \end{equation}
%
The functionals $\widetilde{O}_k^\pm$ \eqref{7-functO} send $\mu_1=\mp q$, $y_1=\pm q$ and
$\mu_j=y_j$ for $j>1$. This means that the function $z(\lambda)$
turns into $z^{(\pm)}(\lambda)$ satisfying the equation
 \begin{equation}\label{10-Int-eq0}
 z^{(\pm)}(\la)=f^{(\pm)}_{\Sg}\!\left(\lambda\mid -i\int\limits_{-q}^q
 K(\om-\la)\, z^{(\pm)}(\om)\,\dd\om\right),
 \end{equation}
with $f^{(\pm)}_{\Sg}$ given by
 \begin{equation}\label{10-fS}
 f^{(\pm)}_{\Sg}(\la\mid x)=\frac1{2\pi i}\log\left[1+\gamma\left(
 e^{\beta +x+\Psi^{(\pm)}(\la)}-1\right) \right],
 \end{equation}
and
 \begin{equation}
 \Psi^{(\pm)}(\omega)
 =\left.\log
 \frac{\sinh(\omega-\mu+i\zeta)\,\sinh(\omega-y-i\zeta)}
      {\sinh(\omega-y+i\zeta)\,\sinh(\omega-\mu-i\zeta)}
 \right|_{\substack{\mu=\mp q\\ y=\pm q}}
 =\mp i\int\limits_{-q}^{q}
 K(\omega-\lambda)\,\dd\lambda\, .
 \label{10-Psi}
 \end{equation}
Similarly the function $p'(\la)$ \eq{8-Int-eq-lin} turns into
$p'_\pm(\la)$ satisfying
 \begin{equation}\label{10-Int-eq-lin}
 p'_\pm(\la)+\int\limits_{-q}^q
 K_\Sg^{(\pm)}(\la,\om)\, p'_\pm(\om) \,\dd\om=p'_0(\la)\, ,
 \end{equation}
 where
 \begin{equation}\label{10-K+}
 K_\Sg^{(\pm)}(\la,\om)
 =i\, K(\la-\om)\,
   \left(f^{(\pm)}_{\Sg}\right)'\!
 \pa{\om\mid  -i \Int{-q}{q}  K(\xi-\om)\,
 z^{(\pm)}(\xi)\,\dd \xi}.
 \end{equation}
Comparing \eq{10-fS} with \eq{8-hatnu1O} for
$\hat\nu^{(\pm)}(\lambda)$ we conclude that
$\hat\nu^{(\pm)}(\lambda)=-z^{(\pm)}(\lambda)$. Thus, we obtain
 \begin{multline}\label{10-functO}
 \widetilde{O}_k^\pm\big[{\mc{F}}^{(D)}_k\big]
 =\frac{(-1)^{k-1}}{(k-1)!}\,\partial^k_\gamma
 \exp\left\{\pm im\bigl[p'_0(q)-p'_0(-q)\bigr]+
 \tilde C[-z^{(\pm)}\mp 1]-\tilde C[-z^{(\pm)}]\right\}\\
 \times
 \prod_{\sg=\pm}\left[m\sinh(2q)\,\hat p'_\pm(\sigma q) \right]^{\mp 2z^{(\pm)}(\sg q)-1}
 \Bigl.\Bigr|_{\gamma=0}\cdot\Phi_2\left(\barr{c}
 \mp q \\ \pm q  \ea \right).
\end{multline}
Finally, similar arguments as in Section~\ref{AD-CLS} (see
\eqref{10-Ie-der} to \eqref{10-resolv}) enable us to show that the
quantity  $\hat p'_\pm(\sg q)$ defined by \eq{8-pOsc},
\eqref{10-Phi} is equal to $p'_\pm(\sg q)$, where $p'_\pm(\la)$ is
the solution of the equation \eq{10-Int-eq-lin}.

Observe that the $k^{\e{th}}$ $\gamma$-derivative acts on an expression which is
independent on $k$. Therefore the sum over $k$ in \eq{8-def-G0+-}, once
extended to infinity, gives the Taylor series
 \begin{multline}\label{10-functO-sum}
 \sum_{k=1}^\infty u_k\, \widetilde{O}_k^\pm\big[{\mc{F}}^{(D)}_k\big]=
 \exp\left\{\pm im\bigl(p'_0(q)-p'_0(-q)\bigr)+
 \tilde C[-z^{(\pm)}\mp 1]-\tilde C[-z^{(\pm)}]\right\}\\
 \times\prod_{\sg=\pm}\left[m\,\sinh(2q)\, p'_\pm(\sigma q) \right]^{\mp 2z^{(\pm)}(\sg q)-1}
 \cdot\Phi_2\left(\barr{c}
 \mp q \\ \pm q  \ea \right).
\end{multline}

Combining now all the above results, we obtain
 \begin{multline}\label{8-S-result+-}
 {G}^{(\pm)}_{1\dots n}(\gamma)=
\exp\left\{
 im\int\limits_{-q}^q \hspace{-1mm}p'_0(\lambda)
 \bigl(z^{(\pm)}(\lambda)\pm1\bigr)\,\dd\lambda\right\}
 \prod_{\sg=\pm}\bigl[m\sinh(2q)\, p'_\pm(\sg q)\bigr]^{-\left(z^{(\pm)}(\sigma q)\pm1\right)^2}\\
 \times
 \f{F^{(\pm)} \pa{\Int{-q}{q} g^{\pa{1}}_\pm(\la)\, z^{(\pm)}(\la) \,\dd
 \la ;\quad \dots}
 e^{\tilde C[-z^{(\pm)}\mp1]}}
 { \ddet{}{ I+iK(\xi-\la)
 \left(f^{(\pm)}_{\Sg}\right)'\!\pa{\xi\mid
 -i\Int{-q}{q} K(\om-\xi)\, z^{(\pm)}(\om)\,\dd\om }
 }} \, .
\end{multline}
Here in complete analogy with \eq{fonction W linearisee} the
functions $F^{(\pm)}$ and $\{g^{\pa{i}}_\pm\}$ are defined from the
linearized form of $\widetilde{W}_{|\ell|+1}$
\begin{equation}\label{8-dirF+-}
\widetilde{W}_{\abs{\ell}+1}
\left(\barr{c} \mp q\; ,\ \underset{s,p}{\cup}\{\la_{s,p}\}
\\ \pm q\; ,\ \underset{s,p}{\cup}\{\la_{s,p}+\eps_{s,p}\}   \ea \right)
\to
 F^{(\pm)}\left(\sul{s,p}{} \eps_{s,p}\, g^{(1)}_\pm(\la_{s,p});\ \dots
\right)  .
\end{equation}
%


\section{Leading asymptotic behavior  of correlation functions}
\label{sec-results}

In the first part of this section, we gather all the previous results for the asymptotic behavior of $\moy{\ex{\beta \mc{Q}_m}}$. This will enable us, in a second part, to obtain the leading asymptotic behavior at large distance of the spin-spin correlation function $\moy{\sg_1^z\, \sg_{m+1}^z}$.

\subsection{Asymptotic expansion for $\moy{\ex{\beta \mc{Q}_m}}$\label{AEebQ}}

Let us indicate explicitly that the functions
${G}^{(\sigma)}_{1\dots n}(\gamma)$ \eq{8-def-G0+-} depend on the
distance $m$ and parameter $\beta$: ${G}^{(\sigma)}_{1\dots
n}(\gamma) ={G}^{(\sigma)}_{1\dots n}(\gamma|\beta,m)$. The remarkable
property of the obtained $\gamma$-equivalent results is that they do
not depend on $n$: ${G}^{(\sigma)}_{1\dots
n}(\gamma|\beta,m)={G}^{(\sigma)}(\gamma|\beta,m)$. Therefore the
series \eq{6-Q1-Gn} is the Taylor series of
${G}^{(\sigma)}(\gamma|\beta,m)$ at $\ga=1$, hence leading to :
\begin{align}
 \moy{\ex{\beta \mc{Q}_m}}
 &=\frac{1}{\det[I+\frac{1}{2\pi}K]}
 \sum_{\sigma=0,\pm}\,\sul{n\geq 0}{}
 \f{\Dp{\ga}^n G^{(\sigma)}(\gamma|\beta,m)}{n!}\Bigl.\Bigr|_{\ga=0}
 +\text{corrections}\nonumber\\
 &=
 \sum_{\sigma=0,\pm}\widehat{G}^{(\sigma)}(\beta,m)\big[1+\e{o}(1)\big]\, ,
\qquad\text{with}\quad
 \widehat{G}^{(\sg)}(\beta,m)
 =\frac{{G}^{(\sigma)}(1 |\beta,m)}{\det[I+\frac{1}{2\pi}K]}\, .
  \label{9-ebQ-GGG}
 \end{align}
%
%
The term $\widehat{G}^{(0)}(\beta,m)$ in \eq{9-ebQ-GGG} gives the leading
non-oscillating asymptotic behavior of $\moy{\ex{\beta \mc{Q}_m}}$, while
$\widehat{G}^{(\pm)}(\beta,m)$ describe its leading oscillating correction. Let us consider
these two terms separately.

\subsubsection{Leading non-oscillating behavior of $\moy{\ex{\beta \mc{Q}_m}}$}

It happens that, at $\ga=1$, {\em the non-linear integral equation for the
function $z(\lambda)$ \eq{8-Int-eq0} degenerates into a linear one}
(taking also into account that, in this case, $\Psi_k=0$):
\begin{equation}\label{6-IEforz-lin}
z(\la) =\frac\beta{2\pi i} -\Int{-q}{q} \f{\dd \mu}{2\pi}\,
K(\la-\mu)\, z(\mu)\, .
\end{equation}
Setting
$z(\la)=\frac{\beta Z(\la)}{2\pi i}$, one gets the {\em integral
equation \eq{0-DC} for the dressed charge}:
\begin{equation}\label{7-DC}
 Z(\la)+\Int{-q}{q} \f{\dd \mu}{2\pi}\, K(\la-\mu)\, Z(\mu)=1 \, .
\end{equation}

The linear equation \eq{8-Int-eq-lin} also simplifies due to the fact that
$f'_{\Sg}=1/2\pi i$ at $\gamma=1$,
 \begin{equation}\label{9-Int-eq-lin}
 p'(\mu)+\frac1{2\pi}\int\limits_{-q}^qK(\mu-\la)\, p'(\lambda) \,\dd\lambda=p'_0(\mu)\,.
 \end{equation}
Comparing the last equation with \eq{0-rho}, we find that
$p'(\lambda)=2\pi\rho(\lambda)$ and, hence, {\em the function
$p(\lambda)$ can be identified with the dressed momentum} (see
\eq{0-Dmom}).

Due to the same property of $f'_{\Sg}$, the Fredholm determinant in
the denominator of \eq{8-S-result} turns into
$\det[I+\frac1{2\pi}K ]$. Thus, we have
 \begin{equation}\label{9-S-result}
 \widehat{G}^{(0)}(\beta,m)=
 e^{m\beta D}
 \bigl[2\pi\rho(q)m\sinh(2q)\bigr]^{\f{\beta^2{\cal Z}^2}{2\pi^2}}
 e^{\tilde C\bigl[{\textstyle\frac{\beta Z}{2\pi
 i}}\bigr]}
 %
 \f{F_{0} \bigg(\frac\beta{2\pi i}\Int{-q}{q} g^{\pa{1}}(\mu)\, Z(\mu) \,\dd
 \mu ;\ \dots\bigg) }
 { \det^2\left[I+\frac1{2\pi}K\right]} \, ,
\end{equation}
where ${\cal Z}=Z(\pm q)$, and we have used that $Z(\lambda)$ and
$\rho(\lambda)$ are even functions.

It remains to pass from the symbolic form of $F_0$ to a more specific
one.
When linearized, the product in the representation \eq{2-W} for the function
$\widetilde{W}_{\abs{\ell}}$ turns into
 \begin{multline}\label{9-W0}
  \left.\prod_{s,p}\prod_{s',p'}
\frac{\sinh(z_{s,p}-\lambda_{s',p'}-i\zeta)\,
      \sinh(\lambda_{s',p'}-z_{s,p}-i\zeta)}
 {\sinh(z_{s,p}-z_{s',p'}-i\zeta)\,
  \sinh(\lambda_{s,p}-\lambda_{s',p'}-i\zeta)}
 \right|_{\{z_{u,v}\}=\{\lambda_{u,v}+\eps_{u,v}\}}\\
 \to\quad
 \exp\left\{\sum_{s,p}\sum_{s',p'}
 \eps_{s,p}\,\eps_{s',p'}\,
 g^{(2)}(\lambda_{s,p},\lambda_{s',p'})\right\} ,
 \end{multline}
where
 \begin{equation}\label{9-g2}
 g^{(2)}(\lambda,\mu)=-\frac12\left[\sinh^{-2}(\lambda-\mu+i\zeta)+
 \sinh^{-2}(\lambda-\mu-i\zeta)\right].
 \end{equation}
After the summation of the continuous Lagrange series, this part
becomes
 \begin{equation}\label{9-W}
 \left.\prod_{s,p}\prod_{s',p'}
\frac{\sinh(z_{s,p}-\lambda_{s',p'}-i\zeta)\,
      \sinh(\lambda_{s',p'}-z_{s,p}-i\zeta)}
 {\sinh(z_{s,p}-z_{s',p'}-i\zeta)\,
  \sinh(\lambda_{s,p}-\lambda_{s',p'}-i\zeta)}
 \right|_{\{z_{u,v}\}=\{\lambda_{u,v}+\eps_{u,v}\}}
\hspace{-2mm}
 \to \quad e^{\frac{\beta^2}{4\pi^2}C_0},
 \end{equation}
where
 \begin{equation}\label{7-C0}
 C_0={\dis\int\limits_{-q}^{q}}
 \frac{Z(\lambda)\, Z(\mu)}{\sinh^2(\lambda-\mu-i\zeta)}\,\dd\lambda\,
 \dd\mu\, .
 \end{equation}

The remaining part of $\wt{W}$ contains the Fredholm determinants
\eq{2-det-lz}, whose kernels depend on the following products (see
\eq{2-U1}, \eq{2-U2})
 \begin{equation}\label{9-linPa}
 \left.P_\alpha=\prod_{s,p}
 \frac{\sinh(w-\lambda_{s,p}+i\alpha\zeta)}{\sinh(w-z_{s,p}+i\alpha\zeta)}
 \right|_{\{z_{s,p}\}=\{\lambda_{s,p}+\eps_{s,p}\}}
 \hspace{-6mm}
 \to \quad
 \exp\left\{\sum_{s,p}\eps_{s,p}\,g^{(1,\alpha)}(\lambda_{s,p})\right\},
 \quad
 \alpha=0,\pm1 \,,
 \end{equation}
where $g^{(1,\alpha)}(\lambda)=\coth(w-\lambda+i\alpha\zeta)$. Let
us introduce the $i\pi$-periodic Cauchy transform of the dressed
charge $Z$,
\begin{equation}\label{9-CauchyZ}
\tilde{z}(\om)= \f{1}{2\pi i} \Int{-q}{q} \coth(\la-\om)\, Z(\lambda)\,
\dd \la\,.
\end{equation}
Then  we obtain
 \begin{equation}\label{9-V-eps}
 P_\alpha\to
 e^{\beta\tilde z(w+i\alpha\zeta)},\qquad
 \alpha=0,\pm1\,.
 \end{equation}
Observe that the function $\tilde z$ has the following properties
 \begin{equation}\label{6-prop-z}
 \begin{aligned}
 &\tilde z(w+i\zeta)- \tilde z(w-i\zeta)=1-Z(w),\\
 &\tilde z_+(w)- \tilde z_-(w)=Z(w), \qquad w\in[-q,q],
 \end{aligned}
\end{equation}
where \eq{0-DC} is used in the first equation, and $\tilde z_\pm$
are limiting values of $\tilde z$ from the upper and lower half
planes in the second equation.

Therefore, the leading non-oscillating part of the
generating function can be written as
\begin{equation}\label{7-Nonosc}
 \widehat{G}^{(0)}(\beta,m)
 ={\cal A}(\beta)\cdot
 e^{\beta mD}
 \big[2\pi\,\sinh(2q)\,\rho(q)\,m\big]^{\frac{\beta^2 {\cal Z}^2}{2\pi^2}}
 G^2\!\left(1,\frac{\beta {\cal Z}}{2\pi i}\right)
 e^{\frac{\beta^2}{4\pi^2}(C_0-C_1)} \, .
 \end{equation}
Here the constant $C_0$  given by \eq{7-C0}, $G(1,z)$ is the product
of the Barnes functions (see \eq{5-Barnes}), and $C_1$ reads
 \begin{equation}\label{7-C_1}
 C_1=\frac12\int\limits_{-q}^{q}
 \frac{Z'(\lambda)Z(\mu)-Z(\lambda)\,Z'(\mu)}{\tanh(\lambda-\mu)}
 \,\dd\lambda \,\dd\mu
 +2{\cal Z}\int\limits_{-q}^{q}
 \frac{{\cal Z}-Z(\lambda)}{\tanh(q-\lambda)}\,\dd\lambda\, .
 \end{equation}
The coefficient ${\cal A}(\beta)$ is
 \begin{equation}\label{7-ev-phiC}
 {\cal A}(\beta)
 =\frac{(e^\beta-1)^2 \cdot
\det\left[I+\frac1{2\pi i}\, U_{\theta_1}^{(\lambda)}(w,w')\right] \cdot \det\left[I+\frac1{2\pi i}\, U_{\theta_2}^{(z)}(w,w')\right]}
{\left(e^{\beta\tilde z(\theta_1+i\zeta)}-e^{\beta+\beta\tilde
 z(\theta_1-i\zeta)}\right)\left(
  e^{-\beta\tilde z(\theta_2-i\zeta)}
 -e^{\beta-\beta\tilde z(\theta_2+i\zeta)}\right)
 \cdot
  \det^2[I+\frac1{2\pi}K]}\, ,
 \end{equation}
%
where
 \begin{equation}\label{7-U1}
 U_{\theta_1}^{(\lambda)}(w,w')
 =-\frac{e^{\beta\tilde z(w)} [K_\kappa(w-w')-K_\kappa(\theta_1-w')]}
 {e^{\beta\tilde z(w+i\zeta)}-e^{\beta+\beta\tilde z(w-i\zeta)}}\, ,
 \end{equation}
and
 \begin{equation}\label{7-U2}
 U_{\theta_2}^{(z)}(w,w')
 =\frac{e^{-\beta\tilde z(w')}[K_\kappa(w-w')-K_\kappa(w-\theta_2)]}
 {e^{-\beta\tilde z(w'-i\zeta)}-e^{\beta-\beta\tilde z(w'+i\zeta)}}\, ,
 \end{equation}
and the kernels $U_{\theta_1}^{(\lambda)}(w,w')$ and
$U_{\theta_2}^{(z)}(w,w')$ act on the contour $\Gamma$ surrounding
the interval $[-q,q]$.
%
%
%


\subsubsection{Leading oscillating behavior of $\moy{\ex{\beta \mc{Q}_m}}$}

Consider for exemple the function $\widehat{G}^{(+)}(\beta,m)$,
the case $\widehat{G}^{(-)}(\beta,m)$ being completely analogous.
Like in the previous case, the function $f^{(+)}_\Sg$ \eq{10-fS} becomes linear
at  $\gamma=1$, and $(f^{(+)}_\Sg)'=1/2\pi i$. Therefore the
Fredholm determinant in \eq{8-S-result+-} is equal to
$\det[I+\frac1{2\pi}K]$ like for the non-oscillating term
$\widehat{G}^{(0)}(\beta,m)$ and the integral equation for $p'(\lambda)$
\eq{9-Int-eq-lin} remains the same.

As for the linear integral equation for $z^{(+)}(\lambda)$, it
reads
 \begin{equation}\label{9-IEforz-lin+}
z^{(+)}(\la)=\frac{\beta+\Psi^{(+)}(\lambda)}{2\pi i}
-\Int{-q}{q} \f{\dd \mu}{2\pi}\, K(\la-\mu)\, z^{(+)}(\mu) \, .
\end{equation}
Using the expression \eq{8-Psi} of $\Psi(\lambda)$, one can write
\eq{9-IEforz-lin+} in the form
 \begin{equation}\label{9-IEforz-lin+1}
z^{(+)}(\la)=\frac{\beta}{2\pi i}
 -\Int{-q}{q} \f{\dd \mu}{2\pi}\,
K(\la-\mu)\, (z^{(+)}(\mu)+1) \, ,
\end{equation}
and the solution is
 \begin{equation}\label{9-solIE+}
 z^{(+)}(\la)=Z(\lambda)\left(\frac\beta{2\pi i}+1\right)-1\,.
 \end{equation}

One should now notice that, everywhere but in the argument of $F^{(+)}$, the function $z^{(+)}(\la)$
enters  in
\eq{8-S-result+-} only through the combination
 \begin{equation}\label{9-combi}
 z^{(+)}(\la)+1=\frac{
 (\beta+2\pi i)\,Z(\lambda)}{2\pi i}
,
 \end{equation}
which means that this part of the result can merely be obtained from the
non-oscillating term via the replacement $\beta\to
\beta+2\pi i$.
It is easy to see that it is enough to make the same shift
$\beta\to\beta+2\pi i
$ in the function $F^{(+)}$: indeed, the
Fredholm determinants in this function contain the following
products
\begin{equation}
 P_\alpha^{(+)}=\frac{\sinh(w+q+i\alpha\zeta)}{\sinh(w-q+i\alpha\zeta)}
 \prod_{s,p}
 \frac{\sinh(w-\lambda_{s,p}+i\alpha\zeta)}{\sinh(w-z_{s,p}+i\alpha\zeta)},\qquad
 \alpha=0,\pm1\, ,
\end{equation}
which, in the linearized limit $\{z_{s,p}\}=\{\lambda_{s,p}+\eps_{s,p}\}$ become
 \begin{align}
 P_\alpha^{(+)}
 &\to
 \exp\left\{\sum_{s,p}\eps_{s,p}\coth(w-\lambda_{s,p}+i\alpha\zeta)+
 \log\frac{\sinh(w+q+i\alpha\zeta)}{\sinh(w-q+i\alpha\zeta)}\right\}
                \nonumber\\
 &\to
 \exp\left\{
 \int\limits_{-q}^q\coth(w-\lambda+i\alpha\zeta)\,
               (z^{(+)}(\lambda)+1)\,\dd\lambda
 \right\}
 =e^{(\beta+2\pi i)\,\tilde z(w+i\alpha\zeta)},
\label{9-V-eps1}
 \end{align}
and one can similarly show that the same shift of $\beta$ has to be done
in the equation \eq{9-W}.

The term $\widehat{G}^{(-)}(\beta,m)$ can be considered in the same way,
and we finally obtain
 \begin{equation}\label{9-G0G+-}
 \widehat{G}^{(\pm)}(\beta,m)=\widehat{G}^{(0)}(\beta\pm 2\pi i,m)\, .
 \end{equation}
It is worth mentioning at this point that the generating function
$\moy{\ex{\beta \mc{Q}_m}}$ is a polynomial of $e^\beta$, which means that the
exact result should be a $2\pi i$-periodic function of $\beta$.
In the asymptotic formula this periodicity may of course be broken.
We see, however, that {\em the leading oscillating part of the asymptotics 
partly restores the original periodicity}. It is therefore very possible
that the more rapidly oscillating corrections to this formula can simply be obtained, at their leading order, by the shifts $\beta\to\beta+2\pi i n$, $n\in\mathbb{Z}^*$, in
$\widehat{G}^{(0)}(\beta,m)$.

\subsubsection{Final result and comments}

We eventually obtain
 \begin{equation}\label{9-ebQ-GGGs}
 \moy{\ex{\beta \mc{Q}_m}}
 =\sum_{\sigma=0,\pm}\widehat{G}^{(0)}(\beta+2\pi i\sigma,m)
 \left[1+\e{o}(1)\right],
 \end{equation}
in which $\widehat{G}^{(0)}(\beta,m)$, given by \eq{7-Nonosc}, corresponds to the leading
non-oscillating term whereas $\widehat{G}^{(0)}(\beta\pm2\pi i,m)$ are
the leading oscillating ones.

To conclude this section, we would like to stress once more that the result \eqref{9-ebQ-GGGs} for
the leading asymptotics of the generating function $\moy{\ex{\beta
\mc{Q}_m}}$ is formulated in terms of the solution of \eq{0-DC}, which is a {\it linear} integral equation, whereas the integral
equations describing the partial sums $G_{1\dots n}(\gamma)$ were
{\em non-linear}. The linearization arises only after taking the last sum
over $n$, in other words, only if we take into account the
contributions of cycle integrals of all possible lengths. In the
framework of our approach, the necessity to consider  cycles of
arbitrary lengths is quite natural. Indeed, the asymptotics of all
cycle  integrals have a rather common form independently on the length
of the cycle. In particular it always contains a term which is linear over the
distance $m$, and one might therefore expect that all cycle
integrals eventually give a contribution of the same order to the
asymptotics of $\moy{\ex{\beta \mc{Q}_m}}$. It is worth mentioning
however that, for cycle integrals of length $\ell>1$, this
linear dependence on the distance shows itself only in the asymptotics,
while  it is explicit for the cycle integral of  length $\ell=1$:
 \begin{equation}\label{5-CI-l1}
 \oint \f{\dd z}{2\pi i}
 \Int{-q}{q}\f{\dd \la}{2\pi i} \;
 \mc{G}_{1}\!\left(\begin{array}{c}\la\\z
 \end{array}\right)
 \f{\ex{i m (p_0(z)-p_0(\la))  }}
 {\sinh^2(z-\la)}
 =\Int{-q}{q}\f{\dd \la}{2\pi i}
 \big[imp'_0(\la)+\partial_\eps\big]\,
 \mc{G}_{1}\!\left(\begin{array}{c}
 \la\\\la+\eps \end{array}\right)\Bigl.\Bigr|_{\eps=0}\, .
 \end{equation}
Therefore, dealing with the series for $\moy{\ex{\beta \mc{Q}_m}}$
in its initial form \eq{4-fin-answ}, we see only the explicit linear
dependence in $m$ generated by each double pole at $z_j=\lambda_j$
(i.e. by the cycle integral of length $\ell=1$), but we do not
see the hidden linear dependence produced by all other cycle
integrals. This may create the fallacious impression that the role of these double poles predominates.

We specify here all these details since they explain the
appearance of a non-linear integral dressing equation in the works
\cite{IzeK85,IzeK84}. Exactly the same non-linear equation can be
obtained from the series \eq{4-fin-answ} after summing up the
contributions of only the double poles. We would like to point out that
this way leads  to a misleading
representation for the generating function $\moy{\ex{\beta
\mc{Q}_m}}$. In particular, using such representation, the
authors of \cite{IzeK85} concluded that the oscillations in the
asymptotics of the two-point correlation function
$\langle\sigma^z_1\sigma^z_{m+1}\rangle$ are exponentially
suppressed at long distance, which is not true.


\subsection{Asymptotic expansion of $\langle\sigma_1^z\sigma_{m+1}^z\rangle$\label{FR}}

Let us now consider the asymptotic behavior of the two-point correlation
function $\langle\sigma_1^z\,\sigma_{m+1}^z\rangle$ which can
be obtained via \eq{0-corr-funct}:
 \begin{equation}\label{7-corr-funct}
 \langle\sigma_1^z\sigma_{m+1}^z\rangle=2D^2_m\left.\frac{\partial^2}{\partial\beta^2}
 \langle e^{\beta \mc{Q}_m}\rangle\right|_{\beta=0}-4D+1.
 \end{equation}
Similarly as for $\moy{e^{\beta \mc{Q}_m}}$, we will consider separately the non-oscillating and oscillating parts.

\subsubsection{Leading non-oscillating behavior of $\langle\sigma_1^z\sigma_{m+1}^z\rangle$}

To obtain the leading non-oscillating term of the asymptotics of
$\moy{\sg_1^z\sg_{m+1}^z}$, we should apply the differential operator
$2 D_m^2\partial_\beta^2$  to the term $\widehat{G}^{(0)}(\beta,m)$ in
\eq{9-ebQ-GGG}. Evidently, in order to obtain non-vanishing
contributions, one has to apply the derivatives over $\beta$  either
to $e^{\beta m D}$ or to the fractional power of $m$, setting simply $\beta=0$ in the
remaining part. This gives us, for
$m\to\infty$,
 \begin{equation}\label{7-non-osc-cf}
 2D_m^2\left.\frac{\partial^2}{\partial\beta^2}
\widehat{G}^{(0)}(\beta,m)\right|_{\beta=0}
 =\left(4D^2-\frac{2{\cal Z}^2}{\pi^2 m^2}\right)\cdot{\cal A}(\beta=0).
 \end{equation}

Consider now the behavior of the factor ${\cal A}$ at $\beta\to 0$.
Let us take, for example, the determinant of the operator
$U^{(\lambda)}$. Due to the factor $e^{\beta\tilde z(w)}$ the kernel
$U_{\theta_1}^{(\lambda)}(w,w')$ has a cut on the interval $[-q,q]$,
therefore, for $\beta$ small enough, the action of the integral
operator on the closed contour $\Gamma$ can be reduced to
 \begin{equation}\label{7-G-qq}
 \left.U^{(\lambda)}\right|_\Gamma \to
  \left.(U^{(\lambda)}_--U^{(\lambda)}_+)\right|_{[-q,q]},
  \end{equation}
where $U^{(\lambda)}_\pm$ are the limiting values of $U^{(\lambda)}$
from the upper (lower) half-planes. Using the equations
\eq{6-prop-z} we obtain
 \begin{equation}\label{7-UU}
 \left.\det\left[I+\frac1{2\pi
 i}U_{\theta_1}^{(\lambda)}(w,w')\right]\right|_\Gamma
 =
  \left.\det\left[I+\frac1{2\pi
 i}\tilde U_{\theta_1}^{(\lambda)}(w,w')\right]\right|_{[-q,q]},
 \end{equation}
where the operator in the r.h.s. acts on $[-q,q]$ and its kernel is
 \begin{equation}\label{7-tUl}
 \tilde U_{\theta_1}^{(\lambda)}(w,w')=-e^{\beta\tilde z_-(w)-\beta\tilde z(w+i\zeta)}
 [K_\kappa(w-w')-K_\kappa(\theta_1-w')]\,.
 \end{equation}
Setting now  $\beta=0$ we obtain
 \begin{equation}\label{7-b0}
 \left.\det\left[I+\frac1{2\pi
 i}U_{\theta_1}^{(\lambda)}(w,w')\right]\right|_{\beta=0}=
 \det\left[I+\frac1{2\pi }[K(w-w')-K(\theta_1-w')]\right].
 \end{equation}
On the other hand,
 \begin{multline}\label{7-neznaju}
 \det\left[I+\frac1{2\pi }[K(w-w')-K(\theta_1-w')]\right]
 =\det\left[I+\frac1{2\pi
 }K(w-w')\right]\cdot\det\left[I-R(\theta_1,w')\right]\\
 =\det\left[I+\frac1{2\pi }K\right]
 \cdot\left[1-\int\limits_{-q}^{q}R(\theta_1,w)\,\dd w\right]
 =Z(\theta_1)\cdot\det\left[I+\frac1{2\pi }K\right],
 \end{multline}
where $R(w,w')$ is the resolvent of the operator $I+\frac1{2\pi}K$.
Thus,
 \begin{equation}\label{7-limUl}
 \left.\det\left[I+\frac1{2\pi i}U_{\theta_1}^{(\lambda)}(w,w')\right]
 \right|_{\beta=0}
 =
Z(\theta_1)\cdot\det\left[I+\frac1{2\pi }K\right].
 \end{equation}

 One can prove similarly that
 \begin{equation}\label{7-limUz}
 \left.\det\left[I+\frac1{2\pi
 i}U_{\theta_2}^{(z)}(w,w')\right]\right|_{\beta=0
 }=
Z(\theta_2)\cdot\det\left[I+\frac1{2\pi
 }K\right].
 \end{equation}

Substituting these equations into \eq{7-ev-phiC} and using again
\eq{6-prop-z}, we find that
 \begin{equation}\label{7-lim-A}
 \lim_{\beta\to 0}{\cal A}(\beta)=1.
 \end{equation}

Thus, taking into account \eq{7-corr-funct} and \eq{7-non-osc-cf}, we
obtain the following leading non-oscillating asymptotic behavior of the correlation function:
 \begin{equation}\label{7-corr-funct-n-osc}
 \langle\sigma_1^z\sigma_{m+1}^z\rangle_{_{\e{non-osc}}}
 =(2D-1)^2-\frac{2{\cal Z}^2}{\pi^2 m^2}+\e{o}\left(\frac{1}{m^2}\right),
 \qquad m\to\infty.
 \end{equation}
%

\subsubsection{Leading oscillating behavior of $\langle\sigma_1^z\sigma_{m+1}^z\rangle$}

As we have seen, in order to obtain the leading oscillating term of the
asymptotics of the correlation function, it is enough to shift
$\beta$ by $\pm 2\pi i$ in \eq{7-Nonosc}. Observe that the constant
${\cal A}(\beta)$ is proportional to $(e^\beta-1)^2$, therefore ${\cal A}(\pm 2\pi i)={\cal A}'(\pm 2\pi i)=0$.
Thus, in order to obtain a non-zero contribution,  one should differentiate
only the factor $(e^\beta-1)^2$ when taking the second derivative with respect to $\beta$, setting $\beta=\pm 2\pi i$ in the
rest of \eq{7-Nonosc}.

Let us consider, for example, the case $\beta\to\beta+2\pi i$.
It gives
 \begin{equation}\label{7-Osc}
 \left.\frac{\partial^2}{\partial\beta^2}
 \widehat{G}^{(0)}(\beta+2\pi i,m)\right|_{\beta=0}
 ={\cal A}''(2\pi i)\cdot
 \frac{e^{2i mp_{{}_F}}\, G^2(1,{\cal Z})\,
 e^{C_1-C_0} }{[2\pi\,\sinh(2q)\,\rho(q)\,m]^{2{\cal Z}^2}}.
 \end{equation}
Here $p_{{}_F}$ is the Fermi momentum $p_{{}_F}=p(q)$ and
 \begin{equation}\label{7-A-dir-b}
 {\cal A}''(2\pi i)
 =\frac{2\left.
 \det\left[I+\frac1{2\pi i}U_{\theta_1}^{(\lambda)}(w,w')\right]
 \cdot
 \det\left[I+\frac1{2\pi i}U_{\theta_2}^{(z)}(w,w')\right]
 \right|_{\beta=2\pi i}}
 {\left[e^{2\pi i\tilde z(\theta_1+i\zeta)}-e^{2\pi i\tilde
 z(\theta_1-i\zeta)}\right]\!
 \left[
  e^{-2\pi i\tilde z(\theta_2-i\zeta)}
 -e^{-2\pi i\tilde z(\theta_2+i\zeta)}\right]
  \det^2[I+\frac1{2\pi}K]}.
 \end{equation}

Recall that $\theta_{1,2}$ in \eq{7-A-dir-b} are arbitrary complex
numbers. Let us set $\theta_1=-q$ and  $\theta_2=q$\footnote[1]{%
The $\theta$-independence of the expression given in \eq{7-A-dir-b} is proven in Appendix \ref{ap-theta}.  This specific choice of $\theta_j$ is the most convenient for the calculation
of the Fredholm determinants in \eq{7-A-dir-b} in the vicinity of
the free fermion point, see Appendix \ref{ap-FF}.}. Then it is easy to
see that, at $\beta=2\pi i$,
 \begin{equation}\label{7-conj-kern}
 \left(\frac1{2\pi i}U_{-q}^{(\lambda)}\right)^\dagger(w,w')=
 \frac1{2\pi  i}U_q^{(z)}(-\bar w,-\bar w'),
 \end{equation}
where $\dagger$ stays for Hermitian conjugation. Hence, the two
determinants in the numerator of \eq{7-A-dir-b} are complex
conjugated.
%
%
It is also easy to show using \eq{6-prop-z} that
 \begin{equation*}
\left[e^{2\pi i\tilde z(-q+i\zeta)}
 -e^{2\pi i\tilde z(-q-i\zeta)}\right]
 \left[e^{-2\pi i\tilde z(q-i\zeta)}-e^{-2\pi i\tilde
  z(q+i\zeta)}\right]
  =-4\sin^2\pi{\cal Z}\cdot e^{2\pi i [\tilde z(-q-i\zeta)
  -\tilde z(q-i\zeta)]}.
 \end{equation*}
Combining \eqref{7-A-dir-b} with the Barnes function in \eqref{7-Osc} and defining $\widetilde{\cal A}\equiv -4\,G^2(1,{\cal Z})\,{\cal A}''(2\pi i)$, we find
 \begin{equation}\label{7-const-A}
 \widetilde{\cal A}=
 \left|
 e^{\pi i[\tilde z(q-i\zeta)-\tilde z(-q-i\zeta)]}\,
 \frac{G(2,{\cal Z})\cdot
 \det\left[I+\frac1{2\pi i}U_{-q}^{(\lambda)}(w,w')\right]}
 {\pi{\cal Z}\cdot\det\left[I+\frac1{2\pi }K\right]}\right|^2,
 \end{equation}
where $G(2,{\cal Z})=G(2+{\cal Z})G(2-{\cal Z})$ and where we have used
\eq{5-Barnes}.

One can transform similarly the expression for ${\cal A}''(-2\pi
i)$.  The result in this case is still given by the same constant $\widetilde{\cal A}$ \eq{7-const-A}.
Then, taking the second lattice derivative, we eventually obtain in the limit
$m\to\infty$,
 \begin{equation}\label{7-corr-funct-osc}
 \langle\sigma_1^z\sigma_{m+1}^z\rangle_{_{\e{osc}}}
 = 8\,\widetilde{\cal A}\,\sin^2p_{{}_F}\cdot
 \frac{\cos(2mp_{{}_F})\, e^{C_1-C_0} }
      {[2\pi\,\sinh(2q)\,\rho(q)\,m]^{2{\cal Z}^2}}
 +\e{o}\big(m^{-2{\cal Z}^2}\big).
 \end{equation}

The obtained amplitude of this leading oscillating term appears to be closely related to a special form factor
of the operator $\sigma^z$. We will comment this relationship in the conclusion.

\subsubsection{Final result and comments}

Thus, at $m\to\infty$, the two-point correlation function of the
third components of spin in the external magnetic field behaves as
 \begin{equation}\label{7-corr-funct-1}
 \langle\sigma_1^z\sigma_{m+1}^z\rangle_{_{\e{leading}}}
=(2D-1)^2-
 \frac{2{\cal Z}^2}{\pi^2 m^2}
 +8\, \widetilde{\cal A}\, e^{C_1-C_0}\, \sin^2p_{{}_F}
 \cdot
 \frac{\cos(2mp_{{}_F}) }{[2\pi\,\sinh(2q)\,\rho(q)\,m]^{2{\cal Z}^2}}.
 \end{equation}
Note that, dependending on the value on $\Delta$ (and thus on $\mathcal{Z}$), the second term of \eqref{7-corr-funct-1} may be dominant compared to the third one or {\em vice-versa}.

Let us discuss this result from the viewpoint of its
comparison with known results and predictions.

As already  mentioned
in the introduction, the asymptotic equation \eq{7-corr-funct-1}
completely confirms the predictions from Luttinger liquid and
 conformal field theory approaches.
In the limit of free fermions $\zeta=\frac\pi2$ (see
Appendix \ref{ap-FF}), this result also agrees 
with the known answer
 \begin{equation}\label{7-corr-funct-ff}
 \langle\sigma_1^z\sigma_{m+1}^z\rangle_{_{\Delta=0}}
 =(2D-1)^2-
 \frac{2}{\pi^2 m^2}\bigl[1-\cos(2mp_{{}_F})\bigr].
 \end{equation}

In the general case, there exists no prediction concerning the value
of the amplitude of the oscillating term, except in the zero
magnetic field limit (see \cite{Luk98,Luk99,LukT03}). However, in
the framework of our approach, the non-zero magnetic field plays the
role of a certain {\em regularization} which ensures the finiteness
of all the constants ($C_0$, $C_1$, $\widetilde{\cal A}$,
$\rho(q)\sinh2q$) entering this amplitude: at $h=0$ all these
constants become divergent. It is nevertheless  possible to show
that the total combination of these constants remains finite. As we
have mentioned already the corresponding proof is highly
non-trivial, therefore we do not present it here. We would just like
to mention that the most complicated part of the calculations is
related to the extraction of the divergent part from the Fredholm
determinant $\det[I+\frac1{2\pi i}U_{-q}^{(\lambda)}(w,w')]$: one
can show that it coincides with the divergent part of
$\det[I-\frac1{2\pi}K]$.

Due to the difficulties that arise when taking this limit $h\to 0$ in our result, the complete comparison of  \eq{7-corr-funct-1} with the
results of \cite{Luk98,Luk99,LukT03} still remains an unsolved problem.
Note nevertheless that, in the limit of free fermions, the formula
\eq{7-corr-funct-1} holds for arbitrary magnetic field including the
case $h=0$. We succeeded moreover to compute explicitly the
limiting value at $h=0$ of the amplitude of the oscillating term in the vicinity of free fermions up to the
second order in $\epsilon=\frac\zeta\pi-\frac12$. Our
result,
 \begin{equation}\label{9-FR}
 \langle\sigma_1^z\sigma_{m+1}^z\rangle_{osc}
 = \frac{2(-1)^m}{\pi^2m^{2{\cal Z}^2}}\exp\left[-4\eps(\log2+\mathbf{C})+8\eps^2\bigl({\textstyle \frac{\pi^2}6}
 -1-\mathbf{C}-\log2\bigr)\right]+\e{O}(\eps^3),
 \end{equation}
$\mathbf{C}$ being the Euler constant, coincides at this order in $\eps$ with the one of
\cite{Luk98,Luk99,LukT03}.


\section{Quantum non-linear Schr\"odinger equation\label{sec-QNLS}}

In this section we briefly explain how to apply the method described above to another model, the quantum one-dimensional Bose gas
(or quantum nonlinear Schr\"odinger equation model).

Recall that the starting point of our derivation is the master equation \eq{1-GME}.
We have
stressed already that such integral representation exists not
only for the XXZ chain, but also for other models possessing the
six-vertex $R$-matrix \cite{KitKMST07} and,
in particular, for the system of one-dimensional
interacting bosons on some finite interval $[0,L]$.

The Hamiltonian of this model is given by
 \begin{equation}\label{8-Ham}
 H=\int_0^L\!\left(
 \partial_x\Psi^\dagger\partial_x\Psi
 +
 c\Psi^\dagger\Psi^\dagger\Psi\Psi-h\Psi^\dagger\Psi\right)\dd x\, .
 \end{equation}
Here $\Psi$ and $\Psi^\dagger$ are Bose-fields possessing canonical
equal-time commutation relations, $c$ is a coupling constant and $h$
a chemical potential. For $c>0$ and $h>0$, the ground state of the
model goes to a Dirac sea in the thermodynamic limit and
can be described by a set of integral equations similar to those
of the XXZ chain.

The analog of the $\mc{Q}_m$ operator is
 \begin{equation}\label{8-Q1}
 \mc{Q}_x=\int_0^x\Psi^\dagger(z)\,\Psi(z)\,\dd z\, ,
 \end{equation}
and the function $\langle e^{\beta \mc{Q}_x}\rangle$ is a generating function for the correlation function of the densities
$j(x)=\Psi^\dagger(x)\Psi(x)$ via
 \begin{equation}\label{8-cor-fun}
 \langle j(x)j(0)\rangle =\left.\frac12\frac{\partial^2}{\partial x^2}
 \frac{\partial^2}{\partial \beta^2}\langle e^{\beta \mc{Q}_x}\rangle \right|_{\beta=0}.
 \end{equation}

The method to compute the asymptotics of \eqref{8-cor-fun} in the case of the
one-dimensional Bose gas literally repeats the steps we have described in details in the case of the XXZ chain.
Let us just point a few peculiarities.

The master equation for $\langle e^{\beta \mc{Q}_x}\rangle$ has the form \eqref{1-GME}, but the
functions $a(\lambda)$, $d(\lambda)$, and $l(\lambda)$ (see
\eq{1-ad}, \eq{1-l}) are now
 \begin{equation}\label{8-adl}
 a(\lambda)=e^{-\frac{iL\lambda}2},\qquad
 d(\lambda)=e^{\frac{iL\lambda}2}, \qquad
 l(\lambda)=e^{-ix\lambda}.
 \end{equation}
In the remaining part of the master equation, all hyperbolic
functions should be replaced by  rational ones, i.e.
$\sinh(\lambda-z\pm i\zeta)\to\lambda-z\pm i\zeta$, {\it etc}. After this
operation one should set $\zeta=c$. The transformations of the
master equation described in Sections \ref{PME} and \ref{EME} are
the same for the one-dimensional bosons.

One of the difference between the quantum one-dimensional Bose gas and
the XXZ model is due to the choice of notations\footnote[1]{%
One could avoid this difference by replacing
$\zeta\to\pi-\zeta$ in the original formulas.}. In particular, the
integral equation for the dressed charge in the Bose gas model
coincides with its XXZ analog up to the sign of the integral
operator:
 \begin{equation}\label{8-DC}
 Z(\lambda)-\frac1{2\pi}
 \int\limits_{-q}^qK(\lambda-\mu)\,Z(\mu)\,\dd\mu=1\, ,
 \qquad K(\lambda)=\frac{2c}{\lambda^2+c^2}\, .
 \end{equation}
One should also pay attention to the fact that the function $l(\lambda)$
\eq{8-adl} can be analytically continued to the lower half-plane,
while its analog \eq{1-l} can be continued to the upper half plane
(in the case of $i\pi$-periodic hyperbolic functions, we can of
course always consider the domain $|\Im\lambda|\le\frac\pi2$).
This difference leads to some differences of signs in several
formulas, but the common strategy of the derivation
remains just the same as in the case of the XXZ chain.

The final
answer for the leading asymptotic behavior of \eqref{8-cor-fun} is
 \begin{equation}\label{8-asy-cor-funct}
 \langle j(x)j(0)\rangle_{_{\e{leading} } }
 =D^2-
 \frac{{\cal Z}^2}{2\pi^2 x^2}
 +2\widetilde{\cal A}\, p^2_{{}_F}\cdot
 \frac{\cos(2xp_{{}_F})\, e^{C_1-C_0} }
 {[4\pi\, q\, \rho(q)\, x]^{2{\cal Z}^2}}\, .
 \end{equation}
Here, just as in the previous case, ${\cal Z}=Z(\pm q)$, $Z(\lambda)$ satisfies
the integral equation \eq{8-DC} and $q$, the value of the spectral
parameter at the Fermi boundary, depends on $c$ and $h$ (see e.g.
\cite{BogIK93L}). In the model of the one-dimensional Bose gas the
spectral density $\rho(\lambda)$ is related to the dressed charge by
$\rho=\frac Z{2\pi}$. The Fermi momentum $p_{{}_F}$ and the average
density $D$ are given by \eq{0-Dens}.

The expressions for the constants $C_0$ and $C_1$ are quite similar
to \eq{7-C0}, \eq{7-C_1},
 \begin{equation}\label{8-C0}
 C_0={\dis
 \int\limits_{-q}^{q}}
 \frac{Z(\lambda)\,Z(\mu)}{(\lambda-\mu-ic)^2}\,\dd\lambda\,
 \dd\mu\, ,
 \end{equation}
 \begin{equation}\label{8-C_1}
 C_1=\frac12\int\limits_{-q}^{q}
 \frac{Z'(\lambda)\, Z(\mu)
 -Z(\lambda)\, Z'(\mu)}{\lambda-\mu}\,\dd\lambda\,
 \dd\mu
 +2{\cal Z}\int\limits_{-q}^{q}
  \frac{{\cal Z}-Z(\lambda)}{q-\lambda}\,\dd\lambda\, .
 \end{equation}
The formula for the constant  $\widetilde{\cal A}$ coincides also basically with \eq{7-const-A}:
 \begin{equation}\label{8-const-A}
 \widetilde{\cal A}=
 \left|e^{\pi i[\tilde z(q-ic)-\tilde z(-q-ic)]}\,
 \frac{G(2,{\cal Z})\cdot
 \det\left[I+\frac1{2\pi i}U_{-q}^{(\lambda)}(w,w')\right]}
 {\pi{\cal Z}\cdot\det\left[I-\frac1{2\pi }K\right]}\right|^2.
 \end{equation}
Here
 \begin{equation}\label{8-tz}
 \tilde z(w)=\frac1{2\pi
 i}\int\limits_{-q}^q\frac{Z(\lambda)}{\lambda-w}\,\dd\lambda\, ,
 \end{equation}
and
 \begin{equation}\label{8-Ul}
 \tilde U_{-q}^{(\lambda)}(w,w')
 =i\frac{e^{2\pi i\tilde z(w)}}
 {e^{2\pi i\tilde z(w+ic)}-e^{2\pi i\tilde
 z(w-ic)}}\left[ K(w-w')-K(q+w')\right].
 \end{equation}

It is not difficult to check that the result \eq{8-asy-cor-funct}
has the correct free fermion limit ($c=\infty$). One can also show that
it reproduces at first order in $c^{-1}$ the asymptotic behavior
obtained in \cite{IzeK84}.


\section*{Conclusion\label{C-n}}

We have described a method to compute the asymptotic
behavior of correlation functions of quantum integrable systems.
Our study is based on the multiple integral
representation \eq{1-GME}, called the master equation.
Such master equation can be obtained via the algebraic Bethe ansatz
for a rather wide class of integrable models and different
correlation functions. In the present paper, we have considered only
one specific correlation function for the XXZ chain and
one-dimensional bosons. To conclude, we would like to discuss
possible further developments.

First of all, it should be possible to apply this method to compute the
asymptotic behavior of other correlation functions,
such as
$\langle\sigma_1^+\sigma_{m+1}^-\rangle$ for the XXZ chain or $\langle\Psi^\dagger(x)\Psi(0)\rangle$ for
the Bose gas. It would also be interesting to apply this approach
to compute the long-distance correlations in the massive phase $\Delta>1$ of the
XXZ chain. In that case there is no power law corrections
in the asymptotic behavior of the cycles. This should lead to an
exponential decay of the correlations.

The case of the XXX model is also of special interest. For
non-zero magnetic field the result \eq{7-corr-funct-1} apparently
remains valid. However, at $h=0$, one can expect logarithmic
contributions to the asymptotic expression \cite{Aff98,BarA99}. It would be
interesting to see how such terms can appear in the framework of
our approach. However, the case of zero magnetic field appears to be
the most complicated one, even for the  XXZ chain. Therefore before
studying the XXX model it would be desirable to understand completely what happens for the XXZ chain at $h=0$, and in particular to compute explicitely the (finite) limiting value for the amplitude of the oscillating term.

We hope that our approach can be used for the asymptotic analysis of
the temperature dependent correlation functions. At least in the case of the one-dimensional Bose gas, for which there is no bound states in
the spectrum of the Hamiltonian for $c>0$, the
generalization of our method looks quite straightforward. For $T>0$,
the integrals over the spectral parameters should be taken over
the whole real axis. This leads to the absence of power law corrections
in the asymptotic behavior of the cycle integrals, just as in the
massive regime of the XXZ chain. As a result, the asymptotic
behavior of the correlation functions should decrease exponentially
with the distance.

We would like finally to discuss the nice relation that exists between
the amplitude of the leading oscillating term of the asymptotics
\eq{7-corr-funct-1} and a special form factor of the operator
$\sigma^z$, namely the matrix element
 \begin{equation}\label{9-mat-el}
 F_\sigma=\left(\frac M{2\pi}\right)^{{\cal Z}^2}
 \cdot\frac{\langle\psi(\{\mu\})|\sigma_k^z|\psi(\{\lambda\})\rangle}
 {\|\psi(\{\mu\})\| \cdot \| \psi(\{\lambda\})\|},
 \end{equation}
where $|\psi(\{\lambda\})\rangle$ is the $N$-particle ground state
and $\langle\psi(\{\mu\})|$ is an excited state containing one
particle and one hole at the different boundaries of the Fermi
sphere, for instance, $\lambda_p=q$, $\lambda_h=-q$. We recall that
in \eq{9-mat-el} $M$ denotes the length of the chain. Using the
determinant representations of form factors obtained in
\cite{KitMT99}, one can show \cite{KitKMST08b} that, in the
thermodynamic limit $|F_\sigma|^2$ is equal to,
 \begin{equation}\label{9-F-sigma}
 \lim_{M\to\infty} |F_\sigma|^2
 = 4\widetilde{\cal A}\,\sin^2p_{{}_F}\,
 e^{C_1-C_0}\, [2\pi\, \sinh(2q)\, \rho(q)]^{-2{\cal Z}^2} .
 \end{equation}
Thus, the amplitude of the leading oscillating term is equal to
the square of the norm of the form factor \eq{9-mat-el}:
 \begin{equation}\label{10-corr-funct}
 \langle\sigma_1^z\sigma_{m+1}^z\rangle_{_{\e{leading} }}
 =(2D-1)^2-
 \frac{2{\cal Z}^2}{\pi^2 m^2}+2|F_\sigma|^2\cdot
 \frac{\cos(2mp_{{}_F}) }{m^{2{\cal Z}^2}}\, .
 \end{equation}
The coefficient $2$ is due to the fact that there are two such form factors. This observation perfectly agrees with the conformal
field theory approach. Namely, if we consider the two-point
correlation function as a sum of form factors, then the main
contribution to the asymptotics comes from the terms corresponding to the
excitations in a vicinity of the Fermi boundaries. Hereby, the leading
oscillating term in the asymptotics is produced by the excited
states having particles and holes on the different sides of the
Fermi zone. Thus there is an opportunity to  improve drastically  the
form factor approach. This, in turn, would  open a path towards the
computation of dynamical correlation functions.


\section*{Acknowledgements}
J. M. M., N. S. and V. T. are supported by CNRS. N. K., K. K. K.,
J. M. M. and V.~T. are supported by the ANR program GIMP
ANR-05-BLAN-0029-01. N. K. and V. T. are supported by the ANR program
MIB-05 JC05-52749. N. S. is supported by the French-Russian Exchange
Program in Theoretical and Mathematical Physics, the Program of RAS Mathematical Methods of the Nonlinear
Dynamics, RFBR-08-01-00501a, Scientific Schools 795.2008.1. N. K. and
N. S. would like to thank the Theoretical Physics group of the
Laboratory of Physics at ENS Lyon for hospitality, which makes this
collaboration possible. N. K. would like also to thank LPTHE for hospitality.


\appendix

\section{Determinant representations \label{ap-DT}}

\subsection{Proof of Proposition \ref{Extract-Cauchy}}

If the
parameters $\lambda_1,\dots,\lambda_N$ satisfy the system of Bethe
equations \eq{1-BE_Y},  then  $d(\lambda_j)$ can be expressed in terms of
$a(\lambda_j)$, and the determinant  of $\Omega_\kappa$ can be written as
 \begin{equation}\label{1-Om-tM}
 \det_N\Omega_\kappa(\{z\},\{\lambda\}|\{z\})=
 \prod_{j=1}^N a(\lambda_j)\cdot\prod_{a,b=1}^N\sinh(z_a-\lambda_b-i\zeta)\cdot
 \det_N\tilde M_\kappa(\{z\}|\{\lambda\}),
 \end{equation}
where
 \begin{equation}\label{1-tM-la}
 \bigl(\tilde M_\kappa\bigr)_{jk}= t(z_k,\lambda_j)+\kappa\,t(\lambda_j,z_k)
 \cdot V_+(\lambda_j)\, V_-^{-1}(\lambda_j).
 \end{equation}
In order to prove \eqref{2-l-rep}, we shall prove that
 \begin{multline}\label{A1-l-rep}
 \det_N\tilde M_\kappa
 =\det_N\left[\frac1{\sinh(\lambda_j-z_k)}\right]\cdot
 \prod_{j=1}^N\left[\kappa \frac{V_+(\lambda_j)}{V_-(\lambda_j)}-1\right]
  \num
 \times
 \frac{1-\kappa}{V_+^{-1}(\theta)-\kappa
 V_-^{-1}(\theta)}\cdot \det_N\left[\delta_{jk}+U_{jk}^{(\lambda)}(\theta)\right] .
 \end{multline}

Consider the following transformation of the l.h.s. of \eq{A1-l-rep}
 \begin{equation}\label{A1-transf} \det_N\tilde M_\kappa = \frac{\det_N(\tilde M_\kappa A)}{\det_NA},
 \end{equation}
with
\begin{equation}\label{A1-Ajk}
 A_{jk}=\frac{\prod\limits_{a=1}^{N}\sinh(z_j-\lambda_a)}
 {\prod\limits_{\substack{a=1\\ a\ne j}}^{N}\sinh(z_j-z_a)}\times\left\{
 \begin{array}{ll}
 \coth(z_j-\lambda_k)\quad &\text{for}\quad k\ne N,\num
 1\qquad &\text{for}\quad k=N.
 \end{array}\right.
 \end{equation}

The determinant of $A$ can be easily computed. Indeed,
 \begin{equation}\label{A1-detA1}
 \det_NA=\frac{\prod\limits_{a,b=1}^{N}\sinh(z_a-\lambda_b)}
 {\prod\limits_{\substack{a,b=1\\ a\ne b}}^{N}\sinh(z_a-z_b)}\cdot
 \det_N\left(\begin{array}{c}\coth(z_1-\lambda_k)\\ \vdots\\ \coth(z_N-\lambda_k)\end{array}
 \right| \left.\begin{array}{c}1\\ \vdots\\ 1\end{array}   \right).
 \end{equation}
Subtracting the last line  from all others, we reduce the remaining
determinant to a Cauchy determinant, which gives
 \begin{equation}\label{A1-detA2}
 \det_NA=\frac{\prod\limits_{a=1}^{N}\sinh(z_a-\lambda_N)}
 {\prod\limits_{a=1}^{N-1}\sinh(\lambda_a-\lambda_N)}\cdot
 \prod_{a>b}^{N}\frac{\sinh(\lambda_a-\lambda_b)} {\sinh(z_a-z_b)}.
 \end{equation}

The product of matrices $\tilde M_\kappa A$ can also be explicitly
calculated
 \begin{equation}\label{A1-Bjk}
 (\tilde M_\kappa A)_{jk}=\delta_{jk}\;
 \frac{\prod\limits_{\substack{a=1\\ a\ne j}}^{N}
 \sinh(\lambda_j-\lambda_a)}{\prod\limits_{a=1}^{N}\sinh(\lambda_j-z_a)}
 \left[\kappa\frac{V_{+}(\lambda_j)}{V_{-}(\lambda_j)}-1
 \right]
 +K_\kappa(\lambda_j-\lambda_k)\, V_{+}(\lambda_j),\qquad
 k<N,
 \end{equation}
and
 \begin{equation}\label{A1-BjN}
 (\tilde M_\kappa A)_{jN}=(1-\kappa)\, V_{+}(\lambda_j),\qquad
 k=N.
 \end{equation}
Here the function $K_\kappa(\lambda)$ is given by \eq{2-Kk}. Let us
explain how the formulas \eq{A1-Bjk}, \eq{A1-BjN} were obtained. For
instance, in order to obtain \eq{A1-Bjk}, one has to
calculate the sums
 \begin{equation}\label{A1-Gjk}
 G^\pm_{jk}=\sum_{\ell=1}^N
 \frac{-i\, \sin\zeta\, \coth(z_\ell-\lambda_k)}
 {\sinh(z_\ell-\lambda_j)\, \sinh(z_\ell-\lambda_j\pm i\zeta)}
 \cdot
 \frac{\prod\limits_{a=1}^N\sinh(z_\ell-\lambda_a)}
 {\prod\limits_{\substack{a=1\\ a\ne\ell}}^N\sinh(z_\ell-z_a)}.
 \end{equation}
Consider an auxiliary contour integral
 \begin{equation}\label{A1-Cont-Int}
 I_\pm=\frac1{2\pi i}
 \oint\frac{-i\, \sin\zeta\, \coth(w-\lambda_k)\;\dd w}
 {\sinh(w-\lambda_j)\, \sinh(w-\lambda_j\pm i\zeta)}
 \cdot \prod\limits_{a=1}^N
 \frac{\sinh(w-\lambda_a)}{\sinh(w-z_a)},
 \end{equation}
where the integration is taken over the boundary of a horizontal
strip of width $i\pi$. On the one hand, the integral vanishes
$I_\pm=0$ as the integrand is an $i\pi$-periodic function that is
exponentially decreasing for $\Re(w)\to\pm\infty$. On the other
hand, it is equal to the sum of the residues inside of the
integration contour. The sum of the residues at the poles $w=z_a$
gives exactly $G^\pm_{jk}$. Taking into account the contribution of
the additional poles at $w=\lambda_j\mp i\zeta$ and $w=\lambda_j$
(the last one only exists for $j=k$), we arrive at the following
identity
 \begin{equation}\label{A1-identity}
 G^\pm_{jk}\pm\delta_{jk}\;
 \frac{\prod\limits_{\substack{a=1\\ a\ne j}}^N
 \sinh(\lambda_a-\lambda_j)}{\prod\limits_{a=1}^N\sinh(z_a-\lambda_j)}
 \pm\coth(\lambda_j-\lambda_k\mp i\zeta)\, V_\mp(\lambda_j)=0.
 \end{equation}
We have thus computed $G^\pm_{jk}$ and in this way proved the
formula \eq{A1-Bjk}. The result \eq{A1-BjN} can be obtained by a
similar method.

The following transformations are trivial. We can extract the factor
$V_{+}(\lambda_j)$ from each line of $(\tilde M_\kappa A)_{jk}$,
making the elements of the last column equal to $1-\kappa$. After
this, subtracting the last line from all the others and extracting
the coefficients in front of  $\delta_{jk}$ we obtain a new
representation for $\det\tilde M_\kappa$
 \begin{multline}\label{A1-newMt-1}
 \det_N\tilde M_\kappa=\frac{1-\kappa}{V_+^{-1}(\lambda_N)-\kappa
 V_-^{-1}(\lambda_N)}
 \prod\limits_{a=1}^{N}\left[\kappa\frac{V_{+}(\lambda_a)}{V_{-}(\lambda_a)}-1\right]\\
 \times\det_N\left[\frac{1}{\sinh(\lambda_j-z_k)}\right]\cdot
 \det_N\left[\delta_{jk}+U_{jk}^{(\lambda)}(\lambda_N)\right].
 \end{multline}
Thus, we have reproduced the r.h.s. of \eq{A1-l-rep} up to
replacement $\theta\to \lambda_N$.

Now consider
$\det\left[\delta_{jk}+U_{jk}^{(\lambda)}(\theta)\right]$ appearing
in the r.h.s. of equation \eq{A1-l-rep} (see \eq{2-Tl} for its
explicit form). After a simple similarity transformation we obtain
 \begin{equation}\label{A1-Tl}
 U^{(\lambda)}_{jk}(\theta)\to\tilde U^{(\lambda)}_{jk}(\theta)=
 \frac{\prod\limits_{a=1}^N\sinh(z_a-\lambda_k)}
 {\prod\limits_{\substack{a=1\\ a\ne k}}^N\sinh(\lambda_a-\lambda_k)}\cdot
 \frac{K_\kappa(\lambda_{j}-\lambda_{k})-K_\kappa(\theta-\lambda_{k})}{V_+^{-1}(\lambda_k)-\kappa
 V_-^{-1}(\lambda_k)},
 \end{equation}
Let us multiply the first $N-1$ columns by the coefficients $s_k$
 \begin{equation}\label{A1-sj}
 s_k=\frac{V_+^{-1}(\lambda_k)-\kappa V_-^{-1}(\lambda_k)}{V_+^{-1}(\lambda_N)-\kappa V_-^{-1}(\lambda_N)}
 \end{equation}
and add them to the $N$-th column. Then the last column of the
determinant becomes
 \begin{equation}\label{A1-LastL}
 \delta_{jN}+\tilde U^{(\lambda)}_{jN}(\theta)+\sum_{k=1}^{N-1}s_k\left(\delta_{jk}+
 \tilde U^{(\lambda)}_{jk}(\theta)\right)=
 \frac{V_+^{-1}(\theta)-\kappa V_-^{-1}(\theta)}{V_+^{-1}(\lambda_N)-\kappa
 V_-^{-1}(\lambda_N)},
 \end{equation}
(the method of calculation is quite similar to one described in the
formulas \eq{A1-Gjk}--\eq{A1-identity}). Now it is enough to
subtract the last line of the  obtained matrix from all others, and
we arrive at the following identity
 \begin{equation}\label{A1-Ident-det}
 \det_N\left[\delta_{jk}+ \tilde U^{(\lambda)}_{jk}(\theta)\right]
  =
 \frac{V_+^{-1}(\theta)-\kappa V_-^{-1}(\theta)}
 {V_+^{-1}(\lambda_N)-\kappa V_-^{-1}(\lambda_N)}
 \cdot
 \det_N\left[\delta_{jk}+
 \tilde U^{(\lambda)}_{jk}(\lambda_N)\right].
 \end{equation}
In other words, it means that the combination
 \begin{equation}\label{A1-comb}
 \frac{\det_N\left[\delta_{jk}+ U^{(\lambda)}_{jk}(\theta)\right]}
 {V_+^{-1}(\theta)-\kappa V_-^{-1}(\theta)}
  \end{equation}
does not depend on $\theta$. Thus, substituting \eq{A1-Ident-det}
into \eq{A1-newMt-1}, we obtain the r.h.s. of \eq{A1-l-rep}.


The representation \eq{2-z-rep} for $\Omega_{\kappa}$ can be proved
by using the identity
 \begin{equation}\label{A1-hatM}
 \det_N\tilde M_\kappa=\det_N\hat M_\kappa,
 \quad \mbox{where}\quad
 \bigl(\hat M_\kappa\bigr)_{jk}= t(z_k,\lambda_j)+\kappa\,t(\lambda_j,z_k)
 \cdot V_+(z_k)\, V_-^{-1}(z_k),
 \end{equation}
established in \cite{KitKMST07} (see Appendix B of this paper). It
follows from \eq{A1-hatM} that
 \begin{equation}\label{A1-Ident-O}
 \det_N\tilde M_\kappa(\{z\},\{\lambda\})=\kappa^N
  \prod_{a,b=1}^N\frac{\sinh(\lambda_a-z_b-i\zeta)}{\sinh(z_b-\lambda_a-i\zeta)}
  \cdot\det_N\tilde M_{\kappa^{-1}}(\{\lambda\},\{z\}).
  \end{equation}
Then the representation \eq{2-z-rep} follows from \eq{2-l-rep} after
the replacements $z\leftrightarrow \lambda$ and
$\kappa\to\kappa^{-1}$.

\subsection{Fredholm determinant representation}
\label{ap-FDR}

Consider the Fredholm determinants $\det\left[I+\frac1{2\pi i}\hat
U_\theta^{(\lambda,z)}(w,w')\right]$ (see \eq{2-det-lz}). We have
 \begin{equation}\label{A1-log-exp}
 \log\det\left[I+\frac1{2\pi i}\hat
 U_\theta^{(\lambda,z)}\right]=
 \sum_{k=1}^\infty\frac{(-1)^{k+1}}k
 \oint\limits_\Gamma \frac{\dd^k w}{(2\pi i)^k}\,
 \hat U_\theta^{(\lambda,z)}(w_1,w_2)\cdots
 \hat U_\theta^{(\lambda,z)}(w_k,w_1).
 \end{equation}
Computing the multiple integrals by the residues at $w_j=\lambda_\ell$
(respectively at $w_k=z_\ell$), where $\ell=1,\dots,N$, we obtain
 \begin{equation}\label{A1-log-exp1}
 \log\det\left[I+\frac1{2\pi i}\hat
 U_\theta^{(\lambda,z)}\right]=
 \sum_{k=1}^\infty\frac{(-1)^{k+1}}k\sum_{\ell_1,\dots,\ell_k=1}^N
  U^{(\lambda,z)}_{\ell_1\ell_2}(\theta)\cdots
  U^{(\lambda,z)}_{\ell_k\ell_1}(\theta),
 \end{equation}
what is exactly the expansion for
$\log\det_N\left[\delta_{jk}+U^{(\lambda,z)}_{jk}(\theta) \right]$.

\subsection{Some identities for Fredholm determinants}
\label{ap-theta}
Here we prove $\theta$-independence of Fredholm determinants
considered in Sections~\ref{sec-master} and \ref{sec-results}.

Consider an integral operator $I+V$ acting on some contour
${\Gamma}$.
\begin{prop} Let
 \begin{equation}\label{A1-g}
 g(w)=h(w)+\int_{\Gamma} V(w,w')\, h(w')\, \dd w'.
 \end{equation}
Then, for arbitrary $w_0\in{\cal C}$,  the following identity holds
\begin{equation}\label{A1-ident-FD}
 \det\left[I+V(w,w')\right]
 =\frac{g(w_0)}{h(w_0)}\cdot
 \det\left[I+V(w,w')-\frac{g(w)}{g(w_0)}
 V(w_0,w')\right].
 \end{equation}
\end{prop}

\Proof
Suppose that  the resolvent $R(w,w')$ of the
operator $I+V$ exists (if not, then we can consider some regularization
$V_\epsilon(w,w')$ of the original kernel, such that the
corresponding resolvent $R_\epsilon$ exists). Then
\begin{multline}\label{A1-proof}
 \frac{g(w_0)}{h(w_0)}\cdot
 \det\left[I+V(w,w')-\frac{g(w)}{g(w_0)}
 V(w_0,w')\right]\num
 =\frac{g(w_0)}{h(w_0)}
 \cdot \det\left[I+V(w,w')\right]\cdot
 \det\left[I-\frac{g(w)}{g(w_0)}
 R(w_0,w')\right]\num
 =\frac{1}{h(w_0)}\cdot \det\left[I+V(w,w')\right]
 \left(g(w_0)- \int_\Gamma
 R(w_0,w')g(w')\,dw'\right)\num
 =\det\left[I+V(w,w')\right],
 \end{multline}
which ends the proof.
\qed

If the kernel $V(w,w')$ is analytic within some domain ${\cal D}$
containing $\Gamma$, then
the identity \eq{A1-ident-FD} holds for $w_0\in{\cal
D}$.

Let now $\Gamma$ be the contour shown on Fig.~1 and $h(w)$ an
$i\pi$-periodic function that is  holomorphic outside of $\Gamma$
and bounded at $w\to\pm\infty$. Then
 \begin{equation}\label{A1-intK}
 \oint\limits_\Gamma
 \frac{\dd w'}{2\pi i}\big[K_\kappa(w-w')-K_\kappa(\theta-w')\big]
 h(w')
 =
 \kappa h(\theta-i\zeta)-h(\theta+i\zeta)- \kappa h(w-i\zeta)+h(w+i\zeta),
 \end{equation}
where the integral has been computed by the residues lying outside of
the contour $\Gamma$. Compose a kernel $ U^{(h)}_{\theta}(w,w')$  as
 \begin{equation}\label{A1-new-kern}
 U^{(h)}_{\theta}(w,w')
 =\frac{h(w)}{2\pi i\, [\kappa h(w-i\zeta)- h(w+i\zeta)]}
 \cdot\bigl[K_\kappa(w-w')-K_\kappa(\theta-w')\bigr].
 \end{equation}
Let us apply \eq{A1-ident-FD} to the operator $I+
U^{(h)}_{\theta}(w,w')$ with $\theta\in \Gamma$. We have
 \begin{equation}\label{A1-g1}
 g(w)=h(w)+
 \oint\limits_{\Gamma} U^{(h)}_\theta(w,w')\, h(w')\,\dd w'=
 h(\omega)\,\frac{\kappa h(\theta-i\zeta)- h(\theta+i\zeta)}{\kappa h(w-i\zeta)- h(w+i\zeta)}.
 \end{equation}
Substituting this into \eq{A1-ident-FD} we immediately arrive at
 \begin{equation}\label{A1-not-dep}
 \frac{\det\left[I+\frac1{2\pi i}U^{(\lambda)}_{\theta}(w,w')\right]}
 {\kappa h(\theta-i\zeta)- h(\theta+i\zeta)}
 =
  \frac{\det\left[I+\frac1{2\pi i}U^{(\lambda)}_{w_0}(w,w')\right]}
 {\kappa h(w_0-i\zeta)- h(w_0+i\zeta)},
 \end{equation}
for arbitrary $w_0$ such that $|\Im(w_0-w)|<\zeta$, $w\in\Gamma$.

In particular one can take
 \begin{equation}\label{A1-h-part}
 h(w)=\prod_{a=1}^N\frac{\sinh(w-z_a)}{\sinh(w-\lambda_a)}, \qquad
 \mbox{or}\qquad h(w)=e^{\beta\tilde z(w)}.
 \end{equation}
In the first case we obtain the identity for the operator $\hat
U^{(\lambda)}$, in the second for $U^{(\lambda)}$. Similarly one can
prove $\theta$-independence of the kernels $\hat U^{(z)}$ and
$U^{(z)}$.

\section{Free fermions\label{ap-FF}}

Consider the determinant of the matrix $\delta_{jk}+
U^{(\lambda)}_{jk}(\theta)$ in the limit of free fermions
$\zeta=\frac\pi2$. Then $V_+(\mu)=V_-(\mu)$ and the entries
$U^{(\lambda)}_{jk}(\theta)$ are equal to
 \begin{equation}\label{A2-Ujk}
 U^{(\lambda)}_{jk}(\theta)=
 \frac{\prod\limits_{a=1}^N\tanh(z_a-\lambda_j)}
 {\prod\limits_{\substack{a=1\\ a\ne j}}^N\tanh(\lambda_a-\lambda_j)}\cdot
 \bigl[\tanh(\lambda_{j}-\lambda_{k})-\tanh(\theta-\lambda_{k})\bigr],
 \qquad
 \zeta=\frac\pi2.
 \end{equation}
The straightforward calculation of the corresponding determinant
causes serious difficulties. However, there exists a way to avoid
these problems. Due to \eq{A1-l-rep} the determinant of
$\delta_{jk}+ U^{(\lambda)}_{jk}(\theta)$ can be expressed in terms
of $\det\tilde M_\kappa$. One has, for the limit of free fermions,
 \begin{equation}\label{A2-l-rep}
 \det_N\left[\delta_{jk}+U_{jk}^{(\lambda)}(\theta)\right]
 =V_+^{-1}(\theta)\cdot\frac{\det_N\tilde M_\kappa}{ \det_N\left[\frac{1-\kappa}{\sinh(z_k-\lambda_j)}\right]}.
 \end{equation}
On the other hand, the matrix $\tilde M_\kappa$ \eq{1-tM-la} becomes
the Cauchy matrix at $\zeta=\frac\pi2$,
 \begin{equation}\label{A2-tM}
 \bigl(\tilde M_\kappa\bigr)_{jk}=\frac{2(1-\kappa)}{\sinh2(z_k-\lambda_j)},\qquad
 \text{for}\quad
 \zeta=\frac\pi2.
 \end{equation}
Thus, both determinants in the r.h.s. of \eq{A2-l-rep} are
explicitly computable and we arrive at
 \begin{equation}\label{A2-res1}
 \det_N\left[\delta_{jk}+U_{jk}^{(\lambda)}(\theta)\right]
 =\prod_{a=1}^N\frac{\cosh(\theta-z_a)}{\cosh(\theta-\lambda_a)}\cdot
 \frac{\prod\limits_{a>b}^N\cosh(z_a-z_b)\cosh(\lambda_a-\lambda_b)}{\prod\limits_{a,b=1}^N\cosh(z_a-\lambda_b)},
 \quad \zeta=\frac\pi2.
 \end{equation}
Similarly one has
 \begin{equation}\label{A2-res2}
 \det_N\left[\delta_{jk}+U_{jk}^{(z)}(\theta)\right]
  =\prod_{a=1}^N\frac{\cosh(\theta-\lambda_a)}{\cosh(\theta-z_a)}\cdot
 \frac{\prod\limits_{a>b}^N\cosh(z_a-z_b)\cosh(\lambda_a-\lambda_b)}{\prod\limits_{a,b=1}^N\cosh(z_a-\lambda_b)},
 \quad \zeta=\frac\pi2.
 \end{equation}
It is easy to see from the obtained result and the definition \eq{2-W}
of $\widetilde W_N$   that, in the free fermion limit,
 \begin{equation}\label{A2-W}
 \widetilde W_N\left(\begin{array}{c}\{\la\}\\
 \{ z\}\end{array}\right)=1.
 \end{equation}
%

Let us now compute the constant $\widetilde{\cal A}$ in the limit of
free fermions. For $\zeta=\frac\pi2$ we have $Z(\lambda)\equiv1$,
hence,
 \begin{equation}\label{A2-U1}
 U_{-q}^{(\lambda)}(w,w')=-\tanh(w-q)\,\coth(w+q)\,
 \bigl[\tanh(w-w')+\tanh(q+w')\bigr].
 \end{equation}
The kernel \eq{A2-U1} is holomorphic in a vicinity of the interval
$[-q,q]$. We draw the reader's attention on the fact that this property only
holds due to the special choice of $\theta_1=-q$. Otherwise the
kernel $U_{\theta_1}^{(\lambda)}(w,w')$ would have a simple pole at
$w=-q$.

Since the integral operator $U_{-q}^{(\lambda)}(w,w')$ acts on the
closed contour surrounding the interval $[-q,q]$ we obtain that
 \begin{equation}\label{A2-detU}
 \det\left[I+\frac1{2\pi i}U_{-q}^{(\lambda)}(w,w')\right]=1,
 \end{equation}
hence,
 \begin{equation}\label{A2-ev-phiC}
 \widetilde{\cal A} =\frac1{\pi^2}e^{2\pi i\left[\tilde z(q-i\frac\pi2)-\tilde
 z(-q-i\frac\pi2)\right]}.
 \end{equation}

It is also easy to see that for free fermions $C_1=0$ (see
\eq{7-C_1}), and
 \begin{equation}\label{A2-C0}
 2\pi i\left[\tilde z(q-{\textstyle\frac{i\pi}2})-\tilde
 z(-q-{\textstyle\frac{i\pi}2})\right]-C_0=0.
 \end{equation}
Thus, taking into account that
$\rho(\lambda)=\frac1{\pi\cosh(2\lambda)}$ for $\zeta=\frac\pi2$, we
obtain
 \begin{equation}\label{A2-corr-funct-osc}
 \langle\sigma_1^z\sigma_{m+1}^z\rangle_{osc}
 = \left(\frac{\sin p_{{}_F}}{\tanh(2q)}\right)^2\cdot
 \frac{2\cos(2mp_{{}_F})}{\pi^2m^2}.
 \end{equation}
It remains to observe that
 \begin{equation}\label{A2-k_F}
 p_{{}_F}=\pi D=\int\limits_{-q}^q\frac{\dd\lambda}{\cosh2\lambda}=
 \arctan\left(e^{2q}\right)- \arctan\left(e^{-2q}\right).
 \end{equation}
From this we find that $\sin p_{{}_F}=\tanh(2q)$, and  hence
 \begin{equation}\label{A2-corr-funct-osc1}
 \langle\sigma_1^z\sigma_{m+1}^z\rangle_{osc}
 =\frac{2\cos(2mp_{{}_F})}{\pi^2m^2}.
 \end{equation}
%




\section{The Lagrange series and its generalizations\label{ap-LSD}}

In this appendix we consider several generalizations of the Lagrange series (see
e.g. \cite{WhiW}) used in Section~\ref{AD-CLS}.
We give the detailed proof in the standard (scalar)
case. The generalizations then become quite evident.

\subsection{Scalar case}

Let us consider the series
 \begin{equation}\label{A4-SCser}
 G_0=\sum_{n=0}^\infty\frac{1}{n!}\left.\frac{\dd^n}{\dd\epsilon^n}
 \bigl( \phi^n(\epsilon)F(\epsilon)\bigr)\right|_{\epsilon=0},
 \end{equation}
where $F(\eps)$ and $\phi(\eps)$ are holomorphic for  $|\eps|<r_0$.

\begin{prop}\label{PA4-1}
If there exists $r<r_0$ such that $|\phi(\eps)|<r$ for $|\eps|=r$,
then the series \eq{A4-SCser} is absolutely convergent and its sum
is given by
 \begin{equation}\label{A4-SC-res}
 G_0=\frac{ F(z)}{1-\phi'(z)},
 \end{equation}
where $z$ is the root of the equation
 \begin{equation}\label{A4-eq}
 z-\phi(z)=0
 \end{equation}
such that $|z|<r$.
\end{prop}

\Proof
Replacing the $n^{\e{th}}$ derivative by a Cauchy
integral, we obtain
 \begin{equation}\label{A4-SC-Cauchy}
 G_0=\sum_{n=0}^\infty\frac{1}{2\pi i}
 \oint\limits_{|\omega|=r}\frac{
 F(\omega)\,\phi^n(\omega)}{\omega^{n+1}}\,\dd\omega.
 \end{equation}
Since $|\phi(\omega)|<r=|\omega|$ the obtained series is absolutely
convergent, and we arrive at
 \begin{equation}\label{A4-SC-summed}
 G_0=\frac{1}{2\pi i}\oint\limits_{|\omega|=r}\frac{
 F(\omega)}{\omega-\phi(\omega)}\,dz.
 \end{equation}
Due to Rouch\'e's Theorem, the equation $\omega-\phi(\omega)=0$ has
exactly one simple zero $\omega=z$ within the circle $|\omega|<r$.
Taking the residue in this point we obtain the statement of the
proposition.~\qed

\subsection{Matrix case}

The result obtained for the series \eq{A4-SCser} can easily be
generalized to the case of several variables. Consider a multiple
series of the form
 \begin{equation}\label{A4-MCser}
 G_N=\sum_{s_1,\dots,s_N=0}^\infty\frac{1}{s_1!\cdots s_N!}
 \prod_{j=1}^N\frac{\partial^{s_j}}{\partial \epsilon_j^{s_j}}
 \prod_{j=1}^N\phi^{s_j}_j\left(\{\epsilon\}\right)\cdot F(\{\epsilon\})
 \left.\vphantom{\prod}\right|_{\epsilon_j=0},
 \end{equation}
where $F(\{\eps\})$ and $\phi_j(\{\eps\})$ are holomorphic for
$|\eps_k|<r_k^{(0)}$, $k=1,\ldots,N$.
\begin{prop}\label{PA4-2}
If there exist $r_j<r_j^{(0)}$ such that $|\phi_j(\{z\})|<r_j$ for
$|z_j|=r_j$, then the series \eq{A4-MCser} is absolutely convergent
and its sum is given by
 \begin{equation}\label{A4-MC-res}
 G_N=\frac{F(\{z_j\})}{\det_NS_{jk}},
 \end{equation}
where $z_j$ are the roots of the system
 \begin{equation}\label{A4-sys}
 z_j-\phi_j\left(\{z\}\right)=0,
 \end{equation}
and $\det_NS_{jk}$ is the Jacobian of the system \eq{A4-sys}:
 \begin{equation}\label{A4-Jac}
 S_{jk}=\delta_{jk}-\textstyle{\frac\partial{\partial
 z_k}}\phi_j\left(\{z\}\right).
 \end{equation}
\end{prop}

The proof is completely analogous to the one of Proposition~\ref{PA4-1}. It uses the analog of Rouch\'e's Theorem for several
complex variables (see e.g. \cite{AizY79}).

We consider now special cases of the series \eq{A4-MCser}.
Namely, let $\mu_1,\dots,\mu_N$ be a set of complex parameters. Let
 \begin{equation}\label{A4-phij}
 \phi_j(\{\epsilon\})=f\left(\sum_{a=1}^N\epsilon_a\,
 \theta(\mu_a,\mu_j)\right),
 \end{equation}
where $\theta(\lambda,\mu)$ is some smooth function. Let also
 \begin{equation}\label{A4-struct}
 F(\{\epsilon\})=F\Bigg( \sum_{a=1}^N g^{(1)}(\mu_a)\,\epsilon_a;
 \sum_{\substack{a,b=1 \\ a\not=b} }^N g^{(2)}(\mu_a,\mu_b)\,\epsilon_a\, \epsilon_b;\
 \dots\Bigg),
\end{equation}
where $g^{(1)},g^{(2)}\dots$ are smooth
functions. Then the obtained result takes the form
 \begin{equation}\label{A4-MC-result}
 G_N=\frac{F\bigg( \sum\limits_{a=1}^N g^{(1)}(\mu_a)\,z_a;
 \sum\limits_{a,b=1}^N g^{(2)}(\mu_a,\mu_b)\, z_a\, z_b;\ \dots\bigg)}
 {\det_N\left[\delta_{jk}-\theta(\mu_k,\mu_j)\,
  f'\bigg( \sum\limits_{a=1}^N z_a\, \theta(\mu_a,\mu_j)\bigg)\right]},
 \end{equation}
where
 \begin{equation}\label{A4-sys-1}
 z_j-f\left(
 \sum_{a=1}^N z_a\, \theta(\mu_a,\mu_j)\right)=0.
 \end{equation}
%

\subsection{Continuous case}

Let us now consider a series of multiple integrals
 \begin{equation}\label{A4-MI-ser}
 \hat G
 =\sum_{n=0}^\infty\frac{1}{n!}
  \int\limits_{-q}^q\! \dd^n\lambda
  \prod_{j=1}^n
  \frac{\partial}{\partial \varepsilon_j}
  \left.\prod_{j=1}^n
  f\left(\sum_{a=1}^n\varepsilon_a\, \theta(\lambda_a,\lambda_j)\right)
  \cdot
  F\left( \sum_{a=1}^n g^{(1)}(\lambda_a)\, \varepsilon_a;\
          \dots,\right) \right|_{\varepsilon_j=0}
  \hspace{-1mm},
 \end{equation}
where for brevity we have written explicitly only the first argument
of the function $F$. Consider a discreet analog of these multiple
integrals,
 \begin{equation*}
 \hat G(\Delta)
 =\sum_{n=0}^\infty\frac{\Delta^n}{n!}
  \sum_{\lambda_1,\dots,\lambda_n\in\Lambda}
 \left.\prod_{j=1}^n
  \frac{\partial}{\partial \varepsilon_j}
 \prod_{j=1}^n
   f\left(\sum_{a=1}^n\varepsilon_a\, \theta(\lambda_a,\lambda_j)\right)
 \cdot
   F\left( \sum_{a=1}^n g^{(1)}(\lambda_a)\varepsilon_a;\
  \dots\right) \right|_{\varepsilon_j=0}.
 \end{equation*}
Here each $\lambda_j$ independently runs through a finite set of
values $\Lambda=\{\mu_1,\dots,\mu_N\}$, where
$\mu_{k+1}-\mu_k=\Delta$.  Obviously $\hat G(\Delta) \to \hat G$ at
$\Delta\to 0$.

We now reorganize the inner sums over the lattice into sums over the
number of $\lambda$'s equal to some point of the lattice. Namely, we
sum up with respect to all the possible numbers $s_j$ such that
there are $s_j$ $\lambda$'s equal to the lattice point $\mu_j$.
Obviously there exist $\frac{n!}{s_1! \dots s_N!}$ ways of
realizing such a configuration. It is also evident that, for a configuration such that, for each $j$, there are $s_j$
$\lambda$'s equal to  $\mu_j$, then for any function $\Phi$, one has
 \begin{equation}\label{A4-evid}
 \left.\prod_{j=1}^n\frac{\partial}{\partial \varepsilon_j}
 \cdot
 \Phi\left(\sum_{a=1}^n g^{(1)}(\lambda_a)\,\varepsilon_a;\;
 \dots\right)\right|_{\varepsilon_j=0}
 =
 \left.\prod_{j=1}^N\frac{\partial^{s_j}}{\partial \epsilon_j^{s_j}}
 \cdot
 \Phi\left(\sum_{a=1}^N g^{(1)}(\mu_a)\,\epsilon_a;\;\dots\right)
 \right|_{\epsilon_j=0}.
\end{equation}
Hence,
 \begin{equation}\label{A4-REser-1}
 \hat G(\Delta)=\sum\limits_{n=0}^{\infty}\!
 \sum\limits_{\substack{s_i\geq 0\\ \Sg s_i=n }  }{} \!
 \prod_{j=1}^{N} \frac{\Delta^{s_j}}{s_j!}
 \frac{\partial^{s_j}}{\partial\epsilon_j^{s_j}}\bigg|_{\epsilon_j=0}
 \prod_{j=1}^N
  f^{s_j}\hspace{-1mm}\left(\sum_{a=1}^N\epsilon_a \theta(\mu_a,\mu_j)\!\right)
  F\!\left( \sum_{a=1}^N g^{(1)}(\mu_a)\epsilon_a;\dots\! \right)\! . \!
 \end{equation}
Changing the order of summation and re-scaling
$\epsilon_j\to\Delta\epsilon_j$, we obtain the series \eq{A4-MCser}
with $\phi_j$ and $F$ in the form \eq{A4-phij}, \eq{A4-struct},
in which the functions $\theta$ and $g^{(i)}$ should be replaced by $\Delta
\theta(\mu_a,\mu_j)$ and $\Delta g^{(i)}(\mu_a)$. Then the
continuous limit $\Delta\to 0$ is trivial, and we finally obtain
 \begin{equation}\label{A4-RE-res}
 \hat G=\frac{
 F\bigg( \int\limits_{-q}^q g^{(1)}(\mu)\,z(\mu)\,\dd\mu;\ \dots\bigg)
 }{ \det\bigg[\delta(\lambda-\mu)- \theta(\mu,\lambda)\,
 f'\bigg(
 \int\limits_{-q}^q  \theta(\nu,\lambda)\, z(\nu)\,\dd\nu\bigg)\bigg]}\, ,
 \end{equation}
where the function $z(\mu)$ satisfies the integral equation
\begin{equation}\label{A4-RE-Int-eq}
 z(\mu)=f\!\left(
 \int\limits_{-q}^q  \theta(\lambda,\mu)\, z(\lambda)\,\dd\lambda\right).
 \end{equation}
%

\subsection{Multiple series of multiple integrals}\label{ap-LSD-4}

Consider now a multiple series of the form
\begin{multline}\label{A4-MI-mult}
 \hat G_{1\dots n}
 =\sum_{\ell_1,\dots,\ell_n=0}^\infty
  \prod_{s=1}^n\frac{1}{\ell_s!}
  \int\limits_{-q}^q
 \prod_{s=1}^n \prod_{p=1}^{\ell_s}
 \left[\,\dd\lambda_{s,p}\;\frac{\partial}{\partial
 \epsilon_{s,p}}\,\right] \\
 \times
 \prod_{s=1}^n \prod_{p=1}^{\ell_s} f_s\left(\sum_{t=1}^n\sum_{a=1}^{\ell_t}\epsilon_{t,a}\,
 \theta(\lambda_{t,a},\lambda_{s,p})\right)
 \cdot
 F\left( \sum_{t=1}^n\sum_{a=1}^{\ell_t} g^{(1)}(\lambda_{t,a})\,\epsilon_{t,a};\;
 \dots \right) \Bigg|_{\epsilon_{s,p}=0}.
 \end{multline}
Just like in the previous example we can take the lattice
approximation of the integrals. Then it is easy to see that
 \begin{equation}\label{A4-mMI-res}
 \hat G_{1\dots n}(\Delta)
 =\frac1{\det_{nN}S_{jk}}\,
 F\left(\Delta
 \sum_{s=1}^n\sum_{a=1}^Ng^{(1)}(\mu_a)\,z_{s,a};\;\dots\right)\,,
  \end{equation}
where
\begin{equation}\label{A4-RE-Int-eq1}
 z_{s,a}-f_s\!\left(
 \Delta\sum_{s=1}^n\sum_{b=1}^N z_{s,b}\,\theta(\mu_b,\mu_a)\right)=0,
 \end{equation}
and $\det_{nN}S_{jk}$ is the Jacobian of the system
\eq{A4-RE-Int-eq1}. Taking the sum over $s$ in \eq{A4-RE-Int-eq1}, we
obtain
\begin{equation}\label{A4-RE-Int-eq-1}
 z_{a}-f_{\scriptscriptstyle\Sigma_n}\!\left(
 \Delta\sum_{b=1}^N z_{b}\,\theta(\mu_b,\mu_a)\right)=0,
 \qquad\text{where}\quad
 z_{a}=\sum_{s=1}^nz_{s,a},
 \quad f_{\scriptscriptstyle\Sigma_n}= \sum_{s=1}^n f_s.
 \end{equation}
It is also clear that $\det_{mN}S_{jk}=\det_{N}\tilde S_{jk}$, where
 \begin{equation}\label{A4-det-det}
 \tilde S_{jk}
 =\delta_{jk}-\Delta\,
 \theta(\mu_k,\mu_j)\,f'_{\scriptscriptstyle\Sigma_n}\!
 \left(
 \Delta\sum_{b=1}^N z_{b}\,\theta(\mu_b,\mu_j)\right)
 \end{equation}
is the Jacobian of the system \eq{A4-RE-Int-eq-1}. Thus, in the
continuous limit, we have
 \begin{equation}\label{A4-2-RE-res-1}
 \hat G_{1\dots n}=\frac{
 F\bigg( \int\limits_{-q}^q g^{(1)}(\mu)\,z^{(n)}(\mu)\,\dd\mu;\;\dots\bigg)
  }
 { \det\bigg[\delta(\lambda-\mu)
 - \theta(\mu,\lambda)\, f'_{\scriptscriptstyle\Sigma_n}\bigg(
 \int\limits_{-q}^q  \theta(\nu,\lambda)\,z^{(n)}(\nu)\,\dd\nu\bigg)\bigg]},
 \end{equation}
where $f_{\scriptscriptstyle\Sigma_n}= \sum_{s=1}^n f_k$, and
\begin{equation}\label{A4-RE-Int-eq2}
 z^{(n)}(\mu)=f_{\scriptscriptstyle\Sigma_n}\!\left(
 \int\limits_{-q}^q  \theta(\lambda,\mu)\,
 z^{(n)}(\lambda)\,\dd\lambda\right).
 \end{equation}
%


\def\cprime{$'$}

\end{document}